\documentclass[aps,prb,floatfix,showpacs,twocolumn]{revtex4-1}
\usepackage{amsmath}
\usepackage{graphicx}
\usepackage{latexsym}
\usepackage{bbold}
\usepackage{accents}
\usepackage{bm}
\usepackage{color}

\newcommand{\pvec}[1]{\vec{#1}\mkern2mu\vphantom{#1}'}

\begin{document}

\title{Second-principles method including electron and
  lattice degrees of freedom}

\author{ Pablo Garc\'{\i}a-Fern\'andez,$^{1}$ Jacek
  C. Wojde{\l},$^{2}$, Jorge \'I\~niguez,$^{3,2}$ and Javier
  Junquera$^{1}$}

\affiliation{$^{1}$Departamento de Ciencias de la Tierra y F\'{\i}sica
  de la Materia Condensada, Universidad de Cantabria, Cantabria Campus
  Internacional, Avenida de los Castros s/n, 39005 Santander, Spain
  \\ $^{2}$Institut de Ci\`encia de Materials de Barcelona
  (ICMAB-CSIC), Campus UAB, 08193 Bellaterra, Spain 
  \\ $^{3}$Materials Research and Technology Department, 
  Luxembourg Institute of Science and Technology, 
  Avenue des Hauts-Fourneaux 5, L-4362 Esch/Alzette, Luxembourg}


\begin{abstract}
 We present a first-principles-based ({\em second-principles}) scheme
 that permits large-scale materials simulations including both atomic
 and electronic degrees of freedom on the same footing.
 The method is based on a predictive quantum-mechanical theory --
 e.g., Density Functional Theory -- and its accuracy can be
 systematically improved at a very modest computational cost.
 Our approach is based on dividing the electron density of the system
 into a reference part -- typically corresponding to the system's
 neutral, geometry-dependent ground state -- and a deformation part --
 defined as the difference between the actual and reference densities.
 We then take advantage of the fact that the bulk part of the system's
 energy depends on the reference density alone; this part can be
 efficiently and accurately described by a force field, thus avoiding
 explicit consideration of the electrons.
 Then, the effects associated to the difference density can be treated
 perturbatively with good precision by working in a suitably chosen
 Wannier function basis. Further, the electronic model can be
 restricted to the bands of interest.
 All these features combined yield a very flexible and computationally
 very efficient scheme.
 Here we present the basic formulation of this approach, as well as a
 practical strategy to compute model parameters for realistic
 materials.
 We illustrate the accuracy and scope of the proposed method with two
 case studies, namely, the relative stability of various spin
 arrangements in NiO (featuring complex magnetic interactions in a
 strongly-correlated oxide) and the formation of a two-dimensional
 electron gas at the interface between band insulators LaAlO$_3$ and
 SrTiO$_3$ (featuring subtle electron-lattice couplings and screening
 effects).
 We conclude by discussing ways to overcome the limitations of the
 present approach (most notably, the assumption of a fixed bonding
 topology), as well as its many envisioned possibilities and future
 extensions.
\end{abstract}


\maketitle

\section{Introduction}

 Over the past two decades first-principles methods, in particular
 those based on efficient schemes like Density Functional Theory
 (DFT),\cite{kohn_pra65,parryang,martin_book,jensen_book,Kohanoff_book}
 have become an indispensable tool in applied and fundamental studies
 of molecules, nanostructures, and solids.
 Modern DFT implementations make it possible to compute the energy and
 properties (vibrational, electronic, magnetic) of a compound from
 elementary information about its structure and composition.  Hence,
 in DFT investigations the experimental input can usually be reduced
 to a minimum (the number of atoms of the different chemical species,
 and a first guess for the atomic positions and unit cell lattice vectors).
 Further, the behavior of hypothetical materials can be
 readily investigated, which turns the methods into the ultimate
 predictive tool for application, e.g., in materials design problems.

 However, interpreting or predicting the results of experiments
 requires, in many cases, to go beyond the time and length scales that
 the most efficient DFT methods can reach today. This becomes a very
 stringent limitation when, as it frequently happens, the experiments
 are performed in conditions that are out of the comfort zone of DFT
 calculations, i.e., at ambient temperature, under applied
 time-dependent external fields, out of equilibrium, under the
 presence of (charged-) defects, etc.

 The development of efficient schemes to tackle such challenging
 situations, which are of critical importance in areas ranging from
 Biophysics to Condensed Matter Physics and Materials Science,
 constitutes a very active research field. Especially promising are
 QM/MM multi-scale approaches in which different parts of the system
 are treated at different levels of description: the most
 computationally intensive methods [based on Quantum Mechanics (QM),
   as for example DFT itself] are applied to a region involving a
 relatively small number of atoms and electrons, while a large
 embedding region is treated in a less accurate Molecular Mechanics
 (MM) way (e.g., by using one of many available semi-empirical
 schemes).

 Today's multi-scale implementations tend to rely on semi-empirical
 methods -- like tight-binding\cite{slater_pr54,harrison_book} and
 force-field\cite{lifson_jcp68,brown_cr09} schemes -- that were first
 introduced decades ago.
 In some cases, such schemes are designed to retain DFT-like accuracy
 and flexibility as much as possible. One relevant example are the
 self-consistent-charge density-functional tight-binding (DFTB)
 techniques,\cite{porezag_prb95,matthew_prb89,elstner_prb98} and
 related approaches,\cite{harris_prb85,sankey_prb89,lewis_pssb11}
 which retain the electronic description and permit an essentially
 complete treatment of the compounds. Another relevant example are the
 \emph{effective Hamiltonians} developed to describe ferroelectric
 phase transitions and other functional
 effects;\cite{joannopoulos_prl87,zhong_prl94,zhong_prb95} these are
 purely lattice models (i.e., without an explicit treatment of the
 electrons) based on a physically-motivated coarse-grained
 representation of the material, and have been shown to be very useful
 even in non-trivial situations involving chemical
 disorder\cite{bellaiche_prl00} and magnetoelectric
 effects,\cite{rahmedov_prl12} among others.
 Such methods have demonstrated their ability to tackle many important
 problems (see
 e.g. Refs.~\onlinecite{lewis_pssb11,gaus_jctc11,riccardi_jpcb06,giese_tcha12}
 for the DFTB approach), and constitute very powerful tools.
 Nevertheless, they are limited when it comes to treating situations
 in which the key interactions involve minute energy differences (of
 the order of meV's per atom) and a great accuracy is needed, or where
 a complete atomistic description of the material is required.

 Another aspect in which many approximate approaches fail is in the
 simultaneous treatment, at a similar level of accuracy and
 completeness, of electronic and lattice degrees of freedom. Most
 methods in the literature are strongly biased towards either the
 electronic\cite{hubbard_prsl63,spalek_prb88,kugel_jetp73} or the
 lattice\cite{joannopoulos_prl87,zhong_prl94,zhong_prb95,lifson_jcp68,brown_cr09}
 properties. Further, the few schemes that attempt a realistic,
 simultaneous treatment of both types of variables usually involve
 very coarse-grained
 representations.\cite{millis_prl96,stengel_prl11,kornev_prl07}

 Here we introduce a new scheme to tackle the problem of simulating
 both atomic and electronic degrees of freedom on the same footing,
 with arbitrarily high accuracy, and at a modest computational cost.
 Our scheme will be limited to problems in which it is possible to
 identify an underlying lattice or bonding topology that is not broken
 during the course of the simulation.
 As we will show below, such a {\em fixed-topology} hypothesis permits
 drastic simplifications in the description of the system, yielding a
 computationally efficient scheme whose accuracy can be systematically
 improved to match that of a DFT calculation, if needed.
 Note that, while our assumption of an underlying lattice may seem
 very restrictive at first, in fact it is not. There are myriads of
 problems of great current interest -- ranging from electronic and
 thermal transport phenomena to functional (dielectric, ferroelectric,
 piezoelectric, magnetoelectric) effects and most optical properties
 -- that are perfectly compatible with it. Further, this restriction
 can be greatly alleviated by combining our potentials with DFT
 calculations in a multi-scale scheme, a task for which our models are
 ideally suited.

 In essence, our new scheme relies on the usage of a force field to
 treat interatomic interactions, capable of providing a very accurate
 description of the lattice-dynamical properties of the material of
 interest. In particular, the scheme recently introduced by some of us
 in Ref.~\onlinecite{wojdel_jpcm13} constitutes an excellent choice
 for our purposes, as it takes advantage of the aforementioned
 fixed-topology condition to yield physically transparent models whose
 ability to match DFT results can be systematically improved.

 Then, a critical feature of our approach is to identify such a
 lattice-dynamical model with the description of the material in the
 Born-Oppenheimer surface, i.e., with the DFT solution of the neutral
 system in its electronic ground state. Since the force fields of
 Ref.~\onlinecite{wojdel_jpcm13} do not treat electrons explicitly,
 this identification implies that our models will not tackle the
 description of electronic bonding, as DFTB schemes do. In other
 words, we will not be concerned with modeling the interactions
 responsible for the cohesive energy of the material, or for the
 occurrence of a certain basic lattice topology and structural
 features. Within our scheme, all such properties are simply taken for
 granted, and constitute the starting point of our models.

 Instead, our models focus on the description of electronic states
 that differ from the ground state. These are the truly relevant
 configurations for the analysis of excitations, transport, competing
 magnetic orders, etc. By focusing on them, and by adopting a
 description based on material- (and topology-) specific electronic
 wave functions, we can afford a very accurate treatment of the
 electronic part while keeping the models relatively simple and
 computationally light.

 As we will see, while it bears similarities with DFTB schemes, the
 present approach is ultimately more closely related to Hubbard-like
 methods. Yet, at variance with the usual semi-empirical Hubbard
 Hamiltonians, our models are firmly based on a higher-level
 first-principles theory, treating all lattice degrees of freedom, and
 the relevant electronic ones, with similarly high (perfect at the
 limit) accuracy. The term ``second-principles'', used in the title of
 this article, is meant to emphasize such a solid first-principles
 foundation.

 In this article we introduce the general formal framework of our
 approach, and propose a tentative scheme for a systematic calculation
 of the model variables from first principles. We then describe a
 couple of non-trivial applications that were chosen to highlight the
 flexibility of the models, the great physical insight that they
 provide, and their ability to account for complex properties with
 DFT-like accuracy. Note that, while accuracy will be highlighted, in
 these initial applications we have focused on testing the ability of
 our scheme to tackle challenging situations from the physics
 standpoint, and not so much on building complete models. We will also
 give a brief description of the elemental model-construction and
 model-simulation codes that have been developed in the course of this
 work, to stress the computational efficiency of our scheme. Then, the
 development of a systematic -- and automatic -- strategy for the
 construction of models with predefined accuracy is a technically
 challenging task that remains for future work.

 The article is organized as follows. The main body of theory is
 contained in Sec.~\ref{sec:theory}, where we present the basic
 definitions and formulation of the method, and in
 Sec.~\ref{sec:parameters}, where we describe the procedure to
 generate, from first principles, all the information necessary to
 simulate a material.
 Some technical details on the actual implementation of the method in
 the second-principles {\sc scale-up} code are given in
 Sec.~\ref{sec:scaleup}.
 This is followed, in Sec.~\ref{sec:examples}, by an illustration of
 its capabilities with simulations in two non-trivial systems with
 interactions of very different origin: (i) the magnetic Mott-Hubbard
 insulator NiO and (ii) the two-dimensional electron gas (2DEG) that
 appears at the interface between band insulators 
 LaAlO$_3$ and SrTiO$_3$.
 Finally we present our conclusions and a brief panoramic overview of future
 extensions of the method and possible fields of application in
 Sec.~\ref{sec:end}.

\section{Theory}
\label{sec:theory}

\subsection{Basic definitions}
\label{sec:basicnotions}

 As customarily done in most first-principles schemes, we assume the
 Born-Oppenheimer approximation to separate the dynamics of nuclei and
 electrons. Hence, we consider the positions of the nuclei as fixed
 parameters of the electronic Hamiltonian.
 Our approach will give us access to the potential energy surface
 (PES), i.e., for each configuration of the nuclei, the total energy
 of the system will be computed.

 Our goal is to describe the electrons in the system, and the relevant
 electronic interactions, in the simplest possible way. Hence, we will
 typically focus on valence and conduction states, and will thus work
 with a lattice of \emph{ionic cores} comprised by the nuclei and the
 corresponding core electrons, which will not be modeled
 explicitly. Here we use indistinctively the terms atoms, ions, and
 nuclei to refer to such ionic cores.

 Our method relies on the following key concepts:
 the \emph{reference atomic geometry} (RAG henceforth) and
 the \emph{reference electronic density} (RED in the following).  

 As in the recent development of model potentials for
 lattice-dynamical studies,\cite{wojdel_jpcm13} the first step towards
 the construction of our model is the choice of a RAG, that is, one
 particular configuration of the nuclei that we will use as a
 reference to describe any other configuration.
 In principle, no restrictions are imposed on the choice of RAG.
 However, it is usually convenient to employ the ground state
 structure or, alternatively, a suitably chosen high-symmetry
 configuration. Note that these choices correspond to extrema of the
 PES, so that the corresponding forces on the atoms and stresses on
 the cell are zero. Further, the higher the symmetry of the RAG, the
 fewer the coupling terms needed to describe the system and, in turn,
 the number of parameters to be determined from first principles.

 To describe the atomic configuration of the system we shall adopt a
 notation similar to that of Ref.~\onlinecite{wojdel_jpcm13}. In what
 follows, all the magnitudes related with the atomic structure will be
 labeled using Greek subindices.
 For the sake of simplicity, we shall assume a periodic three
 dimensional infinite crystal, with the lattice cells denoted by
 uppercase letters ($\Lambda$, $\Delta$,...) and the atoms in the cell
 by lowercase letters ($\lambda$, $\delta$,...).
 In this manner, the lattice vector of cell $\Lambda$ is
 $\vec{R}_{\Lambda}$, and the reference position of atom $\lambda$ is
 $\vec{\tau}_{\lambda}$.
 In order to allow for a more compact notation, a cell/atom pair will
 sometimes be represented by a lowercase bold subindex, so that
 $\vec{R}_{\Lambda}\lambda\leftrightarrow\bm{\lambda}$.

 Any possible crystal configuration can be specified by expressing the
 atomic positions, $\vec{r}_{\bm{\lambda}}$, as a distortion of the
 RAG, as
 \begin{equation}
    \vec{r}_{\bm{\lambda}} = \left( \mathbb{1} +
    \overleftrightarrow{\eta} \right) \left( \vec{R}_{\Lambda} +
    \vec{\tau}_{\lambda} \right) + \vec{u}_{\bm{\lambda}},
    \label{eq:atomic_geom}
 \end{equation}
 where $\mathbb{1}$ is the identity matrix,
 $\overleftrightarrow{\eta}$ is the homogeneous strain tensor, and
 $\vec{u}_{\bm{\lambda}}$ is the absolute displacement of atom
 $\lambda$ in cell $\Lambda$ with respect to the strained reference
 structure.

 The second step is the definition of a RED, $n_{0}(\vec{r})$, {\em
   for each possible atomic configuration}.
 Our method relies on the fact that, in most cases, the
 self-consistent electron density, $n(\vec{r})$, will be very close to
 the RED, so that changes in physical properties can be described by
 the small {\em deformation density}, $\delta n(\vec{r})$, defined as
 \begin{equation}
    n(\vec{r})=n_0(\vec{r}) + \delta n(\vec{r}).
    \label{eq:dist_n}
 \end{equation}
 Several remarks are in order about Eq.~(\ref{eq:dist_n}).

 First, with $n(\vec{r})$ we denote the electron density that 
 integrates to the number of electrons (i.e. it is positive). 
 It is trivially related with the charge density
 (in atomic units it just requires making it negative 
 due to the sign of the electronic charge).

 Second, this separation of the charge density into reference and
 deformation contributions is similar to what is commonly found in
 DFTB schemes, and even in first-principles methods.\cite{Soler-02}
 However, this parallelism may be misleading.
 Indeed, it is important to note that we make no assumption on the
 form of the RED. In most cases -- e.g., non-magnetic insulators --,
 it will be most sensible to identify the RED with the ground state
 density of the neutral system. Nevertheless, as will be illustrated
 in Sec.~\ref{sec:nio} for Mott insulator NiO, other choices are also
 possible and very convenient in some situations.

 Third, our RED will typically be an actual solution of the
 electronic problem, as opposed to some approximate density -- e.g., a
 sum of spherical atomic-like densities, possibly taken from the
 isolated-atom solution --, as used in some DFTB
 schemes.\cite{lewis_pssb11}

 Fourth, the concepts of RAG and RED are completely independent: In our
 formalism, we define a RED for every atomic structure accessible by
 the system, and not only for the reference atomic geometry.

 Finally, let us remark, in advance to Sec.~\ref{sec:practical}, that our
 method does not require an explicit calculation of $n_0(\vec{r})$ (or
 any other function in space, for that matter), a feature that allows
 us to reduce the computational cost significantly.

 \begin{figure} [h]
    \begin{center}
       \includegraphics[width=1.0\columnwidth]{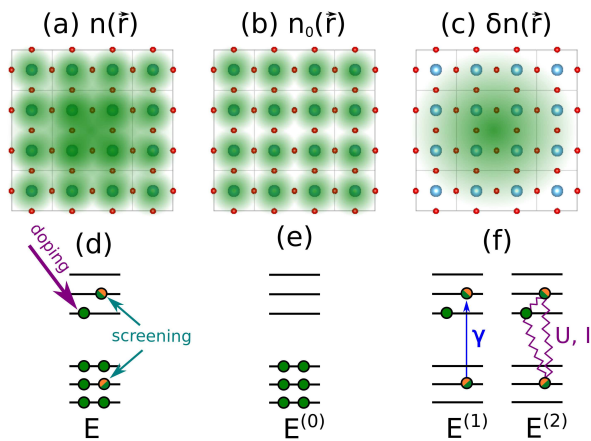}
       \caption{(Color online) Schematic cartoon that represents the key
                physical concepts for the development of the
                second-principles models: the reference atomic structure and
                the reference and deformation electron densities.
                Panels (a)-(c): the meaning of the balls 
                (which represent the position of the
                atoms in a hypothetical semiconductor),
                and the green clouds (which represent charge densities)
                are explained in the main text. Panels (d)-(f): the horizontal
                lines represent the one-electron energy levels obtained
                at the corresponding atomic structures and for the
                reference electronic configuration (neutral ground state). 
                Full green circles represent full occupation of a given state 
                by electrons. Half filled orange/green circles indicate
                partial occupation of a particular level.
                The notations $E^{(0)}$, $E^{(1)}$, and $E^{(2)}$ 
                for the energies are introduced 
                in Sec.~\ref{sec:totalener}.
                The parameters $\gamma$, $U$, and, $I$ are defined in  
                Secs.~\ref{sec:wannierbasis} and \ref{sec:magnetism}.
                Only the case of doping with electrons is sketched.
                Doping with holes would lead to an equivalent picture.
               }
       \label{fig:schemehamil}
    \end{center}
 \end{figure}

 In order to further clarify the concept of RED, let us discuss the
 application of our method to the relevant case of a doped
 semiconductor.
 As sketched in Fig.~\ref{fig:schemehamil}, our hypothetical
 semiconductor is made of two different types of atoms (represented by
 large green and small red balls, respectively) in a square planar
 geometry with a three-atom repeated cell. The RAG corresponds to the
 high-symmetry configuration in which the large atom is located at the
 center of the square, while the small atoms lie at the centers of the
 sides.
 In the neutral (undoped) case, a self-consistent DFT calculation of
 the RAG would yield an electronic configuration with all the valence
 bands occupied and all the conduction bands empty, as illustrated in
 Fig.~\ref{fig:schemehamil}(e). The associated electron density would
 be our RED, $n_{0} (\vec{r})$, represented by the green clouds in
 Fig.~\ref{fig:schemehamil}(b); the associated energy would be
 $E^{(0)}$, using the notation that will be introduced in
 Sec.~\ref{sec:totalener}.

 Now, if we dope the neutral system by adding or removing electrons,
 there will be a response of the electronic cloud, which will tend to
 screen the field caused by the extra charge.
 The doping electron (respectively, hole) will occupy the states at
 the bottom of the conduction band (respectively, top of the valence
 band). The doping-induced charge redistribution can be viewed as
 resulting from an admixture of occupied and unoccupied states of the
 reference neutral configuration. The resulting state, described by
 the total charge density $n(\vec{r})$, is sketched in
 Figs.~\ref{fig:schemehamil}(a) and \ref{fig:schemehamil}(d).
 The difference between the total electronic density and the RED is
 the deformation density $\delta n(\vec{r})$. Such a deformation
 density, which is the key quantity in our scheme, captures both the
 doping and the system's response to it, as sketched in
 Figs.~\ref{fig:schemehamil}(c) and \ref{fig:schemehamil}(f).

\begin{figure} [h]
    \begin{center}
       \includegraphics[width=1.0\columnwidth]{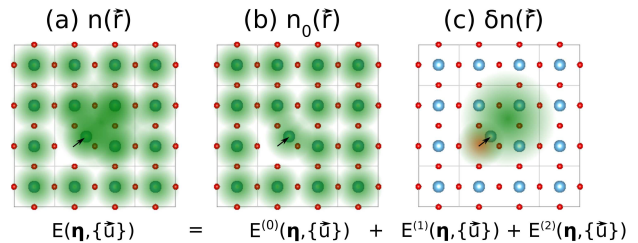}
       \caption{(Color online) Schematic cartoon emphasizing that the
         the division of the electron density into reference and
         deformation parts is carried out for any geometrical
         configuration of the system, as defined by the strain
         $\protect\overleftrightarrow{\eta}$ and atomic displacements 
         $\{ \vec{u}_{\bm{\lambda}} \}$.  The distortion of the reference
         atomic geometry is illustrated by the off-centering of one
         atom (indicated with a black arrow). The atomic distortion
         results in a modified $n_{0}(\vec{r})$ [panel~(b)], as well
         as in additional changes depicted in panel~(c), where the
         green and orange clouds denote positive and negative
         variations in the electronic density.  Symbols have the same
         meaning as in Fig.~\ref{fig:schemehamil}.}
       \label{fig:schemehamil-2}
    \end{center}
 \end{figure}

 Finally, let us further stress the independence between RAG and RED.
 Note that all three quantities $n(\vec{r})$, $n_{0}(\vec{r})$, and
 $\delta n(\vec{r})$ are in fact parametric functions of the atomic
 positions.
 This is illustrated in Fig.~\ref{fig:schemehamil-2}, which sketches a
 case in which one atom is displaced from the RAG.

\subsection{Approximate expression for the energy}
\label{sec:totalener}

 Let us consider an atomic geometry characterized by the homogeneous
 strain tensor, $\overleftrightarrow{\eta}$, and the individual atomic
 displacements, $\{ \vec{u}_{\bm{\lambda}} \}$, as described in
 Eq.~(\ref{eq:atomic_geom}).
 Our main objective is to find a functional form that permits an
 accurate approximation of the DFT total energy at a low computational
 cost. The DFT energy can be written as
 \begin{align}
    E_\text{DFT} = & \sum_{j \vec{k}} o_{j \vec{k}} 
                   \left\langle \psi_{j \vec{k}}
                   \right\vert \hat{t}+v_\text{ext} 
                   \left\vert \psi_{j\vec{k}}\right\rangle
    \nonumber \\ 
                   & +\frac{1}{2}\iint\frac{n(\vec{r})n(\pvec{r})}
                                        {\vert \vec{r}-\pvec{r} \vert}
                   d^3rd^3r^\prime + E_\text{xc}[n] + E_{\text{nn}}.
    \label{eq:dften}
 \end{align}
 In this expression, the first term on the right-hand side includes
 the kinetic energy of a collection of non-interacting electrons as
 computed through the one-particle kinetic energy operator, $\hat{t}$;
 this first term also includes the action of an external potential,
 $v_\text{ext}$, which gathers contributions from the nuclei (or ionic
 cores) and, possibly, other external fields.
 The second term is the Coulomb electrostatic energy, which in the
 context of quantum mechanics of condensed matter systems is also
 referred to as the Hartree term.
 The third term, $E_\text{xc}[n]$, is the so-called exchange and
 correlation functional, which contains the correlation contribution
 to the kinetic energy in the interacting electron system, as well as
 any electron-electron interaction effect beyond the classic Coulomb
 repulsion.
 The last term, $E_{\text{nn}}$, is the nucleus-nucleus electrostatic
 energy.
 Note that Eq.~(\ref{eq:dften}) is written in atomic units, which
 are used throughout the manuscript.
 ($\vert e \vert = m_{e} = \hbar = a_{\rm B} = 1$, where
 $\vert e \vert$ is the magnitude of the electronic charge,
 $m_{e}$ is the electronic mass,
 and $a_{\rm B}$ is the Bohr radius).

 As already mentioned, we assume that the Born-Oppenheimer
 approximation applies, so that the positions of the nuclei can be
 considered as parameters of the Hamiltonian.
 We also assume periodic boundary conditions. (Finite systems can be
 trivially considered by, e.g., adopting a supercell
 approach.\cite{Payne-92})

 Within periodic boundary conditions, the eigenfunctions of the
 one-particle Kohn-Sham equations, $\vert \psi_{j\vec{k}}\rangle$, can
 be written as Bloch states characterized by the wave vector $\vec{k}$
 and the band index $j$, with the occupation of a state given by
 $o_{j\vec{k}}$.
 Note that Eq.~(\ref{eq:dften}) is valid for any geometric structure
 of the system, and we implicitly assume that the total energy
 ($E_\text{DFT}$), the one-particle eigenstates ($\vert
 \psi_{j\vec{k}}\rangle$), and all derived magnitudes (such as the
 electron densities $n$, $n_0$, and $\delta n$) depend on the
 structural parameters $\overleftrightarrow{\eta}$ and $\{
 \vec{u}_{\bm{\lambda}} \}$.

 The total energy of Eq.~(\ref{eq:dften}) is a functional of the
 density which, as described in Eq.~(\ref{eq:dist_n}), can be written
 as the sum of a reference part, $n_{0} (\vec{r})$, and a deformation
 part, $\delta n (\vec{r})$. When we implement this decomposition, the
 linear Coulomb energy term can be trivially dealt with. For the
 non-linear exchange and correlation functional, we follow
 Ref.~(\onlinecite{elstner_prb98}) and expand $E_{\text xc} [n]$
 around the RED as
 \begin{align}
    E_\text{xc}[n] = & E_\text{xc}[n_0] + 
        \int\left.\frac{\delta E_\text{xc}}{\delta n(\vec{r})}\right\vert_{n_0}
            \delta n(\vec{r})d^3r 
        \nonumber \\ 
        & + \frac{1}{2}\iint\left.\frac{\delta^2 E_\text{xc}}
            {\delta n(\vec{r}) \delta n(\pvec{r})}\right\vert_{n_0}
            \delta n(\vec{r}) \delta n(\pvec{r})d^3rd^3r^\prime+ \cdots,
    \label{eq:expxc}
 \end{align}
 where we have introduced functional derivatives of $E_\text{xc}$. In
 principle, Eq.~(\ref{eq:expxc}) is exact if we consider all the
 orders in the expansion. (Expansions like this one are frequently
 found in the formulations of the adiabatic density functional
 perturbation theory.\cite{Gonze-95,Gonze-97})
 In practice, under the assumption of a small deformation density, the
 expansion can be cut at second order.
 As we shall show in Secs.~\ref{sec:magnetism} and 
 \ref{sec:self-consistent-equations}, this
 approximation includes as a particular case the full
 Hartree-Fock-theory; hence, we expect it to be accurate enough for
 our current purposes.

 Within the previous approximation, we can write the total energy as a
 sum of terms coming from the contributions of the deformation density
 at zeroth (reference density), first, and second orders. Formally we
 write
 \begin{equation}
    E_\text{DFT} \approx E = E^{(0)} + E^{(1)} + E^{(2)},
    \label{eq:totale}
 \end{equation}
 where the individual terms have the following form. (A full
 derivation is given in Appendix~\ref{app:a}.)

For the zeroth-order term, $E^{(0)}$, we get
 \begin{align}
    E^{(0)} = & \sum_{j\vec{k}} o_{j\vec{k}}^{(0)} 
                 \left\langle \psi_{j\vec{k}}^{(0)}
                 \right\vert \hat{t} + v_\text{ext} \left\vert
                 \psi_{j\vec{k}}^{(0)}\right\rangle 
    \nonumber \\
              & + \frac{1}{2}\iint\frac{n_0(\vec{r})n_0(\pvec{r})}
                 {\vert \vec{r}-\pvec{r} \vert}d^3rd^3r^\prime + 
                 E_\text{xc}[n_0] + E_{nn}.
    \label{eq:zero}
 \end{align}
 The above equation corresponds, without approximation, to the exact
 DFT energy for the reference density, $n_0$.
 We can choose the RED so that $E^{(0)}$ is the dominant contribution
 to the total energy of the system, and here comes a key advantage of
 our approach: We can compute $E^{(0)}$ by employing a model potential
 that depends only on the atomic positions, where the electrons
 (assumed to remain on the Born-Oppenheimer surface) are integrated
 out.
 This represents a huge gain with respect to other schemes that,
 like the typical DFTB schemes, require an explicit and accurate
 treatment of the electronic interactions yielding the RED {\em as
   well as} solving numerically for $E^{(0)}$ and $n_{0}$ for each
 atomic configuration considered in the simulation.

 The first-order term involves the one-electron excitations 
 as captured by the deformation density,
 \begin{equation}
    E^{(1)} = \sum_{j\vec{k}} \left[ 
               o_{j\vec{k}} \left\langle \psi_{j\vec{k}} \right\vert
               \hat{h}_0 \left\vert \psi_{j\vec{k}}\right\rangle - 
               o_{j\vec{k}}^{(0)} \left\langle \psi_{j\vec{k}}^{(0)}\right\vert 
               \hat{h}_0 \left\vert \psi_{j\vec{k}}^{(0)}\right\rangle
               \right] .
   \label{eq:one}
 \end{equation}
 Here, $\hat{h}_0$ is the Kohn-Sham\cite{kohn_pra65} one-electron
 Hamiltonian defined for the RED,
 \begin{equation}
    \hat{h}_0 = \hat{t} + v_\text{ext} - v_\text{H}(n_0;\vec{r}) + 
                v_\text{xc}[n_0;\vec{r}], 
    \label{eq:h0}
 \end{equation}
 where $v_\text{H}(n_0;\vec{r})$ and $v_\text{xc}[n_0;\vec{r}]$ are,
 respectively, the reference Hartree,
 \begin{equation}\label{eq:pothartree}
    v_\text{H}(n_{0};\vec{r}) = -\int \frac{n_0(\pvec{r})}
                                 {\vert \vec{r}-\pvec{r}\vert} d^3r^\prime,
 \end{equation}
 and exchange-correlation, 
 \begin{equation}\label{eq:potxc}
    v_\text{xc}[n_0;\vec{r}] = \left. \frac{ \delta E_\text{xc} [n] }
                                   {\delta n (\vec{r})}\right\vert_{n_0},
 \end{equation}
 potentials. It is important to note that Eq.~(\ref{eq:one}) is
 different from the one usually employed in DFTB methods (see, for
   example, Refs.~\onlinecite{porezag_prb95} and
   \onlinecite{elstner_prb98}): while typical DFTB schemes include a
 plain sum of one-electron energies, here we deal with the
 \emph{difference} between the value of this quantity for the actual
 system and for the reference one [see sketch in
   Fig.~\ref{fig:schemehamil}(f)]. Such a difference is a much smaller
 quantity, more amenable to accurate calculations.

 Finally, the two-electron contribution from the deformation density,
 $E^{(2)}$, is given by
 \begin{equation}
    E^{(2)} = \frac{1}{2}\int d^3r \int d^3r^\prime g(\vec{r},\pvec{r})
              \delta n(\vec{r}) \delta n(\pvec{r}),
    \label{eq:two}
 \end{equation}
 where the screened electron-electron interaction operator,
 $g(\vec{r},\pvec{r})$, is
 \begin{equation}
     g(\vec{r},\pvec{r}) = \frac{1}
       {\vert\vec{r}-\pvec{r}\vert} + 
       \left.\frac{\delta^2 E_\text{xc} }{\delta n(\vec{r}) 
       \delta n(\pvec{r})}\right\vert_{n_0}.
    \label{eq:scr-ee}
 \end{equation}
 Here, $\delta^2 E_\text{xc}/\delta n (\vec{r}) \delta n
 (\pvec{r})$ captures the effective screening of the
 two-electron interactions due to exchange and correlation.
 The latter magnitude is particularly important in chemistry, as it is
 related to the hardness of a material.\cite{parr_jacs83}

 In summary, in this Section we have shown how a particular splitting
 of the total density, into reference and a deformation parts, allows
 us to expand the DFT energy around $n_{0}$ and as a function of
 $\delta n$.
 This expansion can be truncated at second-order while keeping a high
 accuracy; nevertheless, this approach can be systematically improved
 by including higher-order terms in $\delta n$, in analogy to what is
 done, e.g., in the so-called DFTB3 method.\cite{gaus_jctc11}
 While the general idea is reminiscent of DFTB methods in the
 literature,\cite{porezag_prb95,matthew_prb89,elstner_prb98} our
 scheme has two distinct advantages. On one hand, the zeroth-order
 term can be conveniently parametrized by means of a lattice model
 potential, so that it can be evaluated in a fast and accurate way
 without explicit consideration of the electrons. On the other hand,
 the first-order term is much smaller, and can be calculated more
 accurately, than in usual DFTB schemes, as it takes the form of a
 perturbative correction.

\subsection{Formulation in a Wannier basis}
\label{sec:wannierbasis}

\subsubsection{Choice of Wannier functions}

 Our formulation requires the computation of the matrix elements of
 the Kohn-Sham one-electron Hamiltonian, as defined for the RED, in
 terms of Bloch wavefunctions [Eq.~(\ref{eq:one})], as well as various
 integrals involving the deformation charge density
 [Eq.~(\ref{eq:two})].
 To compute these terms, we will expand the Bloch waves on a basis of
 Wannier-like functions (WFs), $\left\vert \chi_{\bm{a}}\right\rangle$,
 in the spirit of Ref.~\onlinecite{souza_prb01}.
 There are several reasons for our choice of a Wannier basis set over
 the atomic orbitals most commonly used in DFTB
 formulations.\cite{porezag_prb95,matthew_prb89,elstner_prb98,lewis_pssb11}

 First, the Wannier orbitals are naturally adapted to the specific
 material under investigation. In fact, they will be typically
 obtained from a full first-principles simulation of the band
 structure of the target material, which permits a more accurate
 parametrization of the system while retaining a minimal basis set.

 Second, the Wannier functions can be chosen to be spatially
 localized, and several localization schemes are available in the
 literature.\cite{callaway_prb67,teischler_pssb71,satpathy_pssb71,marzari_prb97,souza_prb01,sakura_prb13,wang_prb14}
 The localization will be exploited in our second-principles method to
 restrict the real-space matrix elements to those involving relatively
 close neighbors, as will be explained in Sec.~\ref{sec:parameters}.

 Third, the localized Wannier functions can be chosen to be
 orthogonal.
 Note that methods with non-orthogonal basis functions require the
 calculation of the overlap integrals that have a non-trivial behavior
 as a function of the geometry of the system.
 Moreover, the one-particle Kohn-Sham equations in matrix form become
 a generalized eigenvalue problem, whose solution requires a
 computationally demanding inversion of the overlap matrix.
 The use of orthogonal Wannier functions allows to bypass these
 shortcomings.

 Fourth, the Wannier functions enable a very flexible description of
 the electronic band structure, as they can be constructed to span the
 space corresponding to a specific set of
 bands.\cite{souza_prb01,marzari_rmp12}
 Therefore, the electronic states can be efficiently split into: (i)
 an {\em active} set playing an important role in the properties under
 study; and (ii) a {\em background} set that will be integrated out
 from the explicit treatment.
 For instance, if the problem of interest involves the formation of
 low-energy electron-hole excitons, our active set would be comprised
 by the top-valence and bottom-conduction bands, and we would use the
 corresponding Wannier functions as a basis set.

 Typically, we will start from a set of Bloch-like Hamiltonian
 eigenstates, $\vert \psi_{n \vec{k}}^{(0)} \rangle$, that define a
 manifold of $J$ bands associated to the RED. Then, following e.g.
 the recipe of Ref.~\onlinecite{marzari_prb97}, we have
 \begin{align}
    \vert \chi_{\bm{a}} \rangle \equiv \vert \vec{R}_{A} a \rangle = 
        \frac{V}{\left(2\pi\right)^{3}} \int_{\rm BZ} d \vec{k} \:\:
        e^{-i \vec{k} \cdot \vec{R}_{A}}
        \sum_{m=1}^{J} T_{ma}^{(\vec{k})} \vert \psi_{m \vec{k}}^{(0)} \rangle,
    \label{eq:defWannier}
 \end{align}
 where the Wannier function $\vert \vec{R}_{A} a \rangle$ is labeled
 by the cell $A$ at which it is centered (associated to the lattice
 vector $\vec{R}_{A}$) and by a discrete index $a$.
 Note that we use Latin subindices to label all physical quantities
 related with the electrons; to alleviate the notation, we group in
 the bold symbol $\bm{a}$ both the cell and the discrete index, so
 that $\bm{a}\leftrightarrow \vec{R}_{A}a$.
 In Eq.~(\ref{eq:defWannier}), $V$ is the volume of the primitive unit
 cell, the integral is carried out over the whole Brillouin zone ($\rm
 BZ$), the index $m$ runs over all the $J$ bands of the manifold, and
 the $\bm{T}^{(\vec{k})}$ matrices represent unitary transformations
 among the $J$ Bloch orbitals at a given wavevector.

 \begin{figure*} [t]
    \begin{center}
       \includegraphics[width=2.0\columnwidth]{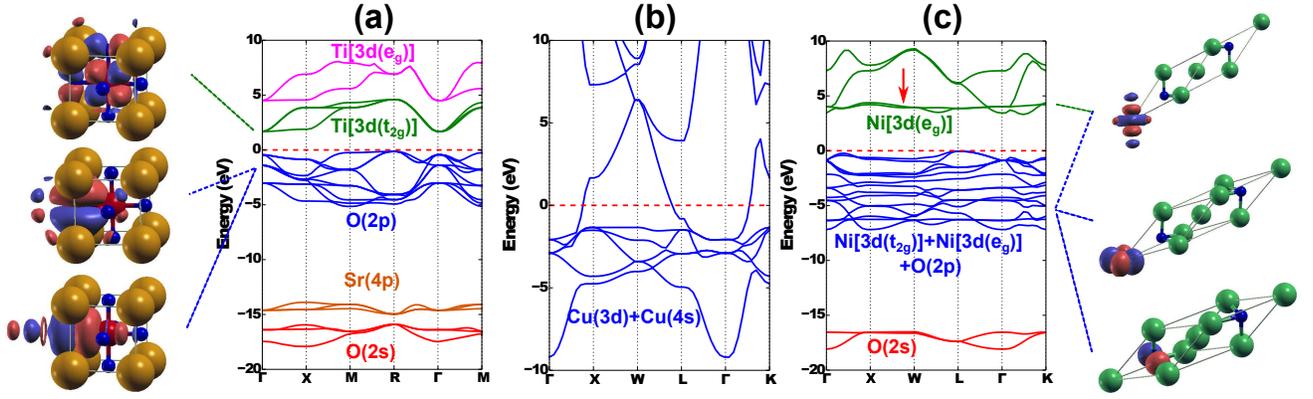}
       \caption{(Color online) Band structure, showing different band
         entanglements, in three archetypal cases: (a) insulating
         SrTiO$_3$, (b) metallic Cu, and (c) antiferromagnetic NiO.
         The groups of entangled bands are separated by energy gaps
         and colored differently. For SrTiO$_3$: bands with dominant
         O($2s$), Sr($4p$), O($2p$), Ti[$3d(t_{2g})$], and
         Ti[$3d(e_{g})$] characters are plotted, respectively in red,
         orange, blue, green, and magenta. For Cu: all the bands are
         entangled. For NiO: some weakly-disperssive bands at the top
         of the valence and bottom of the conduction regions
         (indicated by an arrow) share the Ni[$3d(e_{g})$]
         character. On the leftmost and rightmost edges of the figure,
         the spatial shape of some of the Wannier functions for
         SrTiO$_3$ and NiO are displayed, respectively. Isosurfaces
         corresponding to positive and negative values of the MLWFs
         are plotted with different colors. In these diagrams, golden,
         blue, red, and green spheres represent, respectively, Sr, O,
         Ti, and Ni ions. For NiO, the atoms in the cell used to
         simulate the antiferromagnetic ground state are shown. Dashed
         red lines mark the Fermi energy of the metal and the top of
         the valence bands of the insulators; in all cases such level
         is taken as the zero of energy.}
       \label{fig:entanglement}
    \end{center}
 \end{figure*}

 Figure~\ref{fig:entanglement} shows three paradigmatic examples: a
 non-magnetic insulator [bulk SrTiO$_{3}$, Fig.~\ref{fig:entanglement}(a)], 
 a non-magnetic metal [bulk Cu, Fig.~\ref{fig:entanglement}(b)], 
 and an antiferromagnetic insulator [bulk NiO, Fig.~\ref{fig:entanglement}(c)].

 In the first case, the valence bands are well separated in energy
 from other bands; further, they have a well-defined character
 strongly reminiscent of the corresponding atomic orbitals.
 More precisely, three isolated manifolds corresponding to the
 occupied valence bands -- with dominant O-$2s$, Sr-$4p$, and O-$2p$
 character, respectively -- are clearly visible; these bands are
 centered around 17~eV, 15~eV, and 3~eV below the valence-band top,
 respectively.
 The Bloch eigenstates for these bands can be directly used to compute
 the corresponding localized Wannier functions following the scheme of
 Ref.~\onlinecite{marzari_prb97} or similar ones.
 In contrast, the bottom conduction bands of SrTiO$_{3}$ have a
 dominant Ti-$t_{2g}$ character, but overlap in energy with
 higher-lying (Ti-$e_{g}$) conduction bands.
 The situation is even more complicated in the cases of
 Figs.~\ref{fig:entanglement}(b) and \ref{fig:entanglement}(c), where
 the critical bands -- i.e., those around the Fermi energy in the case
 of Cu, and those comprising the Ni-$3d$ manifold in the case of NiO
 -- are strongly entangled with other states. In such cases, we may
 need to use a disentanglement method -- like e.g. the one proposed in
 Ref.~\onlinecite{souza_prb01} -- to identify a minimal active
 manifold.

 Note that the inverse transformation from Wannier to Bloch functions
 reads
 \begin{equation}
    \vert \psi_{j \vec{k}}^{(0)} \rangle = 
          \sum_{{\bm a}} c_{j a \vec{k}}^{(0)} e^{i \vec{k} \cdot \vec{R}_{A}} 
          \vert \chi_{\bm a} \rangle,
    \label{eq:defBloch}
 \end{equation}
 where the connection between the $c_{j a \vec{k}}^{(0)}$ coefficients
 and the transformation matrices in Eq.~(\ref{eq:defWannier}) is given
 in Appendix~\ref{app:b}.

 The Wannier functions corresponding to the RED
 [Fig.~\ref{fig:schemehamil}(b)] form a complete basis of the Hilbert
 space. Hence, we can use them to represent any {\em perturbed}
 electronic configuration of the system [e.g., the one sketched in
   Fig.~\ref{fig:schemehamil}(a)] as
 \begin{equation}
    \vert \psi_{j \vec{k}} \rangle = 
          \sum_{{\bm a}} c_{j a \vec{k}} e^{i \vec{k} \cdot \vec{R}_{A}} 
          \vert \chi_{\bm a} \rangle ,
    \label{eq:defBlochanyden}
 \end{equation}
 where the sum can be extended to as many bands as needed to
 accurately describe the phenomenon of interest. (As in the example of
 Fig.~\ref{fig:schemehamil}, this might be the addition of an electron
 and the associated screening.)

 Finally, note that the Wannier function basis is implicitly dependent
 on the structural parameters $\overleftrightarrow{\eta}$ and $\{
 \vec{u}_{\bm{\lambda}} \}$, and it should be recomputed for every new
 RED corresponding to varying atomic positions. Ultimately, our models
 will capture all such effects implicitly in the electron-lattice
 coupling terms, whose calculation is described in
 Sec.~\ref{sec:electron-lattice}.

 Also, henceforth we will assume that each and every one of the WFs in
 our basis can be unambiguously associated with a particular atom at
 (around) which it is centered. Further, we will use the notation
 $\bm{a}\in \bm{\lambda}$ to refer to all the WFs associated to atom
 $\lambda$ in cell $\Lambda$, an identification that will be necessary
 when discussing our treatment of electrostatic couplings in
 Sec.~\ref{sec:elect}. 
 
\subsubsection{Equations in a Wannier basis} 
\label{sec:denwann}

 Using Eq.~(\ref{eq:defBlochanyden}), we can write the electron
 density $n (\vec{r})$ in terms of the Wannier functions,
 \begin{align}
    n(\vec{r}) & = \sum_{j\vec{k}} o_{j\vec{k}} 
                   \vert \psi_{j\vec{k}} (\vec{r}) \vert^2
                 = \sum_{j\vec{k}} o_{j\vec{k}} 
                   \psi^{*}_{j\vec{k}} (\vec{r}) \psi_{j\vec{k}} (\vec{r})
    \nonumber \\
    & = \sum_{j \vec{k}} \sum_{\bm{a}\bm{b}}
        o_{j\vec{k}} c_{ja\vec{k}}^{*} c_{jb\vec{k}}
        e^{i\vec{k}(\vec{R}_B-\vec{R}_A)} 
        \chi_{\bm{a}}(\vec{r})\chi_{\bm{b}}(\vec{r})
    \nonumber \\
    & = \sum_{\bm{a}\bm{b}} d_{\bm{a}\bm{b}}          
        \chi_{\bm{a}}(\vec{r})\chi_{\bm{b}}(\vec{r}).
     \label{eq:densW}
 \end{align}
 We can assume we will work with real Wannier
 functions\cite{marzari_rmp12} and therefore drop the complex
 conjugates in our equations.  In Eq.~(\ref{eq:densW}) we have
 introduced a reduced density matrix,
 \begin{equation}
    d_{\bm{a}\bm{b}}= \sum_{j\vec{k}} o_{j\vec{k}} c_{ja\vec{k}}^{*} c_{jb\vec{k}}
         e^{i\vec{k}(\vec{R}_B-\vec{R}_A)}, 
    \label{eq:defdab}
 \end{equation}
 which, following the nomenclature of Ref.~\onlinecite{marzari_prl97},
 will be referred to as the {\em occupation matrix} for the WFs. This
 occupation matrix has the usual properties, including periodicity
 when the Wannier functions are displaced by the same lattice vector
 in real space.

 Equation~(\ref{eq:densW}) can similarly be applied to the RED,
 \begin{equation}
    n_0(\vec{r}) = \sum_{\bm{a}\bm{b}} d_{\bm{a}\bm{b}}^{(0)}
    \chi_{\bm{a}}(\vec{r})\chi_{\bm{b}}(\vec{r}),
    \label{eq:densW0}
 \end{equation}
 where the calculation of the occupation matrix is performed with the
 coefficients of the Bloch functions that define the reference
 electronic density, $c_{j\alpha\vec{k}}^{(0)}$, as in
 Eq.~(\ref{eq:defBloch}).

 In order to quantify the difference between the two densities defined
 in Eqs.~(\ref{eq:densW}) and (\ref{eq:densW0}), we introduce a {\em
 deformation occupation matrix},
 \begin{equation}
    D_{\bm{a}\bm{b}}=d_{\bm{a}\bm{b}}-d_{\bm{a}\bm{b}}^{(0)},
    \label{eq:Ddens}
 \end{equation}
 which will be the central magnitude in our calculations. Now the
 deformation density can be written as
 \begin{equation}
    \delta n(\vec{r}) = \sum_{\bm{a}\bm{b}} D_{\bm{a}\bm{b}}  
     \chi_{\bm{a}}(\vec{r})\chi_{\bm{b}}(\vec{r}).
 \end{equation}
 Using these definitions, we can rewrite the $E^{(1)}$ and $E^{(2)}$
 energy terms. Introducing Eq.~(\ref{eq:Ddens}) into
 Eqs.~(\ref{eq:one}) and (\ref{eq:two}) we get
 \begin{align} 
    E^{(1)} = &  \sum_{j\vec{k}}  \left[ 
               o_{j\vec{k}} \left\langle \psi_{j\vec{k}} \right\vert
               \hat{h}_0 \left\vert \psi_{j\vec{k}}\right\rangle - 
               o_{j\vec{k}}^{(0)} \left\langle 
               \psi_{j\vec{k}}^{(0)} \right\vert 
               \hat{h}_0 \left\vert \psi_{j\vec{k}}^{(0)}\right\rangle
               \right]
    \nonumber \\
            = & \sum_{j \vec{k}} \left[ 
                o_{j \vec{k}} \sum_{\bm{a}\bm{b}} 
                c^{*}_{aj\vec{k}}c_{bj\vec{k}} 
                e^{i\vec{k}(\vec{R}_B-\vec{R}_A)}
                \langle \chi_{\bm{a}} \vert \hat{h}_0 \vert 
                \chi_{\bm{b}}\rangle
                \right.
    \nonumber \\
            & - \left.
                o^{(0)}_{j \vec{k}} \sum_{\bm{a}\bm{b}} 
                \left(c^{(0)}_{aj\vec{k}}\right)^{*} c^{(0)}_{bj\vec{k}} 
                e^{i\vec{k}(\vec{R}_B-\vec{R}_A)}
                \langle \chi_{\bm{a}} \vert \hat{h}_0 \vert 
                \chi_{\bm{b}}\rangle
                \right]
    \nonumber \\
            = & \left[ 
                \sum_{\bm{a}\bm{b}} 
                d_{\bm{a}\bm{b}}
                \langle \chi_{\bm{a}} \vert \hat{h}_0 \vert 
                \chi_{\bm{b}}\rangle
                - \sum_{\bm{a}\bm{b}} 
                d^{(0)}_{\bm{a}\bm{b}}
                \langle \chi_{\bm{a}} \vert \hat{h}_0 \vert 
                \chi_{\bm{b}}\rangle
                \right]
    \nonumber \\
            = & \sum_{\bm{a}\bm{b}} D_{\bm{a}\bm{b}}
                \gamma_{\bm{a}\bm{b}},
    \label{eq:eone}
 \end{align} 
 and
 \begin{align} 
    E^{(2)} & = \frac{1}{2}\int d^3r \int d^3r^\prime g(\vec{r},\pvec{r})
                \delta n(\vec{r}) \delta n(\pvec{r})
    \nonumber \\
            & = \frac{1}{2} \sum_{\bm{a}\bm{b}} 
                            \sum_{\bm{a}^\prime\bm{b}^\prime} 
                D_{\bm{a}\bm{b}}D_{\bm{a}^\prime\bm{b}^\prime}
                \langle \chi_{\bm{a}} \chi_{\bm{a}^\prime} 
                \vert \hat{g} \vert 
                \chi_{\bm{b}} \chi_{\bm{b}^\prime}\rangle
    \nonumber \\
            & = \frac{1}{2} \sum_{\bm{a}\bm{b}} 
                            \sum_{\bm{a}^\prime\bm{b}^\prime} 
                D_{\bm{a}\bm{b}}D_{\bm{a}^\prime\bm{b}^\prime}
                U_{\bm{a}\bm{b}\bm{a}^\prime\bm{b}^\prime},
    \label{eq:etwo}
 \end{align} 
 respectively, where $\gamma_{\bm{a}\bm{b}}$ and
 $U_{\bm{a}\bm{b}\bm{a}^\prime\bm{b}^\prime}$ are the primary
 parameters that define our electronic model.
 These parameters can be obtained, respectively, from the integrals of
 the one- and two-electron operators computed in DFT simulations, as
 \begin{align}
    \gamma_{\bm{a}\bm{b}} & = \left\langle \chi_{\bm{a}} \right\vert 
    \hat{h}_0 \left\vert \chi_{\bm{b}} \right\rangle
    \nonumber \\
    & = \int d^3r \: \chi_{\bm{a}}(\vec{r}) 
                  \: h_0 (\vec{r})
                  \: \chi_{\bm{b}}(\vec{r}),
    \label{eq:gamma}
 \end{align}
 and
 \begin{align}
    U_{\bm{a}\bm{b}\bm{a}^\prime\bm{b}^\prime} 
    & = \langle \chi_{\bm{a}} \chi_{\bm{a}^\prime} 
        \vert \hat{g} \vert 
        \chi_{\bm{b}} \chi_{\bm{b}^\prime}\rangle
    \nonumber \\
    & = \int d^3r \int d^3r^\prime \chi_{\bm{a}}(\vec{r})\chi_{\bm{b}}(\vec{r}) 
        \chi_{\bm{a}^\prime}(\pvec{r})
        \chi_{\bm{b}^\prime}(\pvec{r}) g(\vec{r},\pvec{r}).
    \label{eq:Uelec}
 \end{align}
 Alternatively, they can be fitted so that the model reproduces a
 training set of first-principles data.

\subsection{Magnetic systems}
\label{sec:magnetism}

 The above expressions are valid for systems without spin
 polarization. The procedure to construct the energy for magnetic
 cases is very similar, but there are subtleties pertaining the choice
 of RED.

 In principle, one could use a RED corresponding to a particular
 realization of the spin order, e.g., the anti-ferromagnetic ground
 state for a typical magnetic insulator, or the ferromagnetic ground
 state for a typical magnetic metal. However, such a choice is likely
 to result in a less accurate description of other spin arrangements,
 which would hamper the application of the model to investigate
 certain phenomena (e.g., a spin-ordering transition).

 Alternatively, one might adopt a non-magnetic RED around which to
 construct the model. Such a RED might correspond to an actual {\em
 computable} state: for example, it could be obtained from a
 non-magnetic DFT simulation in which a perfect pairing of spin-up and 
 spin-down electrons is imposed.
 Further, as we will see below for the case of NiO,
 in some cases it is possible and convenient to consider a {\em
 virtual RED} whose character can be inspected \emph{a
 posteriori}. This latter option follows the spirit of the usual
 approach to the construction of spin-phonon effective
 Hamiltonians,~\cite{cazorla13} where the parameters defining the
 reference state cannot be computed directly from DFT, but are
 effectively {\em fitted} by requesting the model to reproduce the
 properties of specific spin arrangements.

 In the following we assume a non-magnetic RED, and present an
 otherwise general formulation.

 The $E^{(0)}$ and $E^{(1)}$ terms thus describe the lattice and
 one-electron energetics corresponding to the non-magnetic RED, and do
 not capture any effect related with the spin polarization. In
 contrast, the screened electron-electron interaction operator
 [Eq.~(\ref{eq:scr-ee})] is spin dependent and equal to
 \begin{equation}
   g( \vec{r}, \pvec{r}, s, s^\prime ) =
       \frac{1}{\vert\vec{r}-\pvec{r}\vert}
   + \left.\frac{\delta^2 E_\text{xc}}
                {\delta n(\vec{r},s)
                 \delta n(\pvec{r},s^\prime)}\right\vert_{n_0}.
   \label{eq:goper}
 \end{equation}
 where $s$ and $s^\prime$ are spin indices that can take ``up'' or
 ``down'' values which we denote, respectively, by $\uparrow$ and
 $\downarrow$ symbols.
 This distinction in the screened electron-electron operator leads us
 to introduce two kinds of $U$ parameters,
 \begin{eqnarray}
    U_{ \bm{a} \bm{b} \bm{a}^\prime \bm{b}^\prime } ^\text{par} & = 
      \int d^3r \int d^3r^\prime \chi_{\bm{a}}(\vec{r}) \chi_{\bm{b}}(\vec{r}) 
      \chi_{\bm{a}^\prime}(\pvec{r}) 
                       \chi_{\bm{b}^\prime}(\pvec{r}) 
      g(\vec{r},\pvec{r},\uparrow,\uparrow) 
    \nonumber\\
      & = 
      \int d^3r \int d^3r^\prime \chi_{\bm{a}}(\vec{r}) \chi_{\bm{b}}(\vec{r}) 
      \chi_{\bm{a}^\prime}(\pvec{r})
                       \chi_{\bm{b}^\prime}(\pvec{r}) 
      g(\vec{r},\pvec{r},\downarrow,\downarrow)
    \nonumber\\
    \label{eq:Upar}
 \end{eqnarray}
 and
 \begin{eqnarray}
    U_{ \bm{a} \bm{b} \bm{a}^\prime \bm{b}^\prime }^\text{anti} & =
      \int d^3r \int d^3r^\prime \chi_{\bm{a}}(\vec{r})\chi_{\bm{b}}(\vec{r}) 
      \chi_{\bm{a}^\prime}(\pvec{r})
                       \chi_{\bm{b}^\prime}(\pvec{r}) 
      g(\vec{r},\pvec{r},\uparrow,\downarrow)
    \nonumber\\
      & = 
      \int d^3r \int d^3r^\prime \chi_{\bm{a}}(\vec{r}) \chi_{\bm{b}}(\vec{r}) 
      \chi_{\bm{a}^\prime}(\pvec{r})
                       \chi_{\bm{b}^\prime}(\pvec{r}) 
      g(\vec{r},\pvec{r},\downarrow,\uparrow),
    \nonumber\\
    \label{eq:Uanti}
 \end{eqnarray}
 which describe, respectively, the interactions between electrons with
 parallel ($U^\text{par}$) and antiparallel ($U^\text{anti}$) spins.
 As a consequence, $E^{(2)}$ in spin-polarized systems is

 \begin{align}
    E^{(2)} & = \sum_{s,s^\prime} \frac{1}{2} \int d^3r \int d^3r^\prime
    g ({r},\pvec{r},s,s^\prime) \delta n(\vec{r},s) \delta n(\pvec{r},s^\prime),
    \label{eq:e2spin}
 \end{align}

 \noindent where 

 \begin{equation}
    \delta n(\vec{r},s) = \sum_{\bm{a}\bm{b}} D_{\bm{a}\bm{b}}^s  
     \chi_{\bm{a}}(\vec{r})\chi_{\bm{b}}(\vec{r}),
    \label{eq:defdenspin}
 \end{equation}

 \noindent and $D_{\bm{a}\bm{b}}^s$ is the deformation occupation matrix for
 the $s$~spin-channel, defined for the up and down spins as
 \begin{align}
       D_{\bm{a}\bm{b}}^\uparrow & = d_{\bm{a}\bm{b}}^{\uparrow} - 
                                          \frac{1}{2} d_{\bm{a}\bm{b}}^{(0)}
       \label{eq:Ddensup}
 \end{align}
 and
 \begin{align}
       D_{\bm{a}\bm{b}}^\downarrow & = d_{\bm{a}\bm{b}}^{\downarrow} - 
                                       \frac{1}{2} d_{\bm{a}\bm{b}}^{(0)},
       \label{eq:Ddensdown}
 \end{align}
 respectively. Replacing Eqs.~(\ref{eq:defdenspin})-(\ref{eq:Ddensdown}) 
 into Eq.~(\ref{eq:e2spin}),

 \begin{align}
    E^{(2)} = & \frac{1}{2} \sum_{s,s^\prime} 
      \sum_{\bm{a}\bm{b}}
      \sum_{\bm{a}^\prime\bm{b}^\prime}
      D_{\bm{a}\bm{b}}^s D_{\bm{a}^\prime\bm{b}^\prime}^{s^{\prime}}
   \nonumber \\
      & \int d^3r \int d^3r^\prime
      \chi_{\bm{a}}(\vec{r})\chi_{\bm{b}}(\vec{r})
      \chi_{\bm{a}^\prime}(\pvec{r})\chi_{\bm{b}^\prime}(\pvec{r})
      g({r},\pvec{r},s,s^\prime) 
   \nonumber \\ 
      = & \frac{1}{2}
      \left\{
      \sum_{\bm{a}\bm{b}}
      \sum_{\bm{a}^\prime\bm{b}^\prime}
       \left[
         D_{\bm{a}\bm{b}}^\uparrow D_{\bm{a}^\prime\bm{b}^\prime}^\uparrow+
         D_{\bm{a}\bm{b}}^\downarrow D_{\bm{a}^\prime\bm{b}^\prime}^\downarrow
       \right]
    \right.
       U_{\bm{a}\bm{b}\bm{a}^\prime\bm{b}^\prime}^\text{par}+
    \nonumber \\
    & 
    \left.    \left[
         D_{\bm{a}\bm{b}}^\uparrow D_{\bm{a}^\prime\bm{b}^\prime}^\downarrow +
         D_{\bm{a}\bm{b}}^\downarrow D_{\bm{a}^\prime\bm{b}^\prime}^\uparrow
       \right]
       U_{\bm{a}\bm{b}\bm{a}^\prime\bm{b}^\prime}^\text{anti}
     \right\}
 \label{eq:e2pol}
 \end{align}

 For physical clarity, and to establish the link of
 Eqs.~(\ref{eq:Upar}) and (\ref{eq:Uanti}) with Eq.~(\ref{eq:Uelec}),
 it is convenient to write $U^\text{par}$ and $U^\text{anti}$ in terms
 of Hubbard- ($U$) and Stoner- ($I$) like parameters:
 \begin{eqnarray}
    U^\text{par}_{ \bm{a} \bm{b} \bm{a}^\prime \bm{b}^\prime}
    & = &
    U_{\bm{a} \bm{b} \bm{a}^\prime \bm{b}^\prime} - 
    I_{\bm{a} \bm{b} \bm{a}^\prime \bm{b}^\prime}
    \label{eq:udef1}\\
    U^\text{anti}_{ \bm{a} \bm{b} \bm{a}^\prime \bm{b}^\prime}
    & = &
    U_{\bm{a} \bm{b} \bm{a}^\prime \bm{b}^\prime} + 
    I_{\bm{a} \bm{b} \bm{a}^\prime \bm{b}^\prime}, 
    \label{eq:udef2}
 \end{eqnarray}
 so
 \begin{eqnarray}
    U_{\bm{a} \bm{b} \bm{a}^\prime \bm{b}^\prime} & = & \frac{1}{2}
    \left( U^\text{par}_{ \bm{a} \bm{b} \bm{a}^\prime \bm{b}^\prime}
         +  U^\text{anti}_{ \bm{a} \bm{b} \bm{a}^\prime \bm{b}^\prime} \right),
    \label{eq:udef3}\\
    I_{\bm{a} \bm{b} \bm{a}^\prime \bm{b}^\prime} & = & \frac{1}{2}
    \left( U^\text{anti}_{ \bm{a} \bm{b} \bm{a}^\prime \bm{b}^\prime}
         -  U^\text{par}_{ \bm{a} \bm{b} \bm{a}^\prime \bm{b}^\prime} \right).
    \label{eq:udef4}
 \end{eqnarray}
 It is also convenient to introduce
 \begin{equation}
      D_{\bm{a}^\prime \bm{b}^\prime}^U =
      D_{\bm{a}^\prime \bm{b}^\prime}^{\uparrow} +
      D_{\bm{a}^\prime \bm{b}^\prime}^{\downarrow} ,
   \label{eq:D_U}                
 \end{equation} 
 and
\begin{equation}
    D_{\bm{a}^\prime \bm{b}^\prime}^I=
    D_{\bm{a}^\prime \bm{b}^\prime}^{\uparrow}-
    D_{\bm{a}^\prime \bm{b}^\prime}^{\downarrow},
  \label{eq:D_I}                
\end{equation} 
 so that Eq.~(\ref{eq:e2pol}) can be rewritten as:
 \begin{align}
    E^{(2)} = 
      \frac{1}{2} \sum_{\bm{a}\bm{b}}
                  \sum_{\bm{a}^\prime\bm{b}^\prime} 
    &
    \left\{
           D_{\bm{a}\bm{b}}^U D_{\bm{a}^\prime\bm{b}^\prime}^U 
       U_{\bm{a}\bm{b}\bm{a}^\prime\bm{b}^\prime} - 
       D_{\bm{a}\bm{b}}^I  D_{\bm{a}^\prime\bm{b}^\prime}^I
       I_{\bm{a}\bm{b}\bm{a}^\prime\bm{b}^\prime}
    \right\} .
    \label{eq:etwoui}
 \end{align}
 Note that the value of $U$ in Eqs.~(\ref{eq:udef1}) and
 (\ref{eq:udef2}) is consistent with the one in Eq.~(\ref{eq:Uelec})
 if we consider a non-spin-polarized density
 ($D_{\bm{a}\bm{b}}^\downarrow=D_{\bm{a}\bm{b}}^\uparrow$). In
 addition, note that the newly introduced constant
 $I_{\bm{a}\bm{b}\bm{a}^\prime\bm{b}^\prime}$ only plays a role in
 spin-polarized systems and is necessarily responsible for magnetism.

\subsubsection*{Connection with other schemes} 

 The two-electron interaction constants -- $U$ and $I$ defined in
 Eqs.~(\ref{eq:udef1}) and (\ref{eq:udef2}), respectively -- are
 formally similar to the four-index integrals typically found in
 Hartree-Fock theory\cite{jensen_book} and can be chosen to completely
 match this approach.

 However, one should note that the electron-electron interaction in
 our Hubbard-like and Stoner-like constants is not the bare one, but
 is screened by the exchange-correlation potential associated to the
 reference density, $n_0$ [see Eq.~(\ref{eq:goper})]. This fact brings
 our formulation closer to the so-called DFT+{\sl U}
 \cite{anisimov_jpcm97,liechtenstein_prb95} and
 GW\cite{hedin_pra65,aryasetiawan_rpp98} methods.

 Looking in more detail at our expressions for $U$ and $I$, 
 \begin{align}
    U_{ \bm{a} \bm{b} \bm{a}^\prime \bm{b}^\prime } & =
      \int d^3r \int d^3r^\prime \chi_{\bm{a}}(\vec{r})\chi_{\bm{b}}(\vec{r}) 
      \chi_{\bm{a}^\prime}(\pvec{r})
                       \chi_{\bm{b}^\prime}(\pvec{r}) 
      g_U(\vec{r},\pvec{r}) \label{eq:u_integral}\\
    I_{ \bm{a} \bm{b} \bm{a}^\prime \bm{b}^\prime } & =
      \int d^3r \int d^3r^\prime \chi_{\bm{a}}(\vec{r})\chi_{\bm{b}}(\vec{r}) 
      \chi_{\bm{a}^\prime}(\pvec{r})
                       \chi_{\bm{b}^\prime}(\pvec{r}) 
      g_I(\vec{r},\pvec{r}), \label{eq:i_integral}     
 \end{align}
 we find that they are very similar to those of $U^\text{par}$
 [Eq.~(\ref{eq:Upar})] and $U^\text{anti}$ [Eq.~(\ref{eq:Uanti})],
 except that the operator involved in the double integral is,
 respectively,

 \begin{align}
    g_U(\vec{r},\pvec{r}) & = \frac{1}{\vert\vec{r}-\pvec{r}\vert} +
    \nonumber \\
    & \frac{1}{2}
      \left[
      \left. \frac{\delta^2 E_\text{xc}}
          {\delta n(\vec{r},\uparrow) \delta n(\pvec{r},\uparrow)}\right\vert_{n_0}+
      \left.\frac{\delta^2 E_\text{xc} }
          {\delta n(\vec{r},\uparrow) \delta n(\pvec{r},\downarrow)}\right\vert_{n_0}
      \right],
    \label{eq:gu}
 \end{align}
 and
 \begin{align}
    g_I(\vec{r},\pvec{r})& =
    \nonumber \\
    & \frac{1}{2} 
    \left[
    \left.\frac{\delta^2 E_\text{xc} }
        {\delta n(\vec{r},\uparrow) \delta n(\pvec{r},\downarrow)}\right\vert_{n_0}-
    \left.\frac{\delta^2 E_\text{xc} }
        {\delta n(\vec{r},\uparrow) \delta n(\pvec{r},\uparrow)}\right\vert_{n_0}
    \right].
    \label{eq:gi}
 \end{align}
 Thus, we see that $U$ contains the classical Hartree interactions,
 screened by exchange and correlation. Moreover, from
 Eq.~(\ref{eq:etwoui}) we see that $U$, as used here, is related with
 the deformation occupation matrix $D^{U}$, that captures the total
 change of the electron density (i.e., the sum of the deformation
 occupation matrix for both components of spins).
 Therefore, it is consistent with the usual definition $U=d^2 E/d
 n^2$, i.e., it quantifies the energy needed to add or remove
 electrons.

 On the other hand,
 Stoner's\cite{stoner_prsa36,gunnarson_jpf76,stohr_book} $I$ only
 includes terms with quantum origin. In particular $g_I$ provides the
 difference in interaction between electrons with parallel and
 antiparallel spins.

\subsection{Electrostatics}
\label{sec:elect}

 \begin{figure} [h]
    \begin{center}
       \includegraphics[width=1.0\columnwidth]{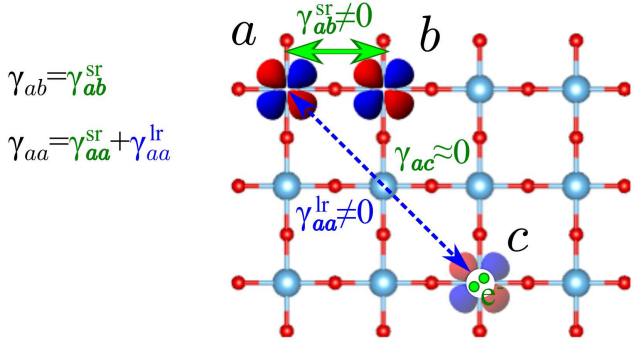}
       \caption{(Color online) Schematic representation of the near-
         and far-field interactions. The shape of the orbitals
         (represented here by two $t_{2g}$-like WFs labeled $\bm a$
         and $\bm b$) is important in the determination of the
         short-range part of the $\gamma$ and $U$ interactions. In
         addition, the diagonal terms like $\gamma_{\bm{a}\bm{a}}$
         and $U_{\bm{a}\bm{a}\bm{b}\bm{b}}$ also include far-field 
         effects due to charges and dipoles at distant regions 
         of the material (see WF $\bm{c}$ in the figure). 
         As regards the far-field interactions, the precise shape of
         the charge distributions generating the potential is not 
         critical (illustrated by the diffuse orbital at point $\bm c$),
         and can be approximated by a multipole expansion.}
       \label{fig:interactions}
    \end{center}
 \end{figure}

\subsubsection{One-electron parameters}
\label{sec:oneelecint}

 The matrix element $\gamma_{\bm{a}\bm{b}}$ [Eq.~(\ref{eq:gamma})]
 gathers Coulomb interactions associated to the electrostatic
 potential created by both electrons and nuclei, which acts on the WFs
 $\chi_{\bm{a}}$ and $\chi_{\bm{b}}$. Note that these are the only
 long-ranged interactions in the system, since all other contributions
 (kinetic, exchange-correlation, external applied fields) can be
 considered local or semi-local. In the following we discuss the
 detailed form of this electrostatic part of $\gamma_{\bm{a}\bm{b}}$,
 which we denote $\gamma_{\bm{a}\bm{b}}^{\rm elec}$.

 Let us first consider the part of $\gamma_{\bm{a}\bm{b}}^{\rm elec}$
 associated to the electrostatic potential created by the
 electrons, $\gamma_{\bm{a}\bm{b}}^\text{elec,e}$. We have
 \begin{align}
    \gamma_{\bm{a}\bm{b}}^\text{elec,e} 
    & \equiv -\left\langle \chi_{\bm{a}} \right\vert 
      v_{\rm H} (n_{0};\vec{r}) \left\vert \chi_{\bm{b}} 
    \right\rangle
    \nonumber \\
    & = \int \chi_{\bm{a}} ( \vec{r} )  
        \left(
            \int \frac{n_0(\pvec{r})}
                  {\vert \vec{r}-\pvec{r}\vert} d^3r^\prime 
        \right) 
        \chi_{\bm{b}} (\vec{r} ) d^3r
    \nonumber \\
    & = \int \chi_{\bm{a}} ( \vec{r} )  
        \left(
           \int \frac{\sum_{\bm{c}} o_{\bm{c}}^{(0)} 
                   \vert \chi_{\bf c} (\pvec{r}) \vert^{2}}
                  {\vert \vec{r}-\pvec{r}\vert} d^3r^\prime 
        \right) 
        \chi_{\bm{b}} (\vec{r} ) d^3r.
    \label{eq:gammalg}
 \end{align}
 The expression of the reference electron density in terms
 of the occupation of Wanniers, $o_{\bm{c}}^{(0)}$,
 and squares of Wannier functions in the reference state 
 will be described in more detail in Sec.~\ref{sec:practical}.
 Following the criteria of Ref.~\onlinecite{Demkov-95}, the
 one-electron matrix elements related with the Coulomb
 electron-electron interaction can be split into two categories:
 (i) the near-field regime, where the two WFs ($\bm {a}$ and $\bm{b}$)
 significantly overlap with the third WF ($\bm{c}$) that creates the
 electrostatic potential, and (ii) the far-field regime, where this
 overlap is negligible.

 In the far-field regime, the electrostatic potential outside the
 region where a source charge $\chi_{\bm c}$ is located can be
 expressed as a multipole expansion (see Chapter~4 of
 Ref.~\onlinecite{jackson_book}). More precisely, we can write the
 far-field (FF) potential created by the charge distribution given by
 $\chi_{\bm c}$ as
 \begin{equation}
   v_{{\rm FF}, {\bm c}}^{\rm e} (n_{0}; \vec{r}) = 
         -o_{\bm c}^{(0)} \int 
               \frac{\vert \chi_{\bm c} (\pvec{r}) \vert^{2}}
                 {\vert \vec{r}-\pvec{r}\vert} d^3r^\prime ,
 \end{equation}
 which applies to $\vec{r}$ points for which $\chi_{\bm c} (\vec{r})
 \approx 0$. Now, let $\bm{\lambda}$ label the atom -- located at the
 RAG reference position $\vec{\tau}_{\bm \lambda} = \vec{R}_{\Lambda}
 + \vec{\tau}_{\lambda}$ -- around which $\chi_{\bm c}$ is
 centered. It is convenient to shift the origin in the integral,
 $\pvec{r}'=\pvec{r}-\vec{\tau}_{\bm\lambda}$, to write
 \begin{equation}
   v_{{\rm FF}, {\bm c}}^{\rm e} (n_{0}; \vec{r}) = 
       -o_{\bm c}^{(0)} \int 
             \frac{\vert \chi_{\bm c} (\pvec{r}' + \vec{\tau}_{\bm
                 \lambda}) \vert^{2}}{\vert \vec{r}-\vec{\tau}_{\bm
                 \lambda}-\pvec{r}'\vert} d^3r^{\prime\prime} .
   \label{eq:classical_pot_wan}
 \end{equation}
 Then, assuming that 
 $\vert \pvec{r}' \vert \ll \vert \vec{r}-\vec{\tau}_{\bm \lambda} \vert$ 
 and using the superscript $T$ to indicate the transpose operation, 
 as necessary to compute inner dot products, we get
 \begin{equation}
   \frac{1}{\vert \vec{r}-\vec{\tau}_{\bm
                 \lambda}-\pvec{r}'\vert} \approx 
                   \frac{1}{\vert 
                            \vec{r}-\vec{\tau}_{\bm \lambda}
                            \vert} 
                 + \frac{(\vec{r}-\vec{\tau}_{\bm\lambda})^T\pvec{r}'}
                        {\vert 
                         \vec{r}-\vec{\tau}_{\bm \lambda}
                         \vert^{3}}                        
                 +\ldots 
   \label{eq:multipole_exp}
 \end{equation}
 Now, substituting Eq.~(\ref{eq:multipole_exp}) into 
 Eq.~(\ref{eq:classical_pot_wan}) we obtain the multipole 
 series
 \begin{align}
   v_{{\rm FF}, {\bm c}}^{\rm e} (n_{0}; \vec{r}) & \approx 
       -o_{\bm c}^{(0)} \left[
             \frac{\int
                  \vert \chi_{\bm c} (\pvec{r}' + \vec{\tau}_{\bm
                      \lambda}) \vert^{2}
                  d^3 r^{\prime\prime}}
                  {\vert \vec{r}-\vec{\tau}_{\bm \lambda} \vert}
              +
                  \right. 
                  \nonumber\\
       &     \left.
                \frac{(\vec{r}-\vec{\tau}_{\bm\lambda})^T
                      \int
                      \vert \chi_{\bm c} (\pvec{r}' + \vec{\tau}_{\bm
                      \lambda}) \vert^{2}
                      \pvec{r}'  
                d^3 r^{\prime\prime}}
                     {\vert 
                      \vec{r}-\vec{\tau}_{\bm \lambda}
                      \vert^{3}}                                       
                + \ldots               
              \right] 
              \nonumber \\
       & = 
    \frac{q_{\bm c}}{\vert \vec{r} - \vec{\tau}_{\bm \lambda} \vert}  +
    \frac{  (\vec{r}- \vec{\tau}_{\bm \lambda})^T\vec{p}_{\bm c}}
      {\vert \vec{r} - \vec{\tau}_{\bm \lambda} \vert^{3}} 
      + \ldots .
    \label{eq:multipole}
 \end{align}
 The coefficient of the first term is the total charge
 (i.e., the monopole), and it is given by
 \begin{equation}
    q_{\bm c} = -o_{\bm c}^{(0)}
                 \int
                    \vert 
                       \chi_{\bm c} 
                          (\pvec{r}' + 
                           \vec{\tau}_{\bm \lambda}) 
                    \vert^{2}
                    d^3 r^{\prime\prime}
              =-o_{\bm c}^{(0)}.
    \label{eq:monopole}
 \end{equation}
 The coefficient of the second term is the electric dipole moment
 associated to $\chi_{\bm c}$, which amounts to
 \begin{align}
    \vec{p}_{\bm c} & = -o_{\bm c}^{(0)}
                \int
                      \vert \chi_{\bm c} (\pvec{r}' + \vec{\tau}_{\bm
                      \lambda}) \vert^{2}
                      \pvec{r}'  
                d^3 r^{\prime\prime} 
                \nonumber \\
 &    = -o_{\bm c}^{(0)} \int \vert \chi_{\bm c} (\pvec{r}) \vert^{2}
                              (\pvec{r} - \vec{\tau}_{\bm \lambda})  
                              d^3r^\prime 
    \nonumber\\
 &    = -o_{\bm c}^{(0)} (\vec{r}_{\bm c} - \vec{\tau}_{\bm \lambda}),
    \label{eq:dipole}
 \end{align}
 where $\vec{r}_{\bm c}$ represents the centroid of $\chi_{\bm c}$. 
 Quadrupole and higher-order moments follow in the expansion,
 but here we assume they can be neglected.
 Finally, the full FF potential created by the electrons at point
 $\vec{r}$ is simply given by
 \begin{equation}
    v_{{\rm FF}}^{\rm e} (n_{0}; \vec{r}) = \sum_{\bm c}{}^{'}
      v_{{\rm FF}, {\bm c}}^{\rm e} (n_{0}; \vec{r}) ,
 \end{equation}
 where the prime indicates that we sum only over WF's such that
 $\chi_{\bm c} (\vec{r}) \approx 0$.

 Let us now consider the part of $\gamma_{\bm{a}\bm{b}}^{\rm elec}$
 associated to the potential created by the nuclei, which we call
 $\gamma_{\bm{a}\bm{b}}^{\rm elec,n}$. In analogy with the electronic
 case, we write the FF electrostatic potential created by the nuclei
 at point $\vec{r}$ as
 \begin{equation}
    v_{\rm FF}^{\rm n} (\vec{r}) \approx  \sum_{\bm \lambda}{}^{'}
    \frac{Z_{\bm \lambda}}{\vert \vec{r}-\vec{\tau}_{\bm \lambda}\vert} +
    \sum_{\bm \lambda}{}^{'} \frac{(\vec{r} - \vec{\tau}_{\bm \lambda})^T
     Z_{\bm \lambda} \vec{u}_{\bm \lambda}}{\vert
      \vec{r}-\vec{\tau}_{\bm \lambda}\vert^{3}} + \ldots ,
    \label{eq:multipole_nuc}
 \end{equation}
 where the primed sums run only over atoms $\bm{\lambda}$ whose
 associated WFs $\bm{a}\subset \bm{\lambda}$ satisfy $\chi_{\bm a}
 (\vec{r}) \approx 0$.

 Then, adding all far-field contributions to
 $\gamma_{\bm{a}\bm{b}}^{\rm elec}$, and assigning each WF to its
 associated nucleus, we get
 \begin{align}
    v_{\rm FF} (n_{0}; \vec{r}) = 
        \sum_{\bm \lambda}{}^{'} \frac{q_{\bm \lambda} }{\vert \vec{r} -
        \vec{\tau}_{\bm \lambda}\vert} 
      + \sum_{\bm \lambda}{}^{'} \frac{ (\vec{r} -
        \vec{\tau}_{\bm \lambda})^T\vec{p}_{\bm \lambda}}{\vert \vec{r} -
        \vec{\tau}_{\bm \lambda}\vert^{3}} 
      + \ldots    ,
 \end{align}
 where
 \begin{equation}
    q_{\bm \lambda}=Z_{\bm \lambda} + \sum_{{\bm c} \subset {\bm
        \lambda}} q_{\bm c} 
    \label{eq:qlambda}
 \end{equation}
 is the charge of {\em ion} $\bm{\lambda}$, while $\vec{p}_{\bm
   \lambda}$ is the local dipole associated to that very ion. Note
 that we add together the contributions from electrons and nuclei,
 which allows us to talk about {\em ions} in a strict sense.
 We can further approximate this local dipole using the Born charge
 tensor $\overleftrightarrow{Z}_{\bm \lambda}^*$, to obtain
 \begin{equation}
     \vec{p}_{\bm{\lambda}} = Z_{\bm \lambda} \vec{u}_{\bm \lambda} + 
     \sum_{{\bm c}\subset {\bm \lambda}} \vec{p}_{\bm c} 
     \approx \overleftrightarrow{Z}^*_{\bm \lambda} \vec{u}_{\bm{\lambda}} .
     \label{eq:localdip}
 \end{equation}

 In order to get the final expression for the FF potential, we note
 that the electrostatic interactions described above do not take place
 in vacuum, but in the material at its reference electronic density.
 Thus, we need to take into account that the RED will react to screen
 such interactions, and that such a screening can be modelled by the
 high-frequency dielectric tensor of the material at its RED. Thus,
 the far-field potential at the center of WF $\chi_{\bm a}$ is
 \begin{align}
    v_{\rm FF} (n_{0}; \vec{r}_{\bm{a}}) \approx &  \sum_{\bm{\lambda}}{}^{'}
    \left[ \vec{e}_{\bm{\lambda}\bm{a}}^T (\overleftrightarrow{\epsilon_\infty})^{-1}
      \vec{e}_{\bm{\lambda}\bm{a}} \right]
    \frac{q_{\bm{\lambda}}}{\left\vert
      \vec{\tau}_{\bm\lambda}-\vec{r}_{\bm a}\right\vert} \nonumber \\ &+
    \sum_{\bm{\lambda}}{}^{'} \frac{ [\vec{p}_{\bm{\lambda}}^T
        (\overleftrightarrow{\epsilon_\infty})^{-1}\vec{e}_{\bm{\lambda}\bm{a}}
    ]}{\left\vert \vec{\tau}_{\bm\lambda}-\vec{r}_{\bm a}\right\vert^2} ,
    \label{eq:farfieldpot}
 \end{align}
 where $\vec{e}_{\bm{\lambda}\bm{a}}$ is a unitary vector parallel to
 $\vec{\tau}_{\bm\lambda}-\vec{r}_{\bm a}$,
 $\overleftrightarrow{\epsilon_\infty}$ is the high-frequency
 dielectic tensor, and the primed sums are restricted in the usual
 way.

 We can now divide $\gamma_{\bm{a}\bm{b}}$ in long-range (lr) and
 short-range (sr) contributions. Considering that $\chi_{\bm{a}}$ and
 $\chi_{\bm{b}}$ are strongly localized and orthogonal to each other,
 we define $\gamma_{\bm{a}\bm{b}}^{\text{lr}}$ as
 \begin{equation}
    \gamma_{\bm{a}\bm{b}}^{\text{lr}} =  
           -v_{\rm FF} (n_{0}; \vec{r}_{\bm{a}}) \delta_{\bm{a}\bm{b}} .
    \label{eq:gamma-lr}
 \end{equation}
 Then, we effectively define the short-range part of
 $\gamma_{\bm{a}\bm{b}}$ as
 \begin{equation}
   \gamma_{\bm{a}\bm{b}}^\text{sr} = \gamma_{\bm{a}\bm{b}} -
   \gamma_{\bm{a}\bm{b}}^\text{lr} .
   \label{eq:gammashlg}
 \end{equation}
 Note that the short-range interactions defined in this way include
 electrostatic effects as well as others associated to chemical
 bonding, orbital hybridization, etc. These interactions do not have a
 simple analytic form; hence, in order to construct our models, they
 will generally be fitted to reproduce DFT results.

 It is important to note that the above derivation, and decomposition
 in long- and short-range parts, is exact and does not involve any
 approximation, except for: (i) the truncation of the multi-pole
 expansion and (ii) the analytic form introduced for the long-range
 electrostatic interactions, which strictly speaking only applies to
 homogeneous materials with a band gap.
 \footnote{For metallic systems the definition and calculation of these 
 interactions requires an adequate treatment of the screening. 
 This problem will be addressed in future publications\label{Note1}}

 Finally, note that the $\gamma_{\bm{a}\bm{b}}$ couplings can be
 expected to be short in range, as they involve WFs $\chi_{\bm{a}}$
 and $\chi_{\bm{b}}$ that are strongly localized in space and decay
 exponentially as we move away from their centers. Hence, the
 $\bm{\gamma}$ matrix can be expected to be sparse, which will result
 in more efficient calculations. It is important to note that this
 short-ranged character of the $\gamma_{\bm{a}\bm{b}}$ couplings is
 expected {\em despite the fact} that the interactions contributing to
 $\gamma_{\bm{a}\bm{b}}^{\rm \text{lr}}$ are electrostatic and thus long
 ranged.

\subsubsection{Two-electron integrals}

 In a similar vein, we can split $U$ in short- and long-range
 contributions, so that 
 \begin{equation}
        U_{\bm{a}\bm{b}\bm{a}^\prime\bm{b}^\prime}^\text{sr} =
        U_{\bm{a}\bm{b}\bm{a}^\prime\bm{b}^\prime} -
        U_{\bm{a}\bm{b}\bm{a}^\prime\bm{b}^\prime}^\text{lr},
    \label{eq:Ushlg}
 \end{equation}
 where the long-range part will contain the classical FF interaction
 between electrons that can be approximated analytically, while the
 short-range part will contain all other interactions including
 many-body effects.

 As above, we expect that (i) long-range two-electron integrals should
 be very small unless $\bm{a}$ overlaps with $\bm{b}$ and
 $\bm{a}^\prime$ overlaps with $\bm{b}^\prime$, (ii) we can truncate
 the multipolar expansion at the monopole level, and (iii) the
 electrostatic interactions take place in a medium characterized by
 the high-frequency dielectric tensor of the material at the RED.
 Under these conditions we choose
 $U_{\bm{a}\bm{b}\bm{a}^\prime\bm{b}^\prime}^\text{lr}$ to be
 \begin{equation}
    U_{\bm{a}\bm{b}\bm{a}^\prime\bm{b}^\prime}^\text{lr}=
       \left[ \vec{e}_{\bm{a}\bm{a}^\prime}^T (\overleftrightarrow{\epsilon_\infty})^{-1}
       \vec{e}_{\bm{a}\bm{a}^\prime} \right]       
       \frac{1}{\vert\vec{r}_{\bm{a}^\prime}-\vec{r}_{\bm{a}}\vert}
    \delta_{\bm{a}\bm{b}}\delta_{\bm{a}^\prime\bm{b}^\prime}.
    \label{eq:Ulgdelta}
 \end{equation}
 In order to avoid the divergence of this term, we assume that all
 one-body integrals ($\bm{a} = \bm{a}'$) are fully included in the
 short-range part,
 $U_{\bm{a}\bm{b}\bm{a}^\prime\bm{b}^\prime}^\text{sr}$.
 Assigning the Wannier functions to their closest nucleus, and summing
 over all the atoms in the lattice, we find that the total
 two-electron long-range energy adds to
 \begin{equation}
    E^\text{(2),lr}=
    \frac{1}{2}\sum_{\bm{\lambda}\bm{\upsilon}}
    \left[ \vec{e}_{\bm{\lambda}\bm{\upsilon}}^T (\overleftrightarrow{\epsilon_\infty})^{-1}
    \vec{e}_{\bm{\lambda}\bm{\upsilon}} \right]        
    \frac{\Delta q_{\bm{\lambda}} \Delta q_{\bm{\upsilon}}}
    {\vert\vec{\tau}_{\bm{\upsilon}}-\vec{\tau}_{\bm{\lambda}}\vert},
 \end{equation}
 where $\Delta q_{\bm{\lambda}}$ is the change in charge of the atom
 $\bm{\lambda}$ when compared to the RED state
 [Eq.~(\ref{eq:qlambda})].
 Thus, the long-range part of $U$ simply updates the one-electron FF
 potentials due to the charge transfers between atoms.

 We would like to stress again that the separation in long- and short-
 range parts does not involve any approximation; indeed, effects
 usually considered important in many physical phenomena, like
 e.g. the anisotropy of the Wannier orbitals at short
 distances,\cite{kugel_spu82} are included in $U^\text{sr}$.

\subsection{Electron-lattice coupling}
\label{sec:electron-lattice}

 The system's geometry determines the reference density $n_{0}
 (\vec{r})$ as well as the corresponding Hamiltonian. In our scheme,
 such a dependence of the model parameters on the atomic configuration
 is captured by the electron-lattice coupling terms.

\begin{figure} [h]
    \begin{center}
       \includegraphics[width=0.9\columnwidth]{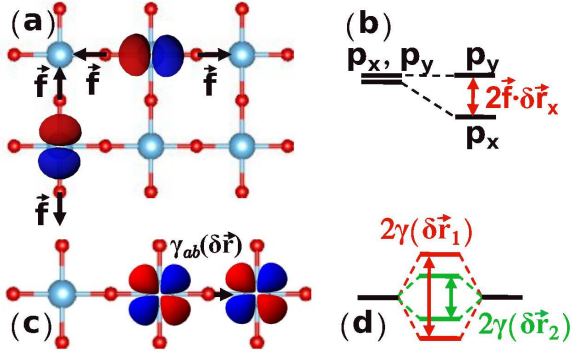}
       \caption{(Color online) Schematic representation of the effects
         of the expansion of $\gamma^{\text{sr}}$ in terms of atomic
         deformations. Panels (a) and (b) illustrate the
         electron-lattice coupling associated to diagonal,
         $\gamma_{\bm{a}\bm{a}}^{\text{sr}}$, matrix elements that control
         the average energy of the corresponding bands. In (a) we
         sketch the forces on the atoms (represented by red spheres)
         as generated by electrons placed on a $p_{x}$ or $p_{y}$-like
         WFs.  In our method, these forces are captured by the tensor
         $\vec{f}$ in Eq.~(\ref{eq:gammadist}).  In (b) we illustrate
         the change in the electronic structure as a consequence of
         the atomic displacement: if the atoms displace along $x$ in the way
         shown in the top atomic chain of panel~(a), 
         then a variation in the position of the
         $p_x$ orbitals is induced, while the $p_y$ level remains
         unaltered.  Panels (c) and (d) illustrate the change in
         non-diagonal $\gamma^{\text{sr}}$ matrix elements between two
         neighboring orbitals when the intermediate atom moves, thus
         altering the band width as illustrated in (d).
         The change in the band width depends on the amount by which
         the atoms are displaced, represented in the cartoon by two different
         displacement vectors $\delta \vec{r}_{1}$ and $\delta \vec{r}_{2}$.
            }
       \label{fig:forces}
    \end{center}
 \end{figure}

 Let us consider the lattice dependence of the one-electron integrals
 $\gamma$ [Eq.~(\ref{eq:gamma})]. In Sec.~\ref{sec:oneelecint}, these
 parameters were split in short- and long-range contributions.
 The explicit dependence of the long-range part with the distortion of
 the lattice is clearly seen in Eq.~(\ref{eq:localdip}), where the
 electric dipole that enters in the multipole expansion of the far
 field potential [Eq.~(\ref{eq:farfieldpot})] depends linearly with
 the atomic displacements, as computed with respect to the RAG.
 As regards the dependence of $\gamma^\text{sr}_{\bm{a}\bm{b}}$ on the
 atomic configuration (see Fig.~\ref{fig:forces}), we include it by
 expanding
 \begin{align}
    \gamma^\text{sr}_{\bm{a}\bm{b}} = \gamma^{\text{RAG, sr}}_{\bm{a}\bm{b}} & +        
       \sum_{\bm{\lambda}\bm{\upsilon}} 
       \left[
       -\vec{f}_{\bm{a}\bm{b},\bm{\lambda}\bm{\upsilon}}^T
        \delta \vec{r}_{\bm{\lambda}\bm{\upsilon}} + 
       \right.
       \nonumber \\
       &+\left.
       \sum_{\bm{\lambda}^\prime\bm{\upsilon}^\prime} 
       \delta \vec{r}_{\bm{\lambda}\bm{\upsilon}}^T       
       \overleftrightarrow{g}_{\bm{a}\bm{b},\bm{\lambda}\bm{\upsilon}\bm{\lambda}^\prime\bm{\upsilon}^\prime}
       \delta \vec{r}_{\bm{\lambda}^\prime\bm{\upsilon}^\prime}+...
       \right], 
    \label{eq:gammadist}       
 \end{align}
 where
 \begin{equation}
    \delta \vec{r}_{\bm{\lambda}\bm{\upsilon}} =
    \overleftrightarrow{\eta}\left(\vec{R}_{\Upsilon}-\vec{R}_{\Lambda}+
    \vec{\tau}_{\upsilon}-\vec{\tau}_{\lambda} \right) +
    \vec{u}_{\bm{\upsilon}}-\vec{u}_{\bm{\lambda}}
 \end{equation}
 quantifies the relative displacement of atoms $\bm{\lambda}$ and
 $\bm{\upsilon}$. In addition, $\vec{\bm{f}}$ and
 $\overleftrightarrow{\bm{g}}$ are the first- and second-rank tensors
 that characterize the electron-lattice coupling, closely related to
 the concept of \emph{vibronic constants}.\cite{jt_book}

\begin{figure} [h]
   \begin{center}
       \includegraphics[width=1.0\columnwidth]{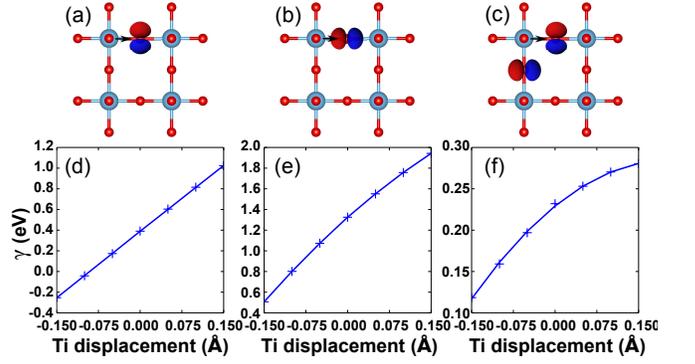}
          \caption{(Color online) Variation of the $\gamma$ matrix
            elements of bulk SrTiO$_3$ that are most sensitive to the
            the displacement of a Ti ion. Panels (a)-(c): the atoms
            at a TiO$_{2}$ plane are represented by big blue (Ti) and small red
            (O) spheres.  The displacement of the Ti$^{4+}$ is marked
            with an arrow.  The Wannier-like functions related with
            the $\gamma$ matrix elements under consideration are
            plotted: (a) a $p$-like orbital with lobules pointing
            perpendicularly to the Ti$^{4+}$ displacement ($\pi$-bonding), (b)
            a $p$-like orbital with lobules pointing parallel to the
            Ti$^{4+}$ displacement ($\sigma$-bonding), and (c) two $p$-like
            orbitals centered on different O atoms with lobules
            pointing perpendicular in one case and parallel in the
            other to the Ti$^{4+}$ displacement.  In (a), and (b), the
            corresponding matrix element whose evolution is studied is
            diagonal (i.e., $\bm a = \bm b$), while in (c) the matrix
            element is off-diagonal ($\bm a \ne \bm b$).  Panels
            (d)-(f) show the variation of these matrix elements with
            respect to the Ti displacement.  First-principles results
            are represented by blue crosses, while the model values
            [see Eq.~(\ref{eq:gammadist})] are represented by
            solid blue lines.  }
       \label{fig:stofit}
   \end{center}
\end{figure}

 We have checked that including quadratic constants is enough to
 describe typical changes in the value of $\gamma$ with the geometry.
 For example, we have inspected the $\gamma$ parameters associated
 with the oxygen $2p$-like WFs of SrTiO$_{3}$ -- i.e., with the
 valence band of the material --, and plotted in Fig.~\ref{fig:stofit}
 the three that are most sensitive to structural deformations: they
 correspond with the diagonal elements of the $\sigma$ and $\pi$
 functions centered on the oxygen ions [see
   Fig.~\ref{fig:entanglement}(a)] and a $\pi-\pi$ off-diagonal term.
 We find that, if we use a quadratic expansion to describe such a
 structural dependence, the errors are smaller than 1\% over a wide
 range of distortions, up to 0.3~\AA.
 Hence, given the strong changes occurring in the hybridization of
 ferroelectric-like materials like SrTiO$_{3}$, we consider that the
 approximation employed in Eq.~(\ref{eq:gammadist}) should be
 reasonable for most systems.

 Moreover, in the cases studied so far, we have found that the
 quadratic constants are typically much smaller than the linear ones;
 further, among the quadratic constants, the diagonal ones are clearly
 dominant. Hence, in Eq.~(\ref{eq:gammadist}) we restrict the
 expansion to two-atom terms, so that
 \begin{equation}
   \overleftrightarrow{g}_{\bm{a}\bm{b},\bm{\lambda}\bm{\upsilon}
      \bm{\lambda}^\prime\bm{\upsilon}^\prime}
   \approx
   \overleftrightarrow{g}_{\bm{a}\bm{b},\bm{\lambda}\bm{\upsilon}
      \bm{\lambda}^\prime\bm{\upsilon}^\prime}
   \delta_{\bm{\lambda}\bm{\lambda}^\prime}
   \delta_{\bm{\upsilon}\bm{\upsilon}^\prime}
   =   \overleftrightarrow{g}_{\bm{a}\bm{b},\bm{\lambda}\bm{\upsilon} } .
 \end{equation}

 The physical meaning of
 $\vec{f}_{\bm{a}\bm{b},\bm{\lambda}\bm{\upsilon}}$ is particularly
 obvious when $\bm{a}=\bm{b}$: it represents the force created by an
 electron occupying the WF $\chi_{\bm{a}}$ over the surrounding atoms
 [see Figs.~\ref{fig:forces}(a) and \ref{fig:forces}(b)]. Such a
 parameter is key to quantify phenomena like the Jahn-Teller effect in
 solids\cite{jt_book} or polaron formation.\cite{emin_polaronbook}

 Off-diagonal terms in $\vec{\bm{f}}$ describe the mixing of two WFs
 upon an atomic distortion, and thus quantify changes in covalency
 [see Figs.~\ref{fig:forces}(c) and \ref{fig:forces}(d)]. They can be
 identified with the pseudo Jahn-Teller vibronic constants and are
 thus relevant to a wide variety of phenomena including
 ferroelectricity,\cite{jt_book} spin-crossover,~\cite{pgf_prl11} and
 spin-phonon coupling.\cite{pgf_prb11}
      
 Finally, the geometrical dependence of the two-electron parameters,
 $U$ and $I$, can be included in our model in a similar
 way. Nevertheless, since these terms are not explicitly dependent on
 the potential created by the ions, their value can be expected to be less
 sensitive to changes in the atomic configuration. Hence, in this
 work, and in analogy to what is customarily done in model Hamiltonian
 and DFT+{\sl U} approaches,\cite{anisimov_jpcm97} we will neglect
 such effects.

 \subsection{Total energy}
 \label{sec:total-energy}

 Replacing the expressions for the one-electron [Eq.~(\ref{eq:eone})],
 and two-electron [Eq.~(\ref{eq:etwoui})] integrals into
 Eq.~(\ref{eq:totale}) for the total energy, we get
 \begin{align}
    E = & E^{(0)} + E^{(1)} + E^{(2)} 
    \nonumber \\
      = & E^{(0)} + 
          \sum_{\bm{a}\bm{b}} 
           D_{\bm{a}\bm{b}}^U \gamma_{\bm{a}\bm{b}} +
    \nonumber \\
      & + \frac{1}{2} \sum_{\bm{a}\bm{b}}
                  \sum_{\bm{a}^\prime\bm{b}^\prime}      
    \left(
           D_{\bm{a}\bm{b}}^U
           D_{\bm{a}^\prime\bm{b}^\prime}^U  
       U_{\bm{a}\bm{b}\bm{a}^\prime\bm{b}^\prime} - 
           D_{\bm{a}\bm{b}}^I 
           D_{\bm{a}^\prime\bm{b}^\prime}^I
       I_{\bm{a}\bm{b}\bm{a}^\prime\bm{b}^\prime}
    \right).
    \label{eq:totalenergydecom}
 \end{align}
 \noindent or, equivalently, in terms of the spin-up
 and spin-down densities,
 \begin{align}
    E = & E^{(0)} + 
          \sum_{\bm{a}\bm{b}} 
            \left(
                  D_{\bm{a}\bm{b}}^\uparrow+D_{\bm{a}\bm{b}}^\downarrow
            \right)
            \gamma^\text{sr}_{\bm{a}\bm{b}} 
    \nonumber \\
    &+ \frac{1}{2}
      \sum_{\bm{a}\bm{b}}
      \sum_{\bm{a}^\prime\bm{b}^\prime} 
    \left\{
         \left(D_{\bm{a}\bm{b}}^\uparrow + D_{\bm{a}\bm{b}}^\downarrow\right)
         \left(D_{\bm{a}^\prime\bm{b}^\prime}^\uparrow + D_{\bm{a}^\prime\bm{b}^\prime}^\downarrow\right)
    \right.
       U_{\bm{a}\bm{b}\bm{a}^\prime\bm{b}^\prime}
    \nonumber \\
    &\left.
        -\left(D_{\bm{a}\bm{b}}^\uparrow - D_{\bm{a}\bm{b}}^\downarrow \right)
         \left(D_{\bm{a}^\prime\bm{b}^\prime}^\uparrow - D_{\bm{a}^\prime\bm{b}^\prime}^\downarrow\right)
       I_{\bm{a}\bm{b}\bm{a}^\prime\bm{b}^\prime}
    \right\}
 \label{eq:totalenergy1updn}
 \end{align}
 Now we introduce the decomposition of the $\gamma$
 [Eqs.~(\ref{eq:gamma-lr}) and (\ref{eq:gammashlg})] and $U$
 [Eqs.~(\ref{eq:Ushlg}) and (\ref{eq:Ulgdelta})] parameters into long
 and short-range parts, and gather together all the long-range terms,
 to obtain,
 \begin{align}
    E = & E^{(0)} + 
          \sum_{\bm{a}\bm{b}} 
            D_{\bm{a}\bm{b}}^U
            \gamma^\text{sr}_{\bm{a}\bm{b}} 
    \nonumber \\
      & + \frac{1}{2} \sum_{\bm{a}\bm{b}}
                  \sum_{\bm{a}^\prime\bm{b}^\prime}  
    \left(
            D_{\bm{a}\bm{b}}^U
            D_{\bm{a}^\prime\bm{b}^\prime}^U
            U^\text{sr}_{\bm{a}\bm{b}\bm{a}^\prime\bm{b}^\prime} -
            D_{\bm{a}\bm{b}}^I
            D_{\bm{a}^\prime\bm{b}^\prime}^I
            I_{\bm{a}\bm{b}\bm{a}^\prime\bm{b}^\prime}
    \right)
    \nonumber \\
     & +  \sum_{\bm{a}} 
          D_{\bm{a}\bm{a}}^U
          \left(
            - v_{\rm FF} (\vec{r}_{\bm{a}})
            + \frac{1}{2} \sum_{\bm{a}^\prime} 
            D_{\bm{a}^\prime\bm{a}^\prime}^U
            U^\text{lr}_{\bm{a}\bm{a}\bm{a}^\prime\bm{a}^\prime}.
          \right)
    \label{eq:totalenergy1}
 \end{align}
 Note that for the case of a non-spin polarized system
 ($D_{\bm{a}\bm{b}}^I = 0$) the
 expression for the total energy reduces to
 \begin{align}
    E = & E^{(0)} + 
          \sum_{\bm{a}\bm{b}} 
           D_{\bm{a}\bm{b}}^U
          \gamma^\text{sr}_{\bm{a}\bm{b}} 
    \nonumber \\
      & + \frac{1}{2} \sum_{\bm{a}\bm{b}}
                  \sum_{\bm{a}^\prime\bm{b}^\prime} 
           D_{\bm{a}\bm{b}}^U D_{\bm{a}^\prime\bm{b}^\prime}^U 
           U^\text{sr}_{\bm{a}\bm{b}\bm{a}^\prime\bm{b}^\prime}
    \nonumber \\
     & +  \sum_{\bm{a}} D_{\bm{a}\bm{a}}^U
          \left(
              - v_{\rm FF} (\vec{r}_{\bm{a}})
              + \frac{1}{2} 
                  \sum_{\bm{a}^\prime} 
                  D_{\bm{a^{\prime}}\bm{a}^\prime}^U
                  U^\text{lr}_{\bm{a}\bm{a}\bm{a}^\prime\bm{a}^\prime}
          \right).                 
    \label{eq:totalenergy3}
 \end{align}

 \subsection{Self-consistent equations}
 \label{sec:self-consistent-equations}

 As it is clearly seen in Eq.~(\ref{eq:totalenergy1}), the total
 energy in our formalism depends on the deformation occupation matrix
 defined in Eq.~(\ref{eq:Ddens}), and later generalized for the case
 of spin-polarized systems in Eqs.~(\ref{eq:Ddensup}) and
 (\ref{eq:Ddensdown}).
 This quantity is directly related with the deformation charge
 density, i.e., with the difference between the total charge density
 and the reference electronic density.
 It can be computed from the coefficients of the Bloch wave functions
 in the basis of Wannier functions, which are thus the only
 variational parameters of the method.

 Solving for the ground state amounts to finding a point at which the
 energy is stationary upon variations in the electronic density,  
 $n( \vec{r})$.
 Following a textbook
 procedure,\cite{parryang,jensen_book,martin_book,Kohanoff_book} we obtain a set of
 self-consistent conditions analogous to the Kohn-Sham
 equations\cite{kohn_pra65}
 \begin{equation}
    \sum_b  h_{ab,\vec{k}}^s\,  c_{jb \vec{k}}^s = 
    \varepsilon_{j\vec{k}}^s\,  c_{ja \vec{k}}^s ,
    \label{eq:eigenproblem}
 \end{equation}
 where, as defined above, $\varepsilon_{j\vec{k}}^s$, is the $j$-th
 band energy at wavevector $\vec{k}$ for the spin channel $s$. The
 corresponding Hamiltonian matrix, $h_{ab,\vec{k}}^s$, is
 \begin{equation}
    h_{ab,\vec{k}}^s =
         \sum_{\vec{R}_B-\vec{R}_A} 
            e^{i\vec{k}\cdot \left( \vec{R}_B-\vec{R}_A \right) } 
            h_{\bm{a}\bm{b}}^s,
    \label{eq:hone}
 \end{equation}
 where $h_{\bm{a}\bm{b}}^s$ is the real-space Hamiltonian
 \begin{align}
     h_{\bm{a}\bm{b}}^s = \gamma_{\bm{a}\bm{b}} + 
       \sum_{\bm{a^\prime b}^\prime} &
            \left[
                \left(
                   D_{\bm{a}^\prime \bm{b}^\prime}^s+
                   D_{\bm{a}^\prime \bm{b}^\prime}^{-s}
                \right)
                U_{\bm{a b a^\prime b}^\prime}-
            \right.
            \nonumber \\
       &
            \left.
                \left(
                   D_{\bm{a}^\prime\bm{b}^\prime}^{s}-
                   D_{\bm{a}^\prime\bm{b}^\prime}^{-s}
                \right)
                I_{\bm{ab a^\prime b}^\prime}
            \right].          
 \label{eq:honeR}
 \end{align}

 Note that this is a mean-field problem fully equivalent to that
 of the Hartree-Fock approach, and it must be solved
 self-consistently. The practical procedure for finding the solution
 is straightforward: given an initial guess for the deformation
 occupation matrix ($D_{\bm{a}\bm{b}}^{s}$), we compute the
 corresponding mean-field Hamiltonian ($h_{\bm{a}\bm{b}}^s$); from the
 diagonalization of this matrix we obtain a new deformation occupation
 matrix, and the procedure is iterated until reaching
 self-consistency. Note that electrostatic effects are accounted for
 by computing the long-range part of the $\gamma$ and $U$ parameters;
 this is our scheme's equivalent to solving Poisson's equation, as
 customarily done in DFT and other approaches.

 Finally, note that in cases in which the system does not present any
 electron excitation -- i.e., whenever the full density is equal to
 the reference density and we have $D_{\bm{a}\bm{b}}^{s}=0$ --, no
 self-consistent procedure is needed to obtain the solution.

 \subsection{Forces and stresses}
 \label{sec:forces}

 Forces and stresses can be computed by direct derivation of the total
 energy [Eq.~(\ref{eq:totalenergydecom})] with respect to the atomic
 positions and cell strains. After some algebra, the result for the
 forces is,
 \begin{align}\label{eq:forces}
     \vec{F}_{\bm{\lambda}} = -\vec{\nabla}_{\bm{\lambda}} E = 
     -\vec{\nabla}_{\bm{\lambda}} E^{(0)} 
     -\sum_{\bm{a}\bm{b}} D_{\bm{a}\bm{b}}^U 
     \vec{\nabla}_{\bm{\lambda}} \gamma_{\bm{a}\bm{b}},
 \end{align}
 where $\bm{\lambda}$ denotes a specific atom in a certain cell; here
 we assume that electron-lattice couplings are restricted to the
 one-electron terms.

 The derivative of $E^{(0)}$ can be computed directly and exactly from the
 force-field on which our model is based.

 The deformation occupation matrix $D_{\bm{a}\bm{b}}^U$
 depends on the eigenvector coefficients and occupations.
 However, its derivative with respect to the atomic displacement 
 is not required, since the energy is stationary with respect to 
 these coefficients and occupations on the Born-Oppenheimer surface,  
 and the Hellman-Feynman theorem guarantees that their first-order
 variation will not modify the total energy, and therefore will not
 affect the forces.
 Moreover, due to the orthogonality of the basis set used, 
 no orthogonality forces need to be included, as it is the case
 when using a basis of non-orthogonal atomic orbitals 
 (see Appendix A of Ref.~\onlinecite{Soler-02}).

 It is interesting to further inspect the similarity between the
 second term in our forces and the Hellmann-Feynman
 result,\cite{martin_book}
 \begin{equation}
   \vec{F}_{\bm{\lambda}} = -\vec{\nabla}E_{nn} 
     -\sum_{j \vec{k}}  
      o_{j \vec{k}} \left\langle \psi_{j \vec{k}} 
                    \right\vert \vec{\nabla}_{\bm{\lambda}} \hat{h}_{0} 
                     \left \vert \psi_{j \vec{k}} \right\rangle,
 \end{equation}
 as (via a Fourier transform) $\vec{\nabla}_{\bm{\lambda}}
 \gamma_{\bm{a}\bm{b}}$ is analogous to $\langle \psi_{j \vec{k}}
 \vert \vec{\nabla}_{\bm{\lambda}} \hat{h}_{0} \vert \psi_{j \vec{k}}
 \rangle$, and $D_{\bm{a}\bm{b}}^U$ plays the role of the occupations
 $o_{j \vec{k}}$.
 This connection should be considered with caution, though, as our
 forces have a dominant contribution from the RED state, which is also
 included in the Hellmann-Feynman expression.

 It is also interesting to note that, if we included the dependence of
 $U$ (and $I$) on the nuclear positions, we would have a Pulay term in
 Eq.~(\ref{eq:forces}),\cite{pulay_mp69} reflecting the change of the
 WF basis set with the atomic displacements.

 Now we calculate the stress tensor in an analogous way.
 We adopt the standard definition\cite{martin_book}
 \begin{equation}
 S_{\alpha\beta}=-\frac{1}{V}\frac{\partial^\prime E}{\partial^\prime
   \eta_{\alpha\beta}} ,
 \end{equation}
 where $V$ is the volume of the real-space cell and $\partial^\prime$ 
 denotes derivative keeping the fractional coordinates of the atoms
 in the system constant. 
 We notice that there are only three terms in the energy that depend
 explicitly on the strain tensor $\overleftrightarrow{\eta}$, namely,
 the RED energy $E^{(0)}$, the short-range one-electron term,
 $\gamma_{\bm{a}\bm{b}}^\text{sr}$, and the electrostatic energy.
 Thus, we have
 \begin{equation}
 S_{\alpha\beta}=-\frac{1}{V}
                 \left[
                     \frac{\partial^\prime E^{(0)}}{\partial^\prime \eta_{\alpha\beta}}+
                     \sum_{\bm{a}\bm{b}} 
                           D_{\bm{a}\bm{b}}^U
                           \frac{\partial^\prime \gamma^{\text sr}_{\bm{a}\bm{b}}}{\partial^\prime \eta_{\alpha\beta}}+
                     \frac{\partial^\prime E^\text{elec}}{\partial^\prime \eta_{\alpha\beta}} 
                 \right],
 \end{equation}
 where $E^\text{elec}$ corresponds with the electrostatic energy  
 as written in the fourth contribution to the total energy in  
 Eq.~(\ref{eq:totalenergy1}).
 As in the case of the forces, the $E^{(0)}$ derivative is computed
 from the force field that describes the RED state.
 Similarly, the calculation of the last, electrostatic term can be
 achieved via Ewald summation techniques (see
 e.g. Ref.~\onlinecite{kawata_jcp01}).
 The only term that requires further manipulation is the derivative of
 $\gamma_{\bm{a}\bm{b}}^\text{sr}$, Eq.~(\ref{eq:gammadist}), with
 respect to the strain, which yields
 \begin{align}
 \frac{\partial^\prime \gamma_{\bm{a}\bm{b}}^\text{sr}}{\partial^\prime \eta_{\alpha\beta}}=&
        \sum_{\bm{\lambda}\bm{\upsilon}} 
       \left[
       -\vec{f}_{\bm{a}\bm{b},\bm{\lambda}\bm{\upsilon}}^T
        \frac{\partial^\prime \left(\delta \vec{r}_{\bm{\lambda}\bm{\upsilon}}\right)}{\partial^\prime \eta_{\alpha\beta}} + 
       \right.
       \nonumber \\
       &+\left.
       \sum_{\bm{\lambda}^\prime\bm{\upsilon}^\prime} 
       \frac{\partial^\prime \left(\delta \vec{r}_{\bm{\lambda}\bm{\upsilon}}^T\right)}{\partial^\prime \eta_{\alpha\beta}}       
       \left(\overleftrightarrow{g}\right)_{\bm{a}\bm{b},\bm{\lambda}\bm{\upsilon}\bm{\lambda}^\prime\bm{\upsilon}^\prime}
       \delta \vec{r}_{\bm{\lambda}^\prime\bm{\upsilon}^\prime}+
       \right.
       \nonumber \\
       &+\left.
       \sum_{\bm{\lambda}^\prime\bm{\upsilon}^\prime} 
       \delta \vec{r}_{\bm{\lambda}\bm{\upsilon}}^T       
       \left(\overleftrightarrow{g}\right)_{\bm{a}\bm{b},\bm{\lambda}\bm{\upsilon}\bm{\lambda}^\prime\bm{\upsilon}^\prime}
       \frac{\partial^\prime \left(\delta \vec{r}_{\bm{\lambda}^\prime\bm{\upsilon}^\prime}\right)}{\partial^\prime \eta_{\alpha\beta}}  
       \right].
 \end{align}
 As in the case of the forces, the similarity between this result and
 the Hellmann-Feynman expression is apparent.

 To end this section, let us stress that only excited electrons/holes,
 which render $D_{\bm{a}\bm{b}}\neq 0$, create forces and stresses not
 included in the underlying force-field described by $E^{(0)}$. In
 fact, in the typical case, the dominant contribution to both forces
 and stresses will come from the derivative of $E^{(0)}$, with
 corrections that are linear in the difference occupation matrix.

\subsection{Practical considerations}
\label{sec:practical}

 So far we have introduced a method for the simulation of materials at
 a large scale.
 We have presented the basic physical ingredients (reference atomic
 geometry, reference electronic density, deformation density, etc.),
 that allow us to approximate the DFT total energy, forces and
 stresses.

 In this section we discuss some practicalities involved in the
 implementation of this method in a computer code to perform actual
 calculations.
 Of course, different implementations are in principle possible; here
 we briefly describe some details pertaining to our specific choices,
 which should be illustrative of the technical issues that need to be
 tackled.
 
 \subsubsection{Definition of the RED}
 \label{sec:defred}
 
 The formulation above is written in terms of differences between the
 actual and reference states of the system in a completely general
 way.  However, from a practical point of view, an appropriate choice
 of the RED, $n_0(\vec{r})$, is a necessary first step towards an
 efficient implementation of our method.

 The most important ingredient to define $n_0$ is the reference
 occupation matrix that, following Eq.~(\ref{eq:defdab}), amounts to
 \begin{equation}
    d_{\bm{a}\bm{b}}^{(0)} = \sum_{j\vec{k}} o_{j\vec{k}}^{(0)} 
      \left[c_{ja\vec{k}}^{(0)}\right]^{*} c_{jb\vec{k}}^{(0)}
      e^{i\vec{k}(\vec{R}_B-\vec{R}_A)},
    \label{eq:densd0}
 \end{equation}
 where $o_{j\vec{k}}^{(0)}$ and $c_{ja\vec{k}}^{(0)}$ characterize the
 RED. While it would be possible to use the $d_{\bm{a}\bm{b}}^{(0)}$
 result computed from first principles to perform second-principles
 simulations, in the following we shall simplify this expression in
 order to obtain a more convenient form.

 Note that the reference occupation matrix satisfies
 $d_{\bm{a}\bm{b}}^{(0)} = 0$ for $\bm{a}$ and $\bm{b}$ belonging to
 different band manifolds.
 (By definition, if $\bm{a}$ and $\bm{b}$ belong to different bands,
 they cannot appear simultaneously in the expansion of a particular
 Bloch state [Eq.~(\ref{eq:defBloch})], and the corresponding
 $d_{\bm{a}\bm{b}}^{(0)}$ [Eq.~(\ref{eq:densd0})] will vanish.)
 It is thus possible to rewrite Eq.~(\ref{eq:densd0}) and split the
 sum over states in two, one over manifolds $\mathcal{J}$ and a second
 one over bands within a manifold.

 After having established this property, we impose that all the bands
 $j$ that belong to the same manifold $\mathcal{J}$ have the same
 occupation in the RED
 \begin{equation}
    o_{j \vec{k}}^{(0)} = o_{\mathcal{J}} \omega_{\vec{k}} = 
      \frac{ o_{\mathcal{J}}}{N_{k}} ,
    \label{eq:occsep}
 \end{equation}
 where $\omega_{\vec{k}}$ is the weight of each $\vec{k}$-point in the BZ.
 As we assume an homogeneous sampling in reciprocal space, 
 $\omega_{\vec{k}} = N_{k}^{-1}$, where $N_{k}$ is the total number of 
 points in our BZ mesh.
 Thus, for example, in a diamagnetic insulator [see
   Fig.~\ref{fig:entanglement}(a)] (where the valence and conduction
 band always belong to different manifolds) we would choose the
 occupation for the reference states so that all the valence bands are
 fully occupied while all conduction bands are completely empty.
 In this way the reference electronic density for a diamagnetic
 insulator is simply the ground state density.

 For metals [Fig.~\ref{fig:entanglement}(b)], and magnetic insulators
 [Fig.~\ref{fig:entanglement}(c)], where a disentanglement
 procedure\cite{souza_prb01} has to be carried out to separate the
 desired bands from others with which they are hybridized in a given
 energy window, the choice is not so simple.
 In such cases we distribute all the electrons of the entangled bands
 equally among the bands in the manifold.
 For example, in the case of metallic copper
 [Fig.~\ref{fig:entanglement}(b)], which has an electronic
 configuration 3$d^{10}$4$s^1$ where the five 3$d$ functions cross
 with the 4$s$-like band, we would distribute the eleven electrons over
 the six bands taking $o_{\mathcal{J}}=11/6$.
 On the other hand, for NiO [Fig.~\ref{fig:entanglement}(c)] some
 Ni(3$d$) bands are occupied and entangled with the O(2$p$)
 ones; at the same time, empty $e_g$-like orbitals are part of the
 conduction band and entangled with other levels there. Here, we
 choose to disentangle the bands with strong Ni(3$d$) and O(2$p$)
 character from the other bands. Further, we assign the occupation by
 distributing the corresponding electrons -- eight 3$d$ electrons of
 Ni$^{2+}$ and the six 2$p$ electrons of O$^{2-}$ -- over the
 corresponding bands -- five 3$d$ bands and three 2$p$ bands --
 yielding $o_{\mathcal{J}}=14/8$.  Taking into account the spin
 polarization, the occupation per spin channel is just
 $o_{\mathcal{J}}=7/8$.
 
 Using Eq.~(\ref{eq:occsep}) to rewrite Eq.~(\ref{eq:densd0}), and
 taking into account the relationship between the coefficients of RED
 Bloch states and the unitary transformations between Bloch and
 Wannier representations, we have
 \begin{align}
    d_{\bm{a}\bm{b}}^{(0)} & = N_{k}^{-1} o_{\mathcal{J}}
      \sum_{\vec{k}} e^{i\vec{k}(\vec{R}_B-\vec{R}_A)}
      \sum_{j} 
         \left[\left( T_{aj}^{(\vec{k})} \right)^{-1}\right]^{*} 
         \left(T_{bj}^{(\vec{k})}\right)^{-1} 
      \nonumber \\
         & = N_{k}^{-1}  o_{\mathcal{J}} 
         \sum_{\vec{k}} e^{i\vec{k}(\vec{R}_B-\vec{R}_A)}
         \sum_{j} 
         T_{ja}^{(\vec{k})} 
         \left(T_{bj}^{(\vec{k})}\right)^{-1} 
      \nonumber \\
         & = N_{k}^{-1} o_{\mathcal{J}} 
         \sum_{\vec{k}} e^{i\vec{k}(\vec{R}_B-\vec{R}_A)}
         \delta_{ab}
      \nonumber \\
         & = N_{k}^{-1} o_{\mathcal{J}} \delta_{ab}         
         \sum_{\vec{k}} e^{i\vec{k}(\vec{R}_B-\vec{R}_A)}         
      \nonumber \\
         & = o_{\mathcal{J}} \delta_{ab}
         \delta_{\vec{R}_A\vec{R}_B}  = o_{\mathcal{J}} \delta_{\bm{a}\bm{b}} ,
    \label{eq:wanocc}
 \end{align}
 where we have
 used the properties of the unitary matrices.  From this expression we
 see that $o_{\mathcal{J}}$ is simply the occupation of the WF
 $\chi_{\bm{a}}$ in the reference state, $o_{\bm{a}}^{(0)}$.
 Finally, inserting Eq.~(\ref{eq:wanocc}) into Eq.~(\ref{eq:densW0}),
 we arrive to the conclusion that
 \begin{align}
   n_0(\vec{r}) &= \sum_{\bm{a} \bm{b}} d^{(0)}_{\bm{a}\bm{b}} 
      \chi_{\bm{a}}(\vec{r}) \chi_{\bm{b}}(\vec{r})
   \nonumber \\
      &= \sum_{\bm{a} \bm{b}} o_{\mathcal{J}} 
         \delta_{\bm{a} \bm{b}}      
         \chi_{\bm{a}}(\vec{r}) \chi_{\bm{b}}(\vec{r})
   \nonumber \\
      &= \sum_{\bm{a}} o_{\mathcal{J}}
      \vert \chi_{\bm{a}}(\vec{r}) \vert^{2} 
   \nonumber \\
      & = \sum_{\bm{a}} o_{\bm{a}}^{(0)} \vert \chi_{\bm{a}}(\vec{r})\vert^2 ,
   \label{eq:n0diag}
 \end{align}
 where we have used the fact that
 $o_{\bm{a}}^{(0)} = o_{\mathcal{J}}$ for all WFs in the manifold.

 We would like to stress that this approach simply allows for a more
 efficient computational method since we can still retrieve the full
 electron distribution in space by substituting the first-principles
 WFs into Eqs.~(\ref{eq:densW}) and (\ref{eq:densW0}).

 \subsubsection{Deformation electron density}
 \label{sec:defden}

 From the definition of the charge density in terms of the density
 matrix, Eq.~(\ref{eq:densW}), and the orthogonality of the Wannier
 basis functions, it trivially follows that the total number of
 electrons in the system is the trace of the density matrix,
 \begin{align}
    N & = \int n (\vec{r}) d^{3} r =  
        \int \sum_{\bm{a}\bm{b}} d_{\bm{a}\bm{b}} 
             \chi_{\bm{a}}(\vec{r})\chi_{\bm{b}}(\vec{r}) d^{3}r
   \nonumber \\
      & = \sum_{\bm{a}\bm{b}} d_{\bm{a}\bm{b}}
          \int \chi_{\bm{a}}(\vec{r})\chi_{\bm{b}}(\vec{r}) d^{3}r
   \nonumber \\
      & = \sum_{\bm{a}\bm{b}} d_{\bm{a}\bm{b}}
          \delta_{\bm{a}\bm{b}}
   \nonumber \\
      & = \sum_{\bm{a}} d_{\bm{a}\bm{a}}.
   \label{eq:tracedensmat}
 \end{align}
 Therefore, the trace of the deformation matrix gives the number of
 extra electrons or holes that dope the system.

 From the very definition of the deformation matrix, we can deduce
 that if its diagonal element,
 \begin{equation}
 D_{\bm{a}\bm{a}}=d_{\bm{a}\bm{a}}-d_{\bm{a}\bm{a}}^{(0)},
 \end{equation}
 is negative (positive), that means that we are creating holes 
 (inserting electrons) on that particular state, as illustrated
 in Fig.~\ref{fig:schemehamil}.

 The fact that most of these electron/hole excitations take place
 around the Fermi energy has a very important consequence with regards
 to the efficiency of the method.
 In order to calculate $D_{\bm{a}\bm{a}}$, and the total energy [see
   Eq.~(\ref{eq:totalenergy1})], we do not need to obtain all the
 eigenvalues of the one-electron Hamiltonian, but just those around
 the Fermi energy.
 This opens up the possibility to use efficient diagonalization
 techniques that allow a fast calculation of a few relevant
 eigenvalues, e.g. Lanczos. This approach allows us to speed up the
 calculations in a very significant manner. (The diagonalization of
 the full Hamiltonian matrix is one of the main computational
 bottlenecks in electronic structure methods.)
 Along these lines, the possibility of obtaining linear scaling within
 our method will be discussed in a future publication.
  
\section{Parameter calculation}
\label{sec:parameters}

 The method presented above allows for the simulation of very large
 systems under operation conditions assuming that a few parameters
 describing one-electron and two-electron interactions, as well as the
 electron-lattice couplings, are known beforehand.
 For the sake of preserving predicting power, it is important to
 compute those parameters from first principles.

 All the electronic parameters of our models have well-defined
 expressions [see Eq.~(\ref{eq:gamma}) for $\gamma_{\bm{a}\bm{b}}$,
   Eq.~(\ref{eq:u_integral}) for
   $U_{\bm{a}\bm{b}\bm{a}^\prime\bm{b}^\prime}$ and
   Eq.~(\ref{eq:i_integral}) for
   $I_{\bm{a}\bm{b}\bm{a}^\prime\bm{b}^\prime}$], whose computation
 requires only the knowledge of the Wannier functions, the
 one-electron Hamiltonian, and the operators involved in the double
 integrals, all of them defined for the RED.
 Since the chosen basis functions are localized in space, the required
 calculations could be performed on small supercells. Such a direct
 approach to obtain the model parameters is thus, in principle,
 feasible.

 Note that there has been significant work to calculate related
 integrals from first principles, as can be found e.g. in
 Refs.~\onlinecite{marzari_prb97, mostofi_cpc08,liechtenstein_prb95,
   anisimov_jpcm97, aryasetiawan_rpp98,anisimov_prb91,solovyev_jpcm08,
   solovyev_prb05,gunnarsson_prb89, miyake_prb08,nakamura_prb06,
   vaugier_prb12}.
 Yet, we feel that most of these approaches are too restrictive for
 the more general task that we pursue in this work.
 For instance, the focus in the previous references is placed on
 strongly correlated electrons in a single center, while we are also
 interested in multi-center integrals.

 A significant effort would thus be required to implement the
 calculation of the more complex interactions, including all the
 potentially relevant ones, and developing tools to derive minimal
 models that retain only the dominant parameters and capture the main
 physical effects.
 Note that, in a typical system, the number of potentially relevant
 integrals will be very large.
 In fact, the presence of four-index integrals like
 $U_{\bm{a}\bm{b}\bm{a}^\prime\bm{b}^\prime}$ and
 $I_{\bm{a}\bm{b}\bm{a}^\prime\bm{b}^\prime}$ is the reason why
 Hartree-Fock schemes scale much worse than DFT methods with respect
 to the number of basis functions in the calculation [$\sim
   \mathcal{O}(N^5)$ vs. $\sim \mathcal{O}(N^3)$, respectively].
 Hence, at the present stage we have not attempted a direct
 first-principles calculation of the parameters, which is a challenge
 that remains for the future. 
 Instead, we have devised a practical scheme to {\em fit} our models
 to relevant first-principles data.

\subsection{Parameter fitting}

 Our procedure comprises several steps. 

\subsubsection*{Training set}

 First, we identify a {\em training set} (TS) of representative atomic and
 electronic configurations from which the relevant model parameters
 will be identified and computed.
 For example, the training set for a magnetic system should contain
 simulations for several spin arrangements, so that the mechanisms
 responsible for the magnetic couplings can be captured.
 Additionally, if we want to study a system whose bands are very
 sensitive to the atomic structure, the training set should contain
 calculations for different geometries so that this effect is
 captured.
 Alternatively, if we want to describe how doping affects the physical
 properties of a material, then different DFT simulations on charged
 systems should be carried out,\cite{Makov-95} etc.

 Let us note that it is typically possible to restrict the TS to
 atomic and electronic configurations compatible with small simulation
 boxes. This translates into (and is consistent with) the fact that,
 when expressed in a basis of localized WFs, the non-electrostatic
 interactions in most materials are short ranged.

 We will use $N_{\rm TS}$ to denote the total number of TS elements,
 noting that we will run a single-point first-principles calculation
 for each of them.
 Further, $N_{\rm RAG}$ is the number of TS configurations that
 correspond to the reference atomic geometry.

\subsubsection*{Filtering the training set}

 Let $h_{\bm{a}\bm{b}}^s(i)$ be the Hamiltonian of the $i$-th TS
 configuration, in matrix form and as obtained from a first-principles
 (typically DFT) calculation.
 We denote the whole collection of one-electron Hamiltonians in the
 training set by $\{ h_{\bm{a}\bm{b}}^s (i) \}$.

 These Hamiltonians are expressed in a basis of localized
 WFs. Formally, they can be obtained by inverting Eq.~(\ref{eq:hone})
 (see Sec. VI A of Ref.~\onlinecite{marzari_rmp12}), so that
 \begin{equation}
   h_{\bm{a}\bm{b}}^s =
  \frac{(2\pi)^3}{V}\int_\text{BZ}d^3k
   \left[
      \sum_j \left[ T_{ja}^{s (\vec{k})} \right]^\star \varepsilon_{j\vec{k}}^s
                    T_{jb}^{s (\vec{k})}
   \right]
   e^{i\left(\vec{R}_A-\vec{R}_B\right)\vec{k}},
   \label{eq:honewf}
 \end{equation}
 where the ${\bm T}$ matrices are unitary transformations that convert
 the first-principles eigenstates into Bloch-like waves associated to
 specific (localized) WFs.
 These transformations can be obtained by employing codes like {\sc
   wannier90},\cite{mostofi_cpc08} which implements a particular
 localization scheme, i.e., a particular way to compute optimum ${\bm
   T}$ matrices.\cite{marzari_prb97,marzari_rmp12}

 Once the $h_{\bm{a}\bm{b}}^s(i)$ Hamiltonians are known, we can
 identify the pairs of WFs with a large enough interaction and which
 need to be retained in the fitting procedure.
 In practice, we introduce a cut-off energy $\delta \varepsilon_h$
 such that
 \begin{equation}
   \vert h_{\bm{a}\bm{b}}^s (i) \vert > \delta \varepsilon_h,
   \text{for at least one $i$ in the TS},
 \end{equation}
 defines the Hamiltonian matrix elements to be retained.  (Diagonal
 elements, $h_{\bm{a}\bm{a}}^s$, are always considered independently
 of their value.)  This condition allows us to identify the WF pairs
 $(\bm{a}$, $\bm{b})$ to be included in the fitting procedure,
 regardless of the geometry or spin arrangement.

\begin{figure*} [t]
    \begin{center}
       \includegraphics[width=1.8\columnwidth]{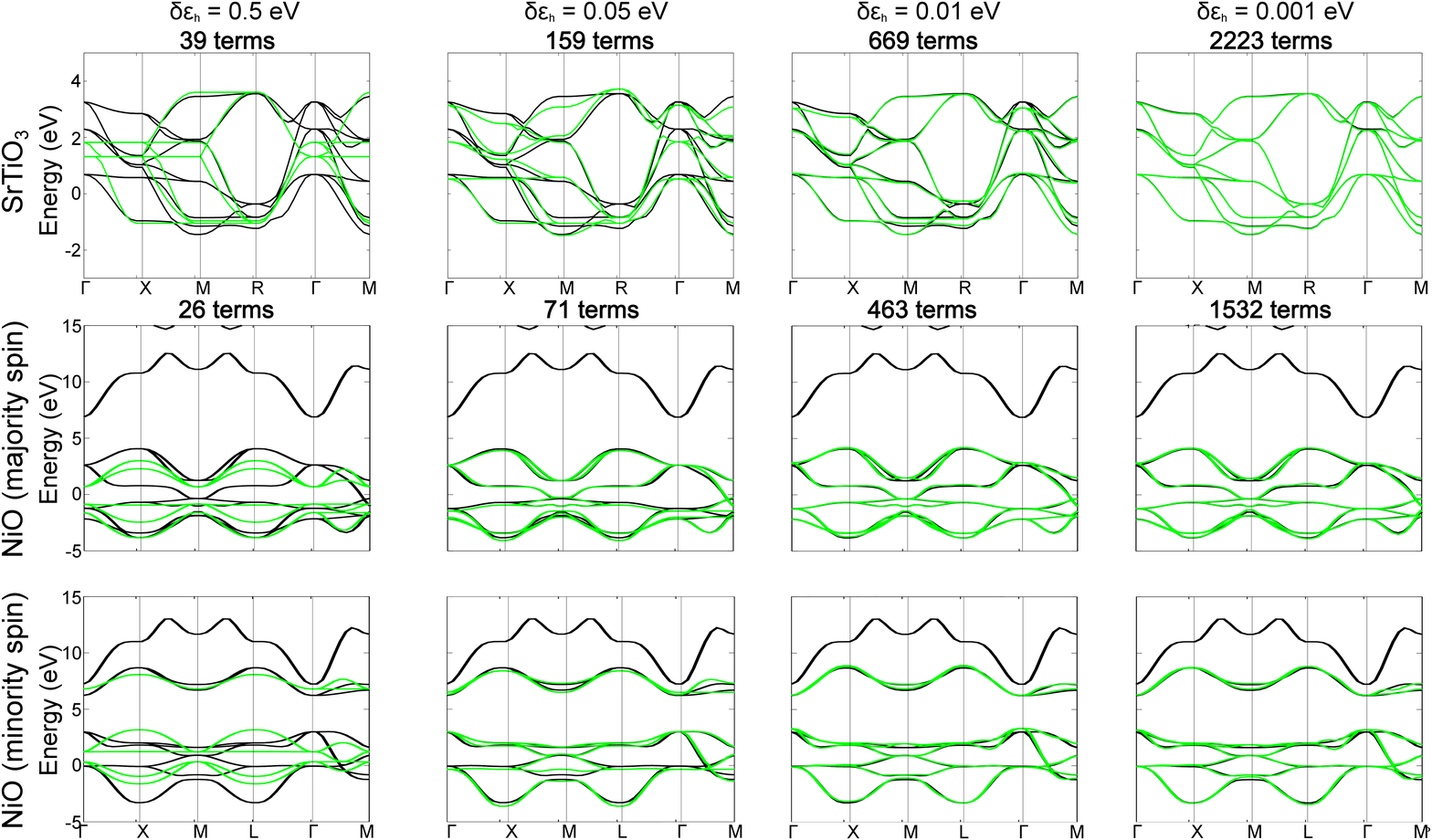}
          \caption{(Color online) Representation of the bands of
            SrTiO$_3$ (top row) and NiO (middle and bottom rows,
            corresponding to the majority and minority spin channels,
            respectively) for various values of the Hamiltonian
            cutoff, $\delta \varepsilon_h$.  Black lines represent the
            bands as obtained from first principles, while the results
            from the second-principles model are shown in green.  On
            top of each plot we indicate the corresponding number of
            $h_{\bm{a}\bm{b}}^s$ matrix elements per primitive cell
            that were included.  }
       \label{fig:cutoff}
    \end{center}
 \end{figure*}

 In Fig.~\ref{fig:cutoff} we compare the full first-principles bands
 for SrTiO$_3$ and NiO with those obtained from models corresponding
 to different energy cut-offs.
 For all the $\delta \varepsilon_{h}$ values considered, we also
 indicate the number of parameters in the resulting models.
 This allows us to estimate the size of the model (and associated
 computational cost) needed to achieve an acceptable description of
 the band structure.
 
 \subsubsection*{Identifying most relevant model interactions}

 Our models, even though we truncate them at second order of the
 expansion in Eq.~(\ref{eq:totale}), contain a daunting number of
 electron-electron interaction parameters. Constructing an actual
 model usually involves further approximations regarding the spatial
 range of the interactions, the maximum number of different bodies
 (WFs) involved, etc. Hence, we need a procedure to identify the
 simplest models that can reproduce the first-principles TS data with
 an accuracy that is sufficient for our purposes.

 The scheme we have implemented is based on a very simple logic: We
 start from a certain {\em complete} model that may contain, in
 principle, all possible one-electron, two-electron, and
 electron-lattice parameters. We can then fit such a model to
 reproduce the one-electron Hamiltonians $\{ h_{\bm{a}\bm{b}}^s (i) \}$ of
 our TS within a certain accuracy. Typically, by doing so, and by
 systematically exploring different combinations of parameters in the
 model, we can identify the simplest interactions (i.e., those that
 are shortest in range, involve fewest WFs, etc.) sufficient to
 achieve the desired level of accuracy; in other words, in this way we
 can identify non-critical couplings that just render the fitting
 problem underdetermined, do not improve the accuracy of the model,
 and can thus be disregarded. Naturally, this split between relevant
 and irrelevant couplings is strongly dependent on the choice of the
 training set, which should be complete enough to capture the physical
 effects of interest.

 To better understand how the scheme works, consider the one-electron
 integrals $\gamma_{\bm{a}\bm{b}}$ in the case of a non-magnetic
 material like SrTiO$_{3}$. These parameters will be the only ones
 entering the description of the band structure of the RAG in the RED
 state. Hence, we can fit them directly by requiring our model to
 reproduce the Hamiltonian $h_{\bm{a}\bm{b}}$ of this particular,
 reference state with a certain accuracy. 

 More generally, the one-electron Hamiltonian corresponding to a TS
 configuration will reflect electronic excitations departing from the
 RED state. More precisely, we can recall Eq.~(\ref{eq:honeR}) to
 write
 \begin{align}
     h_{\bm{a}\bm{b}}^s(i) = \gamma_{\bm{a}\bm{b}} + 
       \sum_{\bm{a^\prime b}^\prime} &
            \left[
                \left(
                   D_{\bm{a}^\prime \bm{b}^\prime}^s(i)+
                   D_{\bm{a}^\prime \bm{b}^\prime}^{-s}(i)
                \right)
                U_{\bm{a b a^\prime b}^\prime}-
            \right.
            \nonumber \\
       &
            \left.
                \left(
                   D_{\bm{a}^\prime\bm{b}^\prime}^s(i)-
                   D_{\bm{a}^\prime\bm{b}^\prime}^{-s}(i)
                \right)
                I_{\bm{ab a^\prime b}^\prime}
            \right],
 \end{align}
 where we restrict ourselves to TS configurations at the RAG, so that
 no electron-lattice term appears in this equation.
 It is convenient to isolate the $U$ contribution by defining 
 \begin{equation}
    h_{\bm{a}\bm{b}}^U (i) =
      \frac{h_{\bm{a}\bm{b}}^\uparrow (i) +h_{\bm{a}\bm{b}}^\downarrow (i) }{2}=
      \gamma_{\bm{a}\bm{b}} + 
         \sum_{\bm{a^\prime b}^\prime} 
                D_{\bm{a}^\prime \bm{b}^\prime}^U (i)
                U_{\bm{a b a^\prime b}^\prime},
    \label{eq:h_U}                
 \end{equation}
 and its average over all the RAG configurations in the training set, 
 \begin{equation}
   \bar{h}_{\bm{a}\bm{b}}^U=\frac{1}{N_{\rm RAG}}\sum_i h_{\bm{a}\bm{b}}^U(i)=
      \gamma_{\bm{a}\bm{b}} +   
         \sum_{\bm{a^\prime b}^\prime} 
               \bar{D}_{\bm{a}^\prime \bm{b}^\prime}^U
                U_{\bm{a b a^\prime b}^\prime} .
   \label{eq:h_U_aver}                
 \end{equation}
 Analogously, the antisymmetrization of the Hamiltonian matrix
 elements with respect to the spin yields
 \begin{equation}
    h_{\bm{a}\bm{b}}^I(i)=
      \frac{h_{\bm{a}\bm{b}}^\uparrow(i)-h_{\bm{a}\bm{b}}^\downarrow(i)}{2}=
       \sum_{\bm{a^\prime b}^\prime} 
                D_{\bm{a}^\prime \bm{b}^\prime}^I(i)
                I_{\bm{a b a^\prime b}^\prime}.
    \label{eq:h_I}                
 \end{equation}

 We expect that the most important $U_{\bm{a}\bm{b a^\prime
     b}^\prime}$ and $I_{\bm{a}\bm{b a^\prime b}^\prime}$ parameters
 will be, respectively, those involving WF pairs whose corresponding
 $h_{\bm{a}\bm{b}}^U(i)$ and $h_{\bm{a}\bm{b}}^I(i)$ are most strongly
 dependent on the TS state.
 Hence, we introduce the two-electron cutoff energy,
 $\delta\varepsilon_{ee}$, and retain only the $(\bm{a},\bm{b})$ pairs
 that satisfy, for at least one TS configuration, at least one of the
 following conditions:
 \begin{equation}\label{eq:cutoff2U}
     \vert h_{\bm{a}\bm{b}}^U(i)-\bar{h}_{\bm{a}\bm{b}}^U \vert = 
       \left \vert
         \sum_{\bm{a^\prime b}^\prime} 
             \left[
                D_{\bm{a}^\prime \bm{b}^\prime}^U(i) - 
                \bar{D}_{\bm{a}^\prime \bm{b}^\prime}^U
             \right]
                U_{\bm{a b a^\prime b}^\prime}
        \right \vert
        > \delta\varepsilon_\text{ee},
 \end{equation}
 or
 \begin{equation}\label{eq:cutoff2I}
    \vert h_{\bm{a}\bm{b}}^I(i)\vert =
       \left \vert
         \sum_{\bm{a^\prime b}^\prime} 
                D_{\bm{a}^\prime \bm{b}^\prime}^I(i)
                I_{\bm{a b a^\prime b}^\prime}
        \right \vert
        > \delta\varepsilon_\text{ee}.
 \end{equation}
 Note that we gauge the $h_{\bm{a}\bm{b}}^U$ matrix elements with
 respect to the $\bar{h}_{\bm{a}\bm{b}}^U$ average values so that the
 corresponding cut-off condition does not depend on the one-electron
 couplings $\gamma$.

 Once we have selected all the $\{(\bm{a},\bm{b})\}$ pairs that
 fulfill such criteria, we can build the list of potentially relevant
 $U$ and $I$ constants to be considered in the fit. 
 Note that the number of free parameters is usually reduced by the
 fact that the $U_{\bm{a}\bm{b}\bm{a}^\prime \bm{b}^\prime}$ and
 $I_{\bm{a}\bm{b}\bm{a}^\prime \bm{b}^\prime}$ integrals are invariant
 upon permutations of the $(\bm{a}, \bm{b}, \bm{a}^\prime,
 \bm{b}^\prime)$ indexes.
 In some cases, and in spite of the reduction of parameters due to
 symmetry, the list of relevant interactions is excessively long and
 needs to be further trimmed to successfully carry out the fitting.
 In such situations we introduce a third cutoff, $\delta D$, that
 operates over the difference occupation matrix to select the
 interactions associated to important changes of the electron density.
 When doing so we only accept $U$ constants for which at least one
 pair of the associated indexes fulfills
 \begin{equation}\label{eq:cutoff_U}
    \vert D_{\bm{a^\prime b}^\prime}^U(i)-
          \bar{D}_{\bm{a^\prime b}^\prime}^U
    \vert > \delta D
 \end{equation}
 and the corresponding expression for $I$
 \begin{equation}\label{eq:cutoff_J}
    \vert D_{\bm{a^\prime b}^\prime}^I(i) \vert > \delta D.
 \end{equation} 
 
 Let us also note that the most relevant $\gamma_{\bm{a}\bm{b}}$
 parameters are trivially identified when we filter the TS
 one-electron Hamiltonians as described above.
 
 \subsubsection*{Fitting the RAG model}

 Once our list of relevant $\gamma$, $U$, and $I$ parameters is
 complete, we fit them to reproduce the $\{h^s_{\bm{a}\bm{b}}\}$
 matrix elements above the $\delta\varepsilon_h$ energy cutoff
 introduced previously.

 We have found it convenient to perform the fit in several steps, so
 that different types of parameters are computed separately. 
 More precisely, we first fit the $U$ parameters by requesting that
 our model reproduces the
 $h_{\bm{a}\bm{b}}^U(i)-\bar{h}_{\bm{a}\bm{b}}^U$ matrices
 [Eqs.~(\ref{eq:h_U}) and (\ref{eq:h_U_aver})].
 Analogously, we obtain the $I$ constants by fitting to the
 $h_{\bm{a}\bm{b}}^I(i)$ matrices [Eq.~(\ref{eq:h_I})].
 Importantly, both of these fits are independent of the one-electron
 integrals, and have typically yielded well-posed, overdetermined
 systems of equations in the cases we have so far considered.
 Finally, we obtain the $\gamma$ parameters from the fitted $U$'s
 directly from Eq.~(\ref{eq:h_U}). 
 Direct comparison of the modeled bands with those obtained from the
 full first-principles $\{h^s_{\bm{a}\bm{b}} (i)\}$ set provides an
 estimate of the goodness of the model (see the example in Section
 \ref{sec:nio} and, particularly, Fig.~\ref{fig:niofit}).

 Note that, alternatively, one could try a direct fit of all the
 $\gamma$, $U$, and $I$ parameters to the real-space Hamiltonians,
 using Eq.~(\ref{eq:honeR}). However, we typically find that this
 strategy leads to nearly-singular problems in which very different
 solutions lead to comparably good results. In the general case, such
 a difficulty may be mitigated by extending the TS. However, here we
 adopted the simple and practical procedure described above, which
 permits a numerically stable method that yields accurate and
 physically sound models.
 
 To end with this section we would like to stress that the
 $\gamma_{\bm{a}\bm{b}}$ constants obtained with this procedure
 contain both the short- and long-range contributions described above
 [Eq.~(\ref{eq:gammashlg})]. In order to isolate
 $\gamma^{\text{sr}}_{\bm{a}\bm{b}}$, we simply subtract the corresponding
 electrostatic contribution [Eq.~(\ref{eq:gamma-lr})] from the
 determined, full $\gamma_{\bm{a}\bm{b}}$ value.
 In order to calculate the electrostatic contribution
 [Eqs.~(\ref{eq:localdip})-(\ref{eq:farfieldpot})] we need
 first-principles results for the Born charge tensor,
 $\overleftrightarrow{Z}_{\bm \lambda}^*$, and the high-frequency
 dielectric tensor, $\overleftrightarrow{\epsilon_\infty}$, that can
 routinely be obtained for systems where the RED is insulating.
 \cite{Note1}

 \subsubsection*{Relevant electron-lattice interactions}
 \label{sec:electronlatticefit}

 As above, we assume that the deviations from the RAG only affect the
 one-electron integrals $\gamma$, and not the $U$ and $I$
 parameters. We further assume that such a dependence on the atomic
 structure is given by the linear and quadratic electron-lattice
 constants $\vec{\bm{f}}$ and $\overleftrightarrow{\bm{g}}$ introduced
 in Eq.~(\ref{eq:gammadist}).

 The selection of the most important electron-lattice couplings is
 performed by observing how much a particular matrix element
 $h^s_{\bm{a}\bm{b}}$ changes with a particular distortion of the
 lattice with respect to the RAG.
 To quantify this change, we need to compare pairs of configurations
 $i$ and $i'$ that correspond to the same electronic state
 (e.g., to the same spin arrangement, to the same amount of
 electron/hole doping, etc.) {\em but} differ in their atomic
 structure. More precisely, one of the configurations must correspond
 to the RAG state ($i$), while the other one ($i'$) is characterized by a
 distortion given by $\{\vec{u}_{\bm \lambda}(i')\}$. For simplicity,
 here we will restrict to distortions involving only one displacement
 component of one atom, so that we only have one specific $u_{{\bm
     \lambda}\alpha}(i') \neq 0$, where $\alpha$ labels the spatial
 direction.
 We then consider that a particular atom $\bm{\lambda}$ participates
 in the electron-lattice affecting the $h_{\bm{a}\bm{b}}^{s}$ element
 if
 \begin{equation}\label{eq:el-latt-cutoff}
   \frac{1}{|u_{{\bm \lambda}\alpha}|} \left\vert
   h_{\bm{a}\bm{b}}^{s}(i') -h_{\bm{a}\bm{b}}^{s}(i)\right\vert >\delta
   f_{e-l},
 \end{equation}
 where $\delta f_{e-l}$ is a new cut-off. Note that, for a large
 enough distortion $|u_{{\bm \lambda}\alpha}|$, this condition
 pertains to both the linear ($\vec{\bm{f}}$) and quadratic
 ($\overleftrightarrow{\bm{g}}$) electron-lattice interactions. Yet,
 since we activate a single atomic displacement at a time, in the case
 of $\overleftrightarrow{\bm{g}}$ we are only probing the diagonal
 elements. Restricting ourselves to the diagonal elements of
 $\overleftrightarrow{\bm{g}}$ is justified by the observation that,
 in the systems we have so far studied, those are the only significant
 ones. At any rate, the scheme can be trivially extended to check a
 possible contribution of off-diagonal terms.

 \section{Implementation of the algorithm: The {\sc scale-up} code}
 \label{sec:scaleup}

 We have implemented this new method in the {\sc scale-up}
 (Second-principles Computational Approach for Lattice and Electrons)
 package, written in Fortran~90 and parallelized using Message Passing
 Interface (MPI).
 Presently, this code can perform single-point calculations, geometry
 optimizations, and Born-Oppenheimer molecular dynamics using either
 full diagonalization or the Lanczos scheme mentioned above.

 The energy of the reference state, $E^{(0)}$, is obtained from model
 potentials like those introduced by Wojde\l\ {\sl et
   al.},\cite{wojdel_jpcm13} which are interfaced with {\sc scale-up}.
 We have also developed an auxiliary toolbox ({\sc modelmaker}) for
 the calculation of all the parameters defining $E^{(1)}$ and $E^{(2)}$, using as
 input DFT results for one-electron Hamiltonians in the format of {\sc
   wannier90}.\cite{mostofi_cpc08} As shown in Sec.~\ref{sec:examples},
 these implementations can be used to create models that match the
 accuracy of the DFT calculations at an enormously reduced
 computational cost, opening the door to large-scale simulations (up
 to tens of thousands of atoms) of systems with a complex electronic
 structure, using modest computational resources.

 The input to the code is based on the flexible data format (fdf)
 library used in {\sc siesta}\cite{Soler-02} and contains several
 python-based tools to plot bands, density of states, geometries and
 other properties.

\section{Examples of application}
\label{sec:examples}

 In order to illustrate the method, we will discuss its application to
 two non-trivial systems with interactions of very different origin.

 The first example consists in the calculation of the energy of a
 Mott-Hubbard insulator, NiO, for different magnetic phases.
 Our goal here is to show that the method can be used to deal
 accurately with complicated electronic structures including phenomena
 like magnetism in transition metal oxides.
 In this example we will also show how the method can tackle rather
 large systems (2,000 atoms) that are at the limit of what can be done
 with first-principles methods today, reducing the computational
 burden by orders of magnitude.

 The second application involves the two-dimensional electron gas (2DEG)
 that appears at the interface between band insulators LaAlO$_3$ 
 and SrTiO$_3$.~\cite{ohtomo_nat04}
 We will not discuss here the origin of the 2DEG, which has been
 treated in great detail in the
 bibliography.\cite{ohtomo_nat04,nakagawa_natmat06,stengel_prl11,Bristowe-14}
 Rather, we will check whether our approach can predict the
 redistribution of the conduction electrons at the LaAlO$_{3}$/SrTiO$_{3}$ 
 interface,
 and the accompanying lattice distortion, as obtained from first principles.
 Thus, this example will showcase the treatment of electron-lattice
 couplings and electrostatics within our approach.

 \subsection{Details of the first-principles simulations}
 \label{sec:compdetail}

 We construct our models following the recipes described in
 Sec.~\ref{sec:parameters}, and the first-principles data are obtained
 from small-scale calculations with the {\sc VASP}
 package.\cite{kresse_prb96,kresse_prb99,blochl_prb94}
 The local density approximation (LDA) to density-functional theory
 is used to create the TS data for SrTiO$_3$. The calculations for NiO
 are also based on the LDA, but in this case an extra Hubbard-$U$ term
 is included to account for the strong electron
 correlations,\cite{dudarev_prb98} as will be discussed below.
 We employ the projector-augmented wave (PAW)
 scheme\cite{blochl_prb94} to treat the atomic cores, solving
 explicitly for the following electrons: Ni's $3s$, $3p$, $3d$, and
 $4s$; O's $2s$ and $2p$; Sr's $3s$, $3p$, and $4s$; and Ti's $3s$,
 $3p$, $3d$, and $4s$.
 The electronic wave functions are described with a plane-wave basis
 truncated at 300~eV for NiO and at 400~eV for SrTiO$_3$. The
 integrals in reciprocal space are carried out using
 $\Gamma$-centered 4$\times$4$\times$4 $k$-point meshes in both cases.

 \subsection{NiO magnetic couplings}
 \label{sec:nio}

 Transition metal oxides are very interesting as they present optical,
 magnetic, and structural properties that are, very often, tightly
 coupled with each other. This fact, together with the large variety
 of functional properties that they can display, makes them a big
 focus of attention in both basic and applied materials science.
 From a theoretical point of view their study is complicated, mostly
 because of the strong correlations associated to the electrons in the
 compact $d$~shell (especially, those of first-row 3$d$ transition
 metal ions) and the frequent presence of many competing magnetic
 phases.
 Naively, one may expect most of these oxides to be metallic due to
 their open-shell nature while, in fact, many are insulators.
 This problem strongly affects computational methods; in fact, most
 common approaches, like DFT with local or semilocal exchange-correlation
 functionals, often fail to correctly reproduce the
 magnetic or insulating properties of these compounds. Simulations at
 this level of theory yield too diffuse states with underestimated
 interactions.
 The key to simulate successfully these materials lies in the way
 electron-electron interactions are handled.
 A panoply of methods have been developed to deal with this issue from
 first principles, ranging from the inclusion of a Hubbard-like term
 in the Hamiltonian in the so-called DFT+{\sl U} methods, to more
 sophisticated schemes with dynamically screened interactions, such as
 the GW approximations.\cite{hedin_pra65,aryasetiawan_rpp98}

 For this first application of our scheme, we have chosen a 
 \emph{simple} transition metal oxide, NiO, a staple example of many new
 electronic structure simulation methods.
 Our goal is to show how our second-principles scheme can handily be
 used to compute the properties of strongly correlated materials,
 based on parameters obtained from first-principles LDA+{\sl U}
 simulations.
 In particular we will study the band structure and magnetism of this
 archetypical binary oxide.
 
 \begin{figure} [h]
    \begin{center}
       \includegraphics[width=1.0\columnwidth]{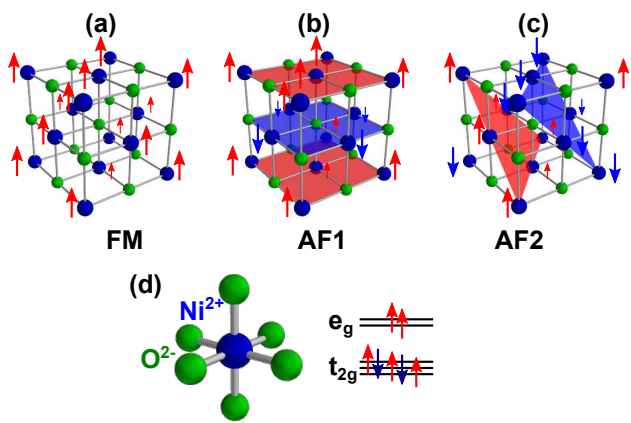}
          \caption{(Color online) Schematic cartoon of different 
                    magnetic configurations of bulk NiO in the 
                    conventional cell of its rocksalt structure: 
                    (a) ferromagnetic phase (FM), 
                    (b) antiferrognatic phase with planes of spin-up 
                    (red arrows)  and 
                    spin-down (blue arrows) polarized nickels 
                    alternating along the 
                    $[001]$ direction (AF1), and
                    (c) antiferrognatic phase with planes of spin-up and 
                    spin-down polarized nickels alternating along the 
                    $[111]$ direction (AF2).
                    (d) Scheme of the $d$-levels associated
                    to an isolated NiO$_6^{10-}$ complex.}
       \label{fig:nioscheme}
    \end{center}
 \end{figure}
 
 \subsubsection{Parameters}
 \label{sec:nio_par}

 The geometric structure of NiO shows some subtle and not fully
 understood distortions associated to its
 magnetism,\cite{cheetham_prb83,schron_prb83} but this issue is beyond
 the scope of the present work. In order to keep the model as simple
 and illustrative as possible, we neglect the lattice degrees
 of freedom in this case.
 As regards the spin order, neutron diffraction
 experiments\cite{roth_pr58} show that the ground state
 corresponds to the so-called AF2 phase, where planes of spin-up and
 spin-down polarized nickels alternate along the $\left\langle 111
 \right\rangle$ direction of the conventional cell [see
 Fig.~\ref{fig:nioscheme}(c)].
 Further experiments\cite{srinivasan_prb83,chatterji_prb09} 
 evidence that the AF2 to paramagnetic transition is of second order
 and occurs at a N\`eel temperature $T_N$~=~524~K.

 The simulations of NiO are carried out in the experimental rocksalt
 cell, with a lattice constant of 4.17~\AA.\cite{schmahl_zac64} This
 cell is compatible with several spin arrangements ranging from fully
 ferromagnetic [FM; Fig.~\ref{fig:nioscheme}(a)]
 to various antiferromagnetic (AF) ones [Fig.~\ref{fig:nioscheme}(b)-(c)]. 
 Our TS includes the ground state spin arrangement (AF2) as well as
 the ferromagnetic one, which we choose because it represents a
 relevant limit case for spin--spin interactions.
 We use the LDA+{\sl U} approach introduced by Dudarev {\sl et
   al}.,\cite{dudarev_prb98} with an effective {\sl U} value of 7~eV,
 applied only on the 3$d$ orbitals of Ni.
 These calculations indicate that the AF2 solution is more stable than
 the FM one by 89~meV per formula unit (f.u.).

 Ligand-field theory predicts the magnetism in this lattice to be the
 result of the half-filled $e_g$ shell of the octahedrally coordinated
 Ni$^{2+}$ ion [see Fig.~\ref{fig:nioscheme}(d)]. Thus, we expect the
 levels around the Fermi energy to have this character.
 After calculating the electronic structure from first principles
 within the LDA+{\sl U}, we find that the top valence and bottom
 conduction bands are composed of several strongly entangled states,
 as shown in Fig.~\ref{fig:entanglement}(c).
 Thus, we project our WFs seeking to disentangle orbitals
 participating in the valence band [Ni($t_{2g}$), Ni($e_{g}$), and
   O($p$)] from others in the conduction band; to do this we use the
 tools provided within the {\sc wannier90} package.
 A graphical representation of the resulting orbitals
 [Fig.~\ref{fig:entanglement}(c)] clearly shows that we are able to
 isolate bands with the expected chemical character: the isosurfaces
 of the maximally-localized WFs (MLWFs) at the right hand side of
 Fig.~\ref{fig:entanglement}(c) clearly resemble the shape of the
 O(${p}$), Ni($d_{xy}$), and Ni($d_{3z^{2}-r^{2}}$) orbitals for the valence
 band, and of the Ni($d_{3z^{2}-r^{2}}$) orbital for the bottom of the conduction
 band.
 Given the strong entanglement of these bands, we use a reference 
 occupation for the construction of our model that is obtained by
 populating equally all of them. As discussed in
 Sec.~\ref{sec:defred}, this amounts to assuming $o_{\mathcal{J}} =
 7/8 = 0.875$ electrons per band and spin channel.

 At this point we start with the analysis of the Hamiltonian as
 described in Sec.~\ref{sec:parameters}, using the
 $\{h^s_{\bm{a}\bm{b}} (i)\}$ set obtained after the disentanglement
 procedure.
 First, we seek to choose a reasonable value of $\delta
 \varepsilon_{h}$ that allows us to describe accurately the system's
 bands without including an excessive number of
 $\gamma_{\bm{a}\bm{b}}$ terms. As can be seen in
 Fig.~\ref{fig:cutoff}, $\delta \varepsilon_h$~=~0.05~eV is a
 reasonable choice; this involves the use of 71 $\gamma$ terms per f.u.

 In order to decide the values of $\delta \varepsilon_{ee}$ and
 $\delta D$ that will determine the $U$ and $I$ parameters considered
 in the fit, we first examine the occupation difference of each of the
 WFs, $D_{\bm{a}\bm{a}}$.
 Let us recall that the diagonal elements of the deformation
 occupation matrices are defined as the difference between the reduced
 density matrix computed at the LDA+{\sl U} for the corresponding
 configuration in the TS and the reference one.
 In the FM phase, the bands with $e_{g}$ character for the majority
 spin channel are expected to be fully occupied ($d_{\bm{a}\bm{a}}$ =
 1), while they should be empty for the minority spin
 ($d_{\bm{a}\bm{a}}$ = 0).  Therefore, for WFs with $e_g$ character,
 we expect to have $D_{\bm{a}\bm{a}}$ = 0.125 for the majority spins
 and $D_{\bm{a}\bm{a}} $ = $-$0.875 for the minority spins, for both
 Ni atoms.
 For the antiferromagnetic configuration, as the spin in one of the Ni
 atoms is reversed, we expect the same behavior for $D_{\bm{a}\bm{a}}$
 as in the FM solution for one of the nickels, while the character of the
 majority and minority spins must be exchanged for the second Ni.
 In both cases (FM and AF2), the $t_{2g}$ bands are fully occupied, so
 that the corresponding $D_{\bm{a}\bm{a}}$ should be 0.125.
 These tendencies are well reproduced in our TS configurations as can
 be seen in Table~\ref{tab:occ}, where the occupations are obtained
 using the $\{h^s_{\bm{a}\bm{b}} (i)\}$ set obtained after applying
 $\delta \varepsilon_h$~=~0.05~eV filter.
 The differences with respect to the ideal ionic values are due to the
 chemical bonding between Ni and O.

 If the results for the FM and AF2 configurations are compared, we can
 see that the only significant change pertains to the occupation of
 the $e_g$-like orbitals of the Ni ion whose spin is flipped: the
 majority (0.125) and minority ($-$0.756) values of the difference
 occupation are basically exchanged as we move from the FM to the AF2
 calculation, as expected from the localization of the magnetic moment
 over these orbitals in Ni$^{2+}$ ions (see Fig.~\ref{fig:nioscheme}).
 This is an indication that, in order to capture the magnetic
 interactions in this system, only electron-electron interactions
 involving $e_g$-type WFs are necessary.
 Moreover, we can observe how, on one hand, the average orbital
 occupations vary very little (essentially by 0.005, 0.001, and
 0.003~electrons for the Ni($e_g$), Ni($t_{2g}$) and O($p$)-like WF,
 respectively, for both spin configurations). This fact translates
 into a similar value of $D^{U}$ [Eq.~(\ref{eq:D_U})].
 As can be seen in Eq.~(\ref{eq:totalenergy1}), if $D^{U}$ is the same
 for the different configurations of the TS, its contribution to the
 total energy is constant, i.e., it does not play any role in the
 calculations of the relevant energy differences.
 On the other hand, the spin-up/spin-down differences of occupation
 are strongly changing between the FM and AF2 configurations.  This
 indicates that only Stoner-type ($I$) interactions are relevant to
 describe the relative stability of the magnetic phases in the
 training set.
 As can be seen in Table-~\ref{tab:nio_res}, by playing with $\delta
 \varepsilon_{ee}$ and $\delta D$ it is straightforward to confirm
 that the $e_g$-$e_g$ interactions drive magnetism in this system,
 while Ni($t_{2g}$)-like and O($p$)-like energy levels have a
 secondary role. Nevertheless, including the latter interactions is
 necessary to accurately describe the bands.

\begin{figure*} [t]
    \begin{center}
       \includegraphics[width=1.5\columnwidth]{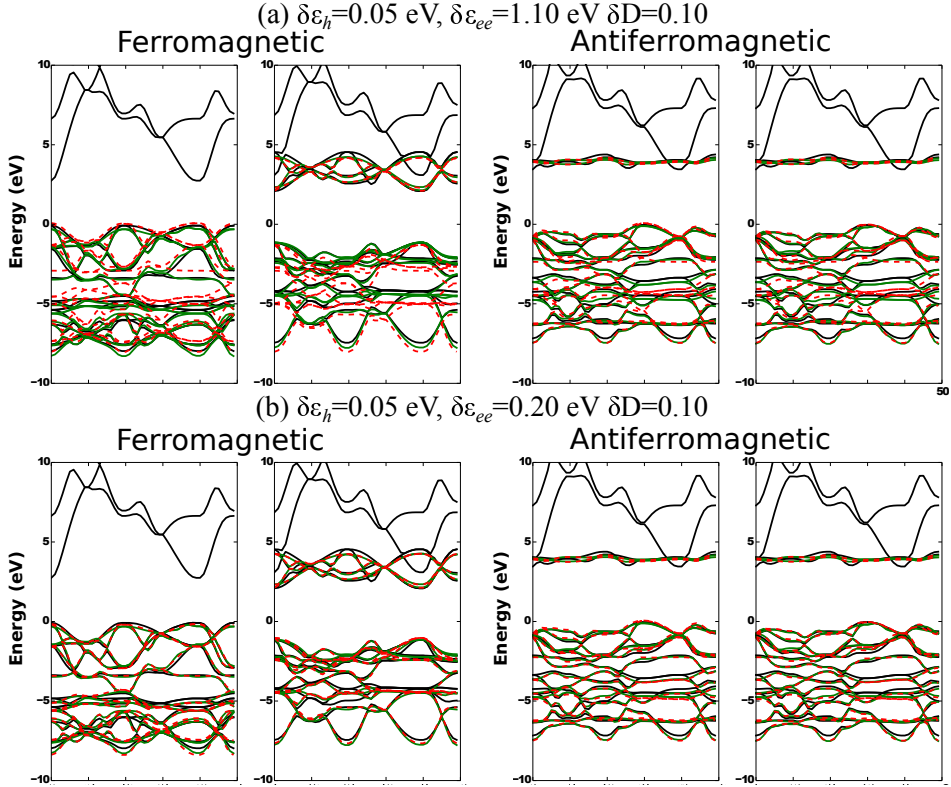}
       \caption{(Color online) Band structure of NiO as obtained for
         two different sets of cutoff parameters.  Solid black lines
         represent first-principles bands obtained from LDA+{\sl U}
         calculations, while solid green lines show the bands as
         obtained after filtering of $h_{\bm{a}\bm{b}}^s$ with a
         cutoff of $\delta\varepsilon_h$=0.05 eV [see
           Sec.~\ref{sec:parameters}].  Dashed red lines represent the
         bands as obtained after a second filtering with (a)
         $\delta\varepsilon_{ee}$=1.10 eV, $\delta D$=0.10 and (b)
         $\delta\varepsilon_{ee}$=0.20 eV, $\delta D$=0.10}
       \label{fig:niofit}
    \end{center}
 \end{figure*}

 In Fig.~\ref{fig:niofit} we show the second-principles computed bands
 for two different set of parameters: (i) the first one, obtained
 after a filtering the electron-electron interactions with a threshold
 of $\delta \varepsilon_{ee}$~=~1.10~eV, was selected to include only
 couplings between Ni($e_g$)-like WFs. The results obtained with this
 set of parameters are labeled SP-LDAU-Ni($e_g$); (ii) the second one,
 obtained with a threshold of $\delta \varepsilon_{ee}$~=~0.20~eV,
 corresponds to a case in which interactions between Ni($e_g$)-like
 and Ni($t_{2g}$)-like WFs at the same atom, as well as between
 Ni($e_g$)-like and nearest-neighboring O($p$)-like WFs, are also
 included.  The results are labeled SP-LDAU-Ni+O.
 As can be seen, the bands for the FM state are better reproduced in
 the second case, because of the correction of the diagonal
 spin-up/spin-down Ni($t_{2g}$) and O($p$) energies.  These diagonal
 Hamiltonian matrix elements, which determine the center of mass of
 the corresponding bands, vary with the Ni($e_g$) occupation as
 expressed in Eq.~(\ref{eq:honeR}).  
 Indeed, we can estimate the maximum error for the
 $h^s_{\bm{a}\bm{b}}$ terms as
 \begin{equation}
    \delta h^s_{\bm{a}\bm{b}} = \max 
       \vert h^s_{\bm{a}\bm{b}}(i) - h^s_{\bm{a}\bm{b}} \vert,
 \end{equation}
 where $h^s_{\bm{a}\bm{b}}(i)$ is a matrix element directly obtained
 from the first-principles TS and $h^s_{\bm{a}\bm{b}}$ is computed 
 from Eq.~(\ref{eq:honeR}) for a given set of parameters.
 This maximum error reduces from 0.651~eV in the SP-LDAU-Ni($e_g$)
 case to 0.132~eV for SP-LDAU-Ni+O.
 We also considered a third, intermediate case with $\delta
 \varepsilon_{ee}$~=~0.5~eV, where Ni($e_g$)-Ni($t_{2g}$) interactions
 are included but those with oxygen orbitals are neglected
 [SP-LDA-Ni(3$d$) in Table \ref{tab:nio_res}].  The associated maximum
 error is 0.29~eV for such a choice.
 
 \begin{table}[h]
    \caption{The difference occupation of the Wannier functions
      ($D_{\bm{a}\bm{a}}$) in the training set used for NiO.  The
      results are presented for the ferromagnetic (FM) and AF2
      antiferromagnetic states, and for the majority (Major) and
      minority (Minor) spin channels.}
    \begin{tabular}{cccccccc}
       \hline
       \hline
       State            & 
       Spin             & 
       Ni$_1$($e_g$)    & 
       Ni$_1$($t_{2g}$) & 
       O$_1$($p$)       & 
       Ni$_2$($e_g$)    & 
       Ni$_2$($t_{2g}$) & 
       O$_2$($p$)       \\
       \hline
       FM               & 
       Major            & 
       0.125            & 
       0.125            & 
       0.125            & 
       0.125            &  
       0.125            &
       0.125            \\
       FM               & 
       Minor            & 
       -0.756           &  
       0.124            & 
       0.047            &
       -0.756           & 
       0.124            &  
       0.047            \\
       AF2              &
       Major            & 
       0.122            &   
       0.125            &  
       0.083            &
       -0.742           & 
       0.125            &   
       0.083            \\
       AF2              & 
       Minor            & 
       -0.742           &  
       0.125            & 
       0.083            & 
       0.123            & 
       0.125            &   
       0.083            \\
       \hline
       \hline
    \end{tabular}
    \label{tab:occ}
 \end{table}

 It is interesting to note that our TS does not contain enough
 information to fit reliably all the Ni($e_g$)-Ni($e_g$) interactions
 compatible with our filters.
 More precisely, we find that there are only two relevant
 $I_{\bm{a}\bm{b},\bm{a}^\prime\bm{b}^\prime}$ constants: one related
 to the self-energy of the $e_g$ states
 ($I_{\bm{a}\bm{a},\bm{a}\bm{a}}$, where $\bm{a}$ is a $e_g$-like
 basis function), and a second one quantifying the interaction between
 the two $e_g$ states in the same atom (essentially, the exchange
 interaction known as Hund's coupling).
 It is clear that our TS is not suitable to distinguish between such
 interactions. In both the FM and AF2 phases, the Ni$^{2+}$ ions
 display a $S=1$ spin configuration; yet, the interplay between
 self-interaction and Hund coupling only appears when trying to
 differentiate between the high- ($S=1$) and low- ($S=0$) spin intra-atomic
 states.

 We checked whether such an indeterminacy affects the energy
 difference between the FM and AF2 phases as obtained from the
 model. To do so, we first add a Hubbard-{\sl U} constant associated
 to the self-energy of the $e_g$ WFs, and make it equal to the
 corresponding $I$ constant (i.e., we impose
 $I_{\bm{aa}\bm{aa}}=U_{\bm{aa}\bm{aa}}$).  In this way the
 self-interaction of an electron placed in one of these orbitals is
 $U-I$~=~0, while the interaction between two electrons that only
 differ in their spin is $U+I=2I=2U$.
 Then, we vary this self-interaction parameter between 0~eV and
 6~eV, optimize the interaction between different $e_g$ orbitals to
 reproduce the bands, and calculate the FM--AF2 energy gap. We
 observe that the energy difference is quite insensitive to the value
 of $I_{\bm{aa}\bm{aa}}$, varying by less than 5\% in the explored
 range. Hence, we simply take $I_{\bm{aa}\bm{aa}}$~=~2~eV to fix the
 indeterminacy in the model.

\subsubsection{Results} 

 Magnetism in rocksalt structures is usually described using a
 Heisenberg Hamiltonian with coupling constants between first- ($J_1$)
 and second- ($J_2$) nearest
 neighbors.\cite{chatterji_prb09,hutchings_prb72,archer_prb11}
 These constants can be obtained from the energy differences between
 different spin arrangements by solving the equation system:
 \begin{alignat}{3}
 E_\text{FM}=&E_\text{ref}-6J_1&-3J_2 \nonumber\\
 E_\text{AF1}=&E_\text{ref}+2J_1&-3J_2 \\
 E_\text{AF2}=&E_\text{ref}&+3J_2 , \nonumber
 \end{alignat}
 which involves the spin arrangements of Fig.~\ref{fig:nioscheme}.
 $E_\text{ref}$ stands for the energy of a reference phase. 

 \begin{figure} [h]
    \begin{center}
       \includegraphics[width=1.0\columnwidth]{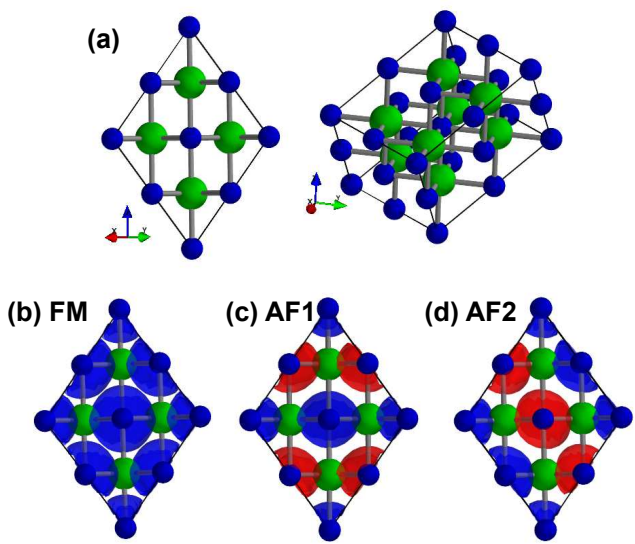}
       \caption{ (Color online) Results for the $D^I$ difference
         occupation matrix in real space.  In panel~(a) we display the
         16-ion supercell employed in the calculation along one of its
         main directions (left) and a general panoramic of the same
         cell (right).  Panels (b), (c) and (d) show the $D^I$
         distribution in real space for the FM (b), AF1 (c), and AF2
         (d) phases, respectively.  Blue and red regions correspond to
         spin-up and spin-down magnetization, respectively.}
       \label{fig:nio_supercell}
    \end{center}
 \end{figure}

 We employ our models to compute the energy of these phases, using a
 2$\times$2$\times$2 supercell containing 16 atoms, as sketched in
 Fig.~\ref{fig:nio_supercell}.
 After converging the calculations, we plot the spatial distribution
 of $D^I$ (using the spatial representation of the WFs in our basis)
 to check that the obtained solutions correctly correspond to the
 FM, AF1, and AF2 spin arrangements. These plots are shown in
 Fig.~\ref{fig:nio_supercell}, where it can be seen that the electron
 distribution in the simulations perfectly matches the spin orderings
 sketched in Fig.~\ref{fig:nioscheme}.
 We can now check the numerical results for the phase energies
 obtained with the three second-principles parameterizations [denoted,
   respectively, SP-LDAU-Ni($e_g$), SP-LDAU-Ni(3$d$), and SP-LDAU-Ni+O
   in Table~\ref{tab:nio_res}] and compare them to our full DFT+{\sl
   U} result and data in the literature (see
 e.g. Ref.~\onlinecite{archer_prb11}).

 \begin{table}[h]
    \caption{Magnetic coupling constants of NiO obtained from various
      experiments, first-principles, and second-principles
      calculations.  The latter have been modeled after the LDA+{\sl
        U} calculations and, as can be seen, converge towards the
      results obtained with this method when reducing $\delta
      \varepsilon_{ee}$.  }
    \begin{tabular}{ccc}
       \hline
       \hline
       Method               &
       $J_1$ (meV)          & 
       $J_2$ (meV)          \\
       \hline
       neutron\cite{hutchings_prb72}       &  1.4  & -19.0 \\
       neutron\cite{shanker_prb73}         & -1.4  & -17.3 \\
       HSE\cite{archer_prb11}              &  2.3  & -21.0 \\
       PSIC\cite{archer_prb11}             &  3.3  & -24.7 \\
       ASIC\cite{archer_prb11}             &  5.2  & -45.0 \\
       GGA+{\sl U}\cite{kotani_jpcm08}     &  1.7  & -19.1 \\
       LDA+{\sl U}                         &  2.6  & -17.5 \\
       SP-LDAU-Ni($e_g$)                   & -0.2  & -19.1 \\
       SP-LDAU-Ni(3$d$)                    & -0.0  & -19.1 \\
       SP-LDAU-Ni+O                        &  3.3  & -17.6 \\
       \hline
       \hline
    \end{tabular}
    \label{tab:nio_res}
 \end{table}

 We find that the coupling constants computed from our models compare
 quite well with the first-principles results. Indeed, we find the
 $J_2$, running along the 180$^{\circ}$ Ni--O--Ni bridge, to be much
 stronger than the $J_1$ coupling along the 90$^{\circ}$ Ni--O--Ni
 path.
 It is worth noting that a parametrization as simple as that of the
 SP-LDAU-Ni($e_{g}$) model captures this essential feature already. 
 Then, when we include $I$ couplings between Ni($e_g$) and
 Ni($t_{2g}$) WFs, we obtain a very similar result, with a very small
 $J_1 = -0.04$~meV. Finally, when we include electron-electron
 couplings with the oxygen orbitals, we get a value for $J_1$ that is
 very close to the first-principles result.

\subsection{Electron gas at the LaAlO$_3$/SrTiO$_3$ interface}
 \label{sec:lao_sto}

 Now we tackle the well-studied electron gas appearing at the
 interface between LaAlO$_3$ (LAO) and SrTiO$_3$ (STO).
 The origin of the 2DEG has been intensively debated in the
 literature.\cite{Hwang-12,Bristowe-14} Here we are going to consider
 an idealized defect-free interface in which the driving force for the
 2DEG is the so-called {\em polar catastrophe} that was proposed
 originally,\cite{ohtomo_nat04,nakagawa_natmat06} which arises from
 the charge discontinuity between LaAlO$_{3}$ and SrTiO$_{3}$ 
 when the bilayer is
 grown along the (001) pseudo-cubic direction of the perovskite
 lattice.\cite{stengel_prb09}
 In such a case, the occurrence of the metallic state strongly depends
 on the electrostatic boundary conditions on each side of the
 interface.\cite{stengel_prl11}
 Let us look at them in some detail to establish the basic elements of
 the calculation.

\begin{figure} [h]
   \begin{center}
          \includegraphics[width=1.0\columnwidth]{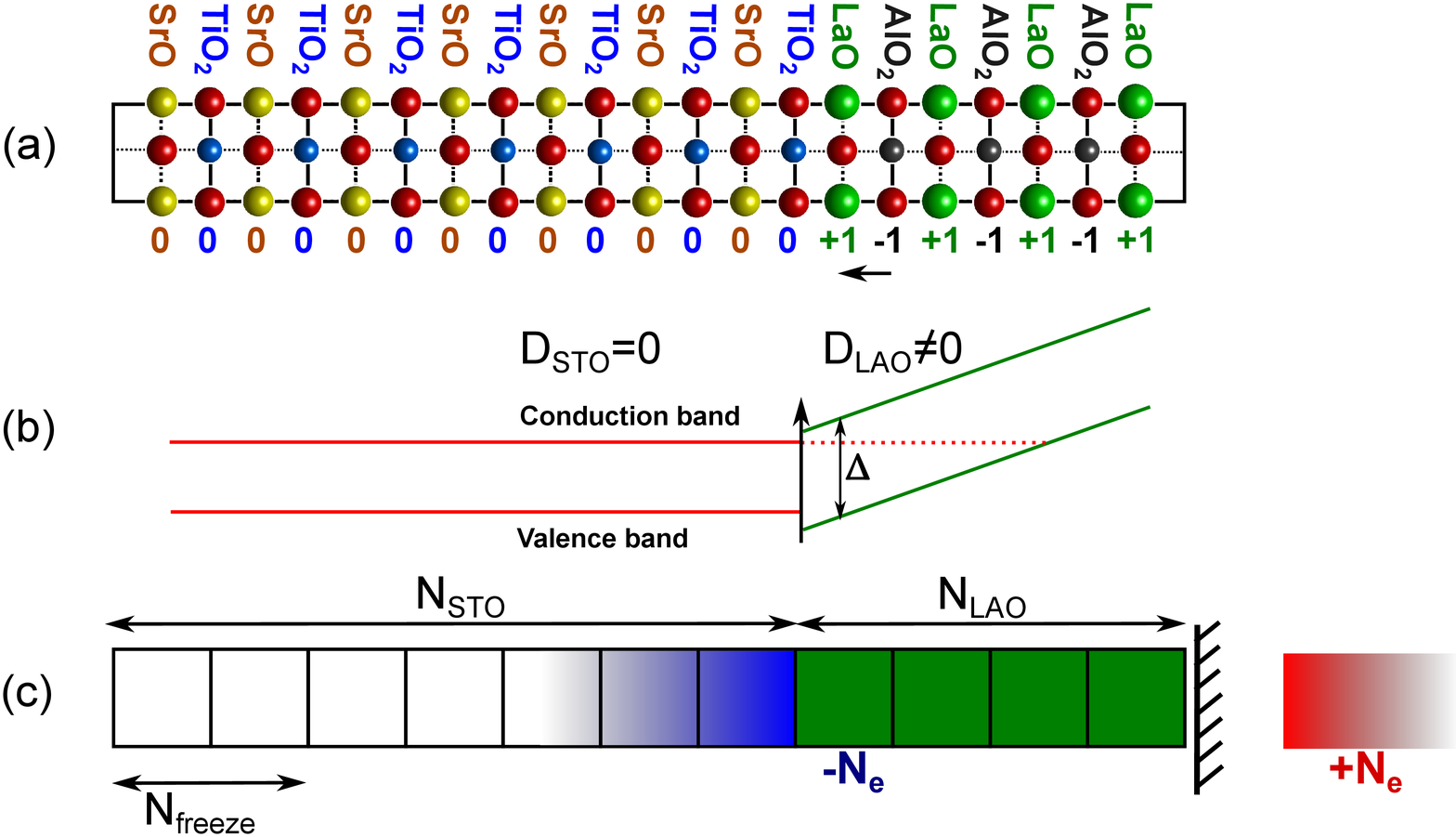}
          \caption{(Color online) Schematic representation of a polar
            (001) SrTiO$_3$/LaAlO$_3$ interface.  Atoms are
            represented by balls: O (red), Ti (blue), Sr (yellow), Al
            (black), and La (green). A free LaO-terminated surface of
            LaAlO$_3$ is assumed. Numbers below each layer indicate
            formal ionic charge.  The built-in polarization of
            SrTiO$_3$ (null) and LaAlO$_3$ is illustrated by black
            arrows.  (b) Schematic representation of the energy bands
            in the case of partial compensation of the polar
            discontinuity at the interface. $\Delta$ represents the
            LaAlO$_3$ gap.  (c) Set-up of the second-principles
            simulation for this interface. White and dark green
            squares represent, respectively, SrTiO$_3$ and LaAlO$_3$
            cells.  The metallic states in SrTiO$_3$, containing $N_e$
            electrons per cell, are represented here by a blue
            gradient. To be consistent with the electrostatic boundary
            conditions we use the charge image method, represented
            here by a positive charge distribution (red gradient) on
            the LaAlO$_3$ side.  The meaning of $N_{\text{STO}}$,
            $N_{\text{LAO}}$ and $N_\text{freeze}$ is explained in the
            text.}
       \label{fig:laosto_setup}
   \end{center}
\end{figure}

 From simple electrostatic arguments we know that
 \begin{equation}
   D_{\text{LAO}}-D_{\text{STO}}=\sigma_\text{free},
   \label{eq:laosto_boun}
 \end{equation}
 where $D_{\text{LAO}}$ and $D_{\text{STO}}$ are the normal components
 of the displacement field in LaAlO$_{3}$ and SrTiO$_{3}$, respectively, and
 $\sigma_\text{free}$ is the free charge density at the interface
 between the materials.
 Hence, depending on the particular values of $D_{\text{LAO}}$ and
 $D_{\text{STO}}$ (which can be controlled in a simulation by varying
 the charges at the open surfaces of the layers\cite{stengel_prl11}),
 a 2DEG appears at the interface according to
 Eq.~(\ref{eq:laosto_boun}).
 Figure~\ref{fig:laosto_setup}(b)
 illustrates the case for a partial compensation ($D_{\text{STO}} = 0$,
 and $D_{\text{LAO}}<-0.5$, both in units of electrons per surface
 unit cell), which correspond to the case of a partial transfer of charge 
 from the LaAlO$_3$ surface to the interface due to the crossing of 
 the top of the valence band of LaAlO$_3$ with the bottom of the conduction 
 band of SrTiO$_3$. 
 (The interface free carriers occupy states in the conduction band of
 SrTiO$_{3}$.)
 When there is no full compensation an electric field is present in the 
 LaAlO$_3$ layer.
 
 This setup is ideal to test our method, since the main physical
 effects are related to the negative doping of SrTiO$_{3}$, such a doping
 being controlled by the electrostatic boundary conditions. Further,
 the properties of the 2DEG (e.g., spatial extension) depend
 essentially on the ability of SrTiO$_{3}$ to screen these additional charges,
 which in turn involves the electron-lattice couplings in our models.

 We simulate the LaAlO$_{3}$/SrTiO$_{3}$ interface 
 [Fig.~\ref{fig:laosto_setup}(c)] by
 considering a slab of $N = N_\text{LAO} + N_\text{STO}$ 5-atom
 perovskite unit cells, where the first $N_\text{LAO}$ cells are
 occupied by LaAlO$_{3}$ and the following $N_\text{STO}$ cells by SrTiO$_{3}$.
 Following Stengel,\cite{stengel_prl11} we do not consider the
 electronic details of the interface, as these were found to be of
 little relevance to describe the main physical features of the 2DEG.
 In fact, as regards the construction of our model, we treat the
 entire slab as if it was made of SrTiO$_{3}$, but introducing the following
 modifications:
 (i) on the LaAlO$_{3}$ side, the levels of the conduction band are shifted
 up (by appropriately modifying the $\gamma_{\bm{a}\bm{a}}$
 self-energies) so that they do not interact with those on the SrTiO$_{3}$
 side, and
 (ii) to account for the large disparity between the 
 LaAlO$_{3}$ and SrTiO$_{3}$
 dielectric constants (the latter is around 25 times larger than the
 former at room temperature) we simply freeze the coordinates of the
 atoms on the LaAlO$_{3}$ layer at the RAG, to prevent atomic displacements
 from screening electric fields.

 As regards the electrostatic boundary conditions, we consider that
 $D_\text{STO}=0$ and $D_\text{LAO}=-N_e/S$, i.e., we have $N_e$
 electrons per unit area $S$ doping the slab.
 To impose such conditions, we first freeze into the centrosymmetric
 structure the atomic positions of
 $N_\text{freeze}$ unit cells at the end of the SrTiO$_3$ side of the
 slab [see Fig.~\ref{fig:laosto_setup}(c)].
 Secondly, we use the image-charge method\cite{jackson_book} 
 [see Fig.~(\ref{fig:laosto_setup})c] to introduce an electric field
 from the LaAlO$_{3}$ side of the interface consistent 
 with $D_\text{LAO}=-N_e/S$.
 
 \subsubsection{Model parameters}
 \label{sec:sto_par}
 
\begin{figure} [h]
   \begin{center}
          \includegraphics[width=1.0\columnwidth]{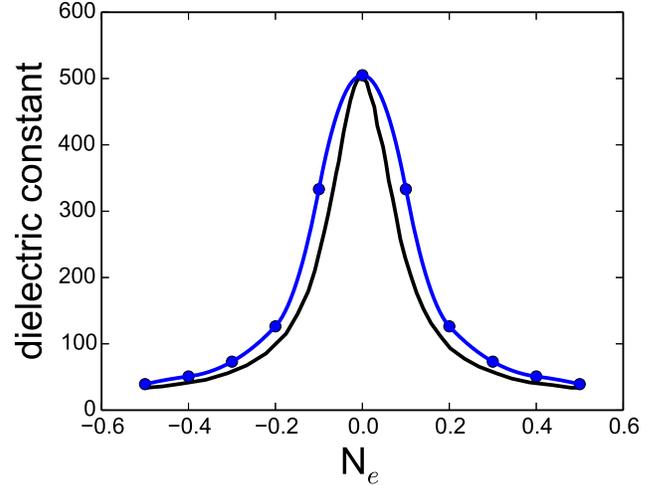}
          \caption{(Color online) Dielectric constant of SrTiO$_3$
            as calculated in LDA~\cite{stengel_prl11} (solid black line)
            and with our second-principles method (solid blue circles 
            and line) as a function of the 
            electric displacement field,
            i.e., as a function of the number of electrons per surface unit
            doping the interface.}
       \label{fig:sto_eps}
   \end{center}
\end{figure}

 We now describe how we obtain the parameters for the SrTiO$_{3}$ layer.
 As already mentioned, SrTiO$_{3}$ is a non-magnetic insulator and the RED
 corresponds to the ground state of the undoped system.
 This allows us to take the lattice potential for pure SrTiO$_{3}$ described
 in Ref.~\onlinecite{wojdel_jpcm13} as the $E^{(0)}$ term of our model
 [see Eqs.~(\ref{eq:totale}) and ~(\ref{eq:zero})], using the
 LDA-relaxed cubic phase as RAG.
 [We slightly modified the force field of
   Ref.~\onlinecite{wojdel_jpcm13}, by tuning the interaction between
   first-nearest-neighbouring Ti and O pairs, to exactly reproduce a
   dielectric constant of 500 for the cubic phase (see
   Fig.~\ref{fig:sto_eps}), as obtained from LDA calculations in
   Ref.~\onlinecite{stengel_prl11}].

 We then extended the model to include the electronic states
 associated with the bottom of the conduction band of SrTiO$_{3}$, which
 present a dominant Ti($t_{2g}$) character [see
   Fig. \ref{fig:entanglement}(a)].
 We followed the recipe in Sec.~\ref{sec:parameters} to extract the
 $\gamma$ parameters describing these bands.

 Note that our focus in this application was to capture the
 electron-lattice effects that determine the properties of the 2DEG,
 and we were not concerned with electron-electron couplings beyond the
 LDA. Thus, we did not include $U$ or $I$ terms in our model, and used
 a TS that contains the RAG and distorted structures (with individual
 atoms displaced by 0.05~\AA, 0.10~\AA, and 0.15~\AA\ from their RAG
 positions), all of which where assumed to be in the RED state.

 We then found all $\gamma$, $\vec{\bm{f}}$ and
 $\overleftrightarrow{\bm{g}}$ parameters [Eq.~(\ref{eq:gammadist})]
 compatible with the choices $\delta \varepsilon_{h}$~=~0.05~eV and
 $\delta f_{e-l}$~=~1.0~eV/\AA.
 \footnote{The first-principles calculations to obtain the
   electron-lattice constants were carried out in a 10-ion
   $\sqrt{2}\times\sqrt{2}\times\sqrt{2}$ supercell. This larger
   simulation cell is necessary to get a reasonable approximation to
   the changes in $\gamma$ due to the displacement of isolated atoms.}
 We observe that the electron-lattice constants associated to diagonal
 one-electron terms, $\gamma_{\bm{a}\bm{a}}$, are much more sensitive
 to the displacement of the ions than the off-diagonal ones.  The
 distortions that induce larger changes involve the Ti--O bond, as
 expected according to the long literature on covalency in
 ferroelectric oxides and related
 materials.\cite{jt_book,cohen_nature92,cohen_prb90,posternak_prb94}
 
 Finally, we took the Born charges and high-frequency dielectric
 tensor used in our lattice model, $E^{(0)}$, to compute the
 electrostatic energy associated to the electronic degrees of freedom.
 
 \subsubsection{Results}
 
\begin{figure*} [t]
   \begin{center}
          \includegraphics[width=1.4\columnwidth]{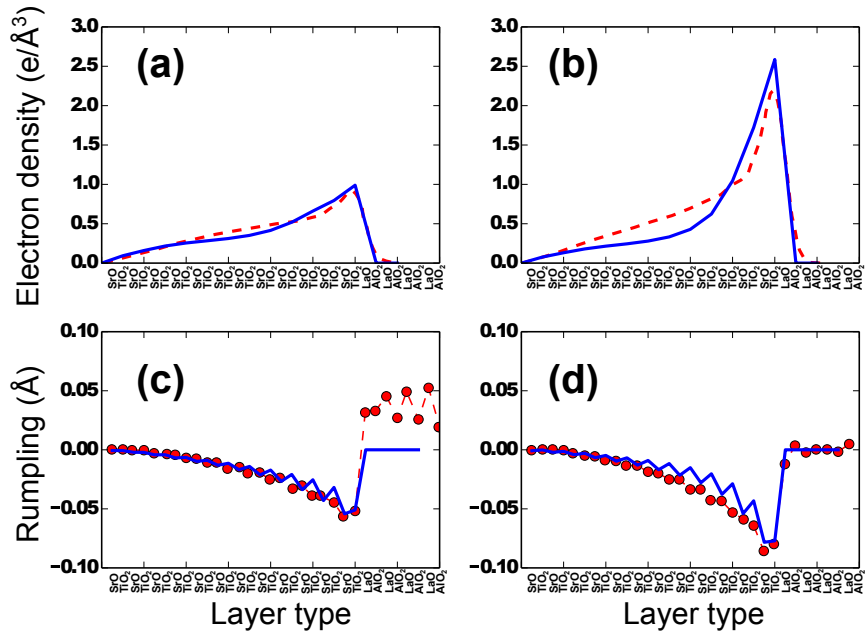}
          \caption{(Color online) Results for the 2DEG at the
            LaAlO$_3$/SrTiO$_3$ interface. The second-principles
            results are indicated with solid blue lines, while the LDA
            results are given by dashed red lines. Panels (a) and (b)
            show the electron density distribution for $N_e$=0.3, and
            $N_e$=0.5, respectively. Panels (c) and (d) show the
            rumpling of the lattice for the same two cases.  }
       \label{fig:stolao_res}
   \end{center}
\end{figure*}

 We now compare the results for the LaAlO$_{3}$/SrTiO$_{3}$ 
 interface obtained with
 the above-described model and the LDA results of
 Ref.~\onlinecite{stengel_prl11}, where the calculations were performed
 for $N_\text{LAO}$~=~5 and $N_\text{STO}$~=~12.
 Using these values, we carry out geometry optimizations with the
 constraints illustrated in Fig.~\ref{fig:laosto_setup}(c).
 The results of these calculations are shown in
 Figs.~\ref{fig:stolao_res}(a) and \ref{fig:stolao_res}(b), where we
 compare the electron densities from first-principles LDA and 
 second-principles simulations for
 $N_e$~=~0.3 and $N_e$~=~0.5, respectively.
 Moreover, in Figs.~\ref{fig:stolao_res}(c) and
 \ref{fig:stolao_res}(d) we show the obtained lattice distortions, in
 terms of the layer-by-layer rumpling, for the same cases.

 We can observe that the second-principles and LDA results match well for both
 atomic and electronic structure. Moreover, following the discussion
 in Ref.~\onlinecite{stengel_prl11}, we checked that our model
 captures correctly the influence that various physical parameters
 (e.g., linear and non-linear response of the lattice, etc.) have in
 the final result.
 As regards relatively small errors in the shape of the electronic
 density profiles, we attribute them to technical differences
 (pseudopotentials, etc.) in the LDA calculations of
 Ref.~\onlinecite{stengel_prl11} and those performed to construct our
 models.

\subsection{Additional considerations}
\label{sec:additional}

 To finish this section we would like to give some estimations 
 of the computer time required to carry out the second-principles
 calculations for the systems discussed in the present work.
 We will focus first on NiO, as we carried out both the DFT and
 second-principles computations for the same unit cells and on the
 same computational platform, so a reliable comparison of timings
 should be feasible.
 In Table~\ref{tab:nio_times} we show the time necessary to perform a
 single-point calculation using a single CPU, using the same reciprocal
 space sampling in the LDA+U and second-principles simulations.
 The values for 4- and 16-ion cells clearly show the very large
 speed-up of our second-principles simulations when compared to
 standard DFT even in small cells.
 Looking at Table~\ref{tab:nio_times} we see that there are very small
 differences between the timing results between the different
 parameterizations associated to different levels of description of
 the electron-electron interactions in NiO.
 Finally, in order to give an estimation of the scaling of the method
 as it stands now (i.e., at an early stage of implementation), we
 carried out a calculation of a 10$\times$10$\times$10 periodic
 supercell that contains 2000 atoms.
 This is approximately the size limit for single-point DFT
 calculations, and it would require significant computational
 resources and a highly parallelized code.
 However, this simulation at the SP-LDAU-Ni($e_g$) level took 6.6
 hours of a single CPU, suggesting that calculations including
 tens of thousands of atoms are within reach using our
 models.

 \begin{table}[h]
    \caption{ Simulation running times on a single CPU for
              the 4-ion 1$\times$1$\times$2, 16-ion 2$\times$2$\times$2
              and 2000-ion 10$\times$10$\times$10 NiO supercells.
              The lower time obtained for the SP-LDAU-Ni(3$d$) 
              with respect to SP-LDAU-Ni($e_g$) is due to the smaller
              number of self-consistent steps required in the former 
              simulation.}
    \begin{tabular}{cccc}
       \hline
       \hline
       Method            & 4 ion (s) & 16 ion (s) & 2000 ion (hours)\\
       \hline
       LDA+U             &  65.0    & 3516.8     &  ---       \\
       SP-LDAU-Ni($e_g$) &   1.4    &   14.5     &  6.63      \\
       SP-LDAU-Ni(3$d$)  &   1.4    &   14.1     &  6.59      \\
       SP-LDAU-Ni+O      &   1.5    &   15.8     &  6.97      \\
       \hline
       \hline
    \end{tabular}
    \label{tab:nio_times}
 \end{table}
 
 Turning now to the simulation of the LaAlO$_{3}$/SrTiO$_{3}$
 interface we note that each of the geometry optimizations involving a
 85-atom supercell took about 13 minutes using a single desktop CPU.
 
\section{Concluding remarks}
\label{sec:end}
 
 In this work we have presented a first-principles-based multi-scale
 method, which we denominate {\em second-principles}, that makes it
 possible to compute the properties of materials at an atomic level,
 with an accuracy essentially equal to DFT, and at a very reduced
 computational cost.

 Our approach is based on dividing the electron density of the system
 into a reference part, usually corresponding to its neutral ground
 state at any geometry, and a deformation part, defined as the
 difference between the actual and reference densities.
 We take advantage of the fact that the largest part of the
 system's energy depends on the reference density, and can be
 efficiently and accurately described by a force field with no
 explicit consideration of the electrons. Then, the effects associated
 to the difference density can be treated perturbatively with good
 precision by working in the Wannier function basis corresponding to
 the reference state. Further, the electronic description can be
 restricted to the bands of interest, which renders a computationally
 very efficient scheme.

 Conceptually, the present approach constitutes a fresh look at the
 problem of how to describe lattice and electronic degrees of freedom
 simultaneously and effectively, introducing a convenient partition of
 the energy that permits an accurate treatment of both types of
 variables and their mutual interactions. In our view, our method
 constitutes a significant step beyond the usual techniques -- ranging
 from molecular-mechanics to tight-binding and
 quantum-mechanics/molecular-mechanics schemes -- towards a more
 unified model.

 As illustrated by the examples described here, the present approach
 allows us to obtain DFT-like accuracy in the analysis of subtle
 physical effects, like those determining the relative stability of
 the magnetic phases of NiO, or those involved in the structural
 relaxations and screening processes associated to the two-dimensional
 electron gas formed at the interface of LaAlO$_3$ and SrTiO$_3$.
 Note that these problems -- which involve electron correlation
 effects, transition-metal ions, etc. -- are usually hard to treat
 within DFTB schemes.\cite{zheng_jctc07}
 
 As currently formulated, our approach has only one essential
 limitation: it is restricted to systems in which it is possible to
 (loosely) define an underlying bonding topology that is to be
 preserved. Hence, while the method allows the system undergo
 significant structural modifications -- e.g., like those involved in
 typical ferroelectric or ferroelastic phase transitions --, it is not
 possible to study full-blown bond breaking directly with it.
 Nevertheless, this limitation can be overcome by using our method in
 multi-scale simulations that permit a more detailed treatment (e.g.,
 with DFT) of the regions of the material in which the
 constant-topology condition is not satisfied.

 It is also important to note that the constant-topology condition is
 perfectly compatible with the study of many structurally non-trivial
 cases, such as nanostructured materials, surfaces,
 chemically-disordered solid solutions, coexistence of different
 structural and electronic phases, etc.
 Hence, the application scope of our scheme is enormous.

 Let us also note that the present method can be extended to cover
 physical effects not mentioned here. For example, it is possible to
 expand it to treat relativistic phenomena (as spin-orbit effects) or
 time-dependent non-equilibrium situations (as resulting from the
 interaction with light) in essentially the same way as the initial
 DFT implementations were extended to do so (by implementing a
 non-colinear treatment of magnetism, time-dependent DFT, etc.).
 Further, since our method provides us with a Hamiltonian for the
 interacting system, one could imagine solving the electronic problem
 in ways that go beyond the mean-field approach adopted here, and thus
 better account for many body effects.
 
 The ability to simulate systems with thousands of atoms, treating
 both lattice and electrons accurately, may permit for the first time
 predictive investigations of a variety of intriguing phenomena --
 e.g., Mott transitions, coupled spin-lattice dynamics, charged and
 conducting domain walls, polaron transport, etc. -- in realistic
 conditions of temperature, applied fields, etc.
 We thus believe that the present method has the potential to
 significantly advance our understanding of some of today's most
 interesting problems in condensed-matter physics and material
 science.

\begin{acknowledgements}
   We thank M. Moreno and J. A. Aramburu for useful discussions.  PGF
   and JJ acknowledge finantial support from the Spanish Ministery of
   Economy and Competitiveness through the MINECO Grant
   No. FIS2012-37549-C05-04.
   PGF also acknowledges funding from the Ram\'on y Cajal Fellowship
   RYC-2013-12515.
   JI is funded by MINECO-Spain Grant MAT2013-40581-P and FNR
   Luxembourg Grant FNR/P12/4853155/Kreisel.

\end{acknowledgements}

\appendix

\section{Expanding the total energy with respect to the deformation density }
\label{app:a}

 The total electronic density of the system can be written as the sum
 of a reference electronic density, $n_{0} (\vec{r})$, 
 and a small deformation electron density,
 $\delta n(\vec{r})$, defined as in Eq.~(\ref{eq:dist_n}).
 Inserting this into the expression for the total DFT energy, 
 Eq.~(\ref{eq:dften}), and considering also the expansion of 
 the exchange and correlation energy upto second order in the
 deformation energy, Eq.~(\ref{eq:expxc}), then
 \begin{align}
    E_\text{DFT} \approx & \sum_{j \vec{k}} o_{j \vec{k}}
                     \left\langle \psi_{j \vec{k}}
                     \right\vert \hat{t} + v_{\text{ext}} 
                     \left\vert \psi_{j\vec{k}}\right\rangle +
    \nonumber \\
                   & + \frac{1}{2}\iint\frac{n_{0}(\vec{r})
                                           n_{0}(\pvec{r})}
                                          {\vert \vec{r}-\pvec{r} \vert}
                          d^3rd^3r^\prime 
    \nonumber \\
                   & +  \frac{1}{2}\iint\frac{\delta n(\vec{r})
                                           \delta n(\pvec{r})}
                                          {\vert \vec{r}-\pvec{r} \vert}
                          d^3rd^3r^\prime 
    \nonumber \\
                   & + \iint\frac{n_{0}(\pvec{r})
                                \delta n(\vec{r})}
                               {\vert \vec{r}-\pvec{r} \vert}
                          d^3rd^3r^\prime + 
                     E_\text{xc}[n_{0}] 
    \nonumber \\
                   & + \int\left.\frac{\delta E_\text{xc}}
                                      {\delta n(\vec{r})}\right\vert_{n_0}
                     \delta n(\vec{r})d^3r 
    \nonumber \\
                   & + \frac{1}{2}\iint\left.\frac{\delta^2 E_\text{xc}}
                                    {\delta n(\vec{r}) \delta n(\pvec{r})}
                                    \right\vert_{n_0}
                     \delta n(\vec{r}) \delta n(\pvec{r})d^3rd^3r^\prime  
    \nonumber \\
                   & +  E_{\text{nn}}.
 \end{align}

 \noindent Adding and substracting $\sum_{j \vec{k}} o^{(0)}_{j \vec{k}}
                     \langle \psi^{(0)}_{j \vec{k}}
                     \vert \hat{t} + v_{\text{ext}}
                     \vert \psi^{(0)}_{j\vec{k}}\rangle$,
 where $\vert \psi^{(0)}_{j\vec{k}}\rangle$ and $o^{(0)}_{j \vec{k}}$
 are the eigenvectors and the occupation numbers for the $j$-band of the
 wavevector $\vec{k}$ that define the reference electron density
 for a given atomic configuration, and grouping together 
 the terms at zero, first and second order in the deformation 
 electronic density, we can write

 \begin{equation}
    E_\text{DFT} \approx E^{(0)} + E^{(1)} + E^{(2)}. 
 \end{equation}

 The zero order term $E^{(0)}$, amounts to 

 \begin{align}
    E^{(0)} = &  \sum_{j \vec{k}} o^{(0)}_{j \vec{k}}
                    \left\langle \psi^{(0)}_{j \vec{k}}
                    \right\vert \hat{t} + v_{\text{ext}}
                    \left\vert \psi^{(0)}_{j\vec{k}}\right\rangle 
    \nonumber \\
                   & + \frac{1}{2}\iint\frac{n_{0}(\vec{r})
                                           n_{0}(\pvec{r})}
                                          {\vert \vec{r}-\pvec{r} \vert}
                          d^3rd^3r^\prime 
    \nonumber \\
                   & + E_{\text{xc}} [n_{0}] + E_{\text{nn}}.
    \label{eq:e0}
 \end{align}

 The first order term $E^{(1)}$ takes the following expression

 \begin{widetext}
 \begin{align}
    E^{(1)} = &  \sum_{j \vec{k}} o_{j \vec{k}}
                    \left\langle \psi_{j \vec{k}}
                    \right\vert \hat{t} + v_{\text{ext}}
                    \left\vert \psi_{j\vec{k}}\right\rangle -
               \sum_{j \vec{k}} o^{(0)}_{j \vec{k}}
                    \left\langle \psi^{(0)}_{j \vec{k}}
                    \right\vert \hat{t} + v_{\text{ext}}
                    \left\vert \psi^{(0)}_{j\vec{k}}\right\rangle 
                     + \iint\frac{n_{0}(\pvec{r})
                                \delta n(\vec{r})}
                               {\vert \vec{r}-\pvec{r} \vert}
                          d^3rd^3r^\prime 
                     + \int\left.\frac{\delta E_\text{xc}}
                                      {\delta n(\vec{r})}\right\vert_{n_0}
                     \delta n(\vec{r})d^3r 
    \nonumber \\
          = &  \sum_{j \vec{k}} o_{j \vec{k}}
                    \left\langle \psi_{j \vec{k}}
                    \right\vert \hat{t} + v_{\text{ext}}
                    \left\vert \psi_{j\vec{k}}\right\rangle -
               \sum_{j \vec{k}} o^{(0)}_{j \vec{k}}
                    \left\langle \psi^{(0)}_{j \vec{k}}
                    \right\vert \hat{t} + v_{\text{ext}}
                    \left\vert \psi^{(0)}_{j\vec{k}}\right\rangle 
                     - \int v_{\text{H}} (n_{0}; \vec{r}) 
                                \delta n(\vec{r}) d^3r
                     + \int v_{\text{xc}} [ n_{0}; \vec{r} ]
                     \delta n(\vec{r}) d^3r 
    \nonumber \\
          = &  \sum_{j \vec{k}} o_{j \vec{k}}
                    \left\langle \psi_{j \vec{k}}
                    \right\vert \hat{t} + v_{\text{ext}}
                    \left\vert \psi_{j\vec{k}}\right\rangle -
               \sum_{j \vec{k}} o^{(0)}_{j \vec{k}}
                    \left\langle \psi^{(0)}_{j \vec{k}}
                    \right\vert \hat{t} + v_{\text{ext}}
                    \left\vert \psi^{(0)}_{j\vec{k}}\right\rangle 
                     + \int 
                       \left[ -v_{\text{H}} (n_{0}; \vec{r}) + 
                              v_{\text{xc}} [n_{0} ; \vec{r}] \right]
                              \left( n(\vec{r}) - n_{0}(\vec{r}) \right) d^3r
    \nonumber \\
          = &  \sum_{j \vec{k}} o_{j \vec{k}}
                    \left\langle \psi_{j \vec{k}}
                    \right\vert \hat{t} + v_{\text{ext}}
                    \left\vert \psi_{j\vec{k}}\right\rangle -
               \sum_{j \vec{k}} o^{(0)}_{j \vec{k}}
                    \left\langle \psi^{(0)}_{j \vec{k}}
                    \right\vert \hat{t} + v_{\text{ext}}
                    \left\vert \psi^{(0)}_{j\vec{k}}\right\rangle 
    \nonumber \\
                   & + \sum_{j \vec{k}} \int 
                       o_{j \vec{k}} \left\langle \psi_{j \vec{k}}
                    \right\vert
                       -v_{\text{H}} (n_{0};\vec{r})  + 
                       v_{\text{xc}} [n_{0};\vec{r}] 
                    \left\vert \psi_{j\vec{k}}\right\rangle d^3r
                    - \sum_{j \vec{k}} \int 
                       o^{(0)}_{j \vec{k}} \left\langle \psi^{(0)}_{j \vec{k}}
                    \right\vert
                       -v_{\text{H}} (n_{0};\vec{r}) + 
                       v_{\text{xc}} [n_{0};\vec{r}] 
                    \left\vert \psi^{(0)}_{j\vec{k}}\right\rangle d^3r
    \nonumber \\
          = &  \sum_{j \vec{k}} o_{j \vec{k}}
                    \left\langle \psi_{j \vec{k}}
                    \right\vert \hat{t} + v_{\text{ext}} 
                    -v_{\text{H}} (n_{0};\vec{r}) +
                    v_{\text{xc}} [n_{0};\vec{r}]
                    \left\vert \psi_{j\vec{k}}\right\rangle  
                    - \sum_{j \vec{k}} \int o^{(0)}_{j \vec{k}}
                    \left\langle \psi^{(0)}_{j \vec{k}}
                    \right\vert \hat{t} + v_{\text{ext}} 
                    -v_{\text{H}} (n_{0};\vec{r}) +
                    v_{\text{xc}} [n_{0};\vec{r}]
                    \left\vert \psi^{(0)}_{j\vec{k}}\right\rangle 
    \nonumber \\
          = &  \sum_{j \vec{k}} o_{j \vec{k}}
                    \left\langle \psi_{j \vec{k}}
                    \right\vert \hat{h}_{0} 
                    \left\vert \psi_{j\vec{k}}\right\rangle 
                    - \sum_{j \vec{k}} \int o^{(0)}_{j \vec{k}}
                    \left\langle \psi^{(0)}_{j \vec{k}}
                    \right\vert \hat{h}_{0} 
                    \left\vert \psi^{(0)}_{j\vec{k}}\right\rangle ,
    \label{eq:e1}
 \end{align}
 \end{widetext}
 
 \noindent where we have defined the Kohn-Sham one electron hamiltonian defined
 for the reference state density $n_{0}$ as 

 \begin{equation}
    \hat{h}_{0} = \hat{t} + v_{\text{ext}} - v_{\text{H}} (n_{0};\vec{r}) + 
                  v_{\text{xc}} [n_{0};\vec{r}],
 \end{equation}
 
 \noindent with the reference Hartree potential 

 \begin{equation}
    v_{\text{H}}(n_{0};\vec{r}) = - \int \frac{n_{0} (\pvec{r})}
               {\vert \vec{r}-\pvec{r} \vert} d^3r^\prime,
 \end{equation}

 \noindent and the exchange and correlation potential 

 \begin{equation}
    v_{\text{xc}}[n_{0};\vec{r}] =  
       \left. \frac{\delta E_{\text{xc}}[n]} {\delta n (\vec{r}) }
                            \right\vert_{n_0} .
 \end{equation}

 The second order term $E^{(2)}$ can be written as
 
 \begin{align}
    E^{(2)} = &  \frac{1}{2}\iint\frac{\delta n(\vec{r})
                                     \delta n(\pvec{r})}
                                    {\vert \vec{r}-\pvec{r} \vert}
                          d^3rd^3r^\prime 
    \nonumber \\
                   & + \frac{1}{2}\iint\left.\frac{\delta^2 E_\text{xc}}
                                    {\delta n(\vec{r}) \delta n(\pvec{r})}
                                    \right\vert_{n_0}
                     \delta n(\vec{r}) \delta n(\pvec{r})d^3rd^3r^\prime  
    \label{eq:e2}
 \end{align}

\section{Inverse transformation from the Wannier to Bloch functions}
\label{app:b}

 We start from the definition of the Wannier functions from an isolated
 manifold of $J$ Bloch orbitals, Eq.~(\ref{eq:defWannier}),
 
 \begin{align}
    \vert \chi_{\bm{a}} \rangle \equiv \vert \vec{R}_{A} a \rangle = 
        \frac{V}{2\pi^{3}} \int_{\rm BZ} d \vec{k} \:\:
        e^{-i \vec{k} \cdot \vec{R}_{A}}
        \sum_{m=1}^{J} T_{ma}^{(\vec{k})} \vert \psi_{m \vec{k}}^{(0)} \rangle,
 \end{align}

 \noindent where the $\bm{T}^{(\vec{k})}$ matrices 
 are the unitary transformations
 that minimize the localization functional.
 The Bloch orbitals are those that define the reference density for a
 given atomic geometry, i.e., those used to construct the Wannier functions.

 Then, multiplying both sides of the equation by 
 $e^{i \vec{k}^{\prime} \cdot \vec{R}_{A}}$ and summing over all the
 lattice sites in real space

 \begin{align}
    & \sum_{\vec{R}_{A}} e^{i \vec{k}^{\prime} \cdot \vec{R}_{A}} 
        \vert \vec{R}_{A} a \rangle =
    \nonumber \\ 
    &  =  \frac{V}{(2\pi)^{3}} \int_{\rm BZ} d \vec{k} \:\:
        \left[ \sum_{\vec{R}_{A}}
        e^{i \left( \vec{k}^{\prime} - \vec{k} \right) \cdot \vec{R}_{A}}\right]
        \sum_{m=1}^{J} T_{ma}^{(\vec{k})} \vert \psi_{m \vec{k}}^{(0)} \rangle
    \nonumber \\ 
        & = \int_{\rm BZ} d \vec{k} \:\:
        \delta(\vec{k}^{\prime}-\vec{k})
        \sum_{m=1}^{J} T_{ma}^{(\vec{k})} \vert \psi_{m \vec{k}}^{(0)} \rangle
    \nonumber \\ 
        & = \sum_{m=1}^{J} 
            \vert \psi_{m \vec{k}^{\prime}}^{(0)} \rangle
            T_{ma}^{(\vec{k}^{\prime})}.
    \label{eq:partialinv}
 \end{align}

 \noindent Finally, multiplying both sides of Eq.~(\ref{eq:partialinv})  
 by the inverse matrix of the unitary transformation

 \begin{align}
    & \sum_{a=1}^{J} \sum_{\vec{R}_{A}} e^{i \vec{k}^{\prime} \cdot \vec{R}_{A}}
        \vert \vec{R}_{A} a \rangle 
        \left( T_{al}^{(\vec{k}^{\prime})}\right)^{-1} = 
    \nonumber \\
        & = \sum_{m=1}^{J} 
            \vert \psi_{m \vec{k}^{\prime}}^{(0)} \rangle
            \sum_{a=1}^{J}
                   T_{ma}^{(\vec{k}^{\prime})} 
                   \left( T_{al}^{(\vec{k}^{\prime})}\right)^{-1}
    \nonumber \\
        & = \sum_{m=1}^{J} 
            \vert \psi_{m \vec{k}^{\prime}}^{(0)} \rangle
            \delta_{ml} 
    \nonumber \\
        & = \vert \psi_{l \vec{k}^{\prime}}^{(0)} \rangle.
 \end{align}

 \noindent This expression can be rewritten as

 \begin{align}
    \vert \psi_{l \vec{k}}^{(0)} \rangle &  
     = \sum_{\bm a} \left( T_{al}^{(\vec{k})} \right)^{-1}
                 e^{i \vec{k} \cdot \vec{R}_{A}}
                 \vert \vec{R}_{A} a \rangle
    \noindent \\
    & = \sum_{{\bm a}} c_{l a \vec{k}}^{(0)} \:\: 
          e^{i \vec{k} \cdot \vec{R}_{A}} 
          \vert  \vec{R}_{A} a \rangle,
 \end{align}

 \noindent that is the same as Eq.~(\ref{eq:defBloch}) with the role of the
 coefficients of the expansion played by the elements of the inverse matrix
 of the unitary transformation.

\end{document}